\documentclass[twocolumn,superscriptaddress,amsmath, amssymb, amsfonts,preprintnumbers,aps,prd,longbibliography,nofootinbib]{revtex4-1}

\renewcommand{\sec}[1]{\textit{#1. --- }}

\usepackage{graphicx}%
\usepackage{dcolumn}%
\usepackage{bm}%
\usepackage{amstext}
\usepackage{amssymb}
\usepackage{subcaption}
\usepackage{ragged2e}
\DeclareCaptionJustification{justified}{\justifying}
\usepackage{amsmath}
\usepackage{graphicx}
\usepackage{color}
\usepackage{bbold}
\usepackage{delimset} %
\usepackage[colorlinks=true]{hyperref}
\usepackage{orcidlink}

\usetikzlibrary{decorations.pathmorphing}
\usetikzlibrary{decorations.markings}
\usetikzlibrary{positioning, shapes, snakes, arrows}
 \usepackage{tikz-feynman}

\tikzset{ 
	graviton/.style={line width=.8pt, -latex,decorate, decoration={snake, segment length=4pt,amplitude=1.8pt, pre length=.1cm, post length=.25cm}},
	worldline/.style={gray, line width=1pt},
	worldlineBold/.style={black, line width=.6pt},
        background/.style={black,dotted,line width=1pt},
	zUndirected/.style={line width=1pt},
	zParticle/.style={line width=1pt,postaction={decorate},decoration={markings,mark=at position .6 with {\arrow[#1]{latex}}}},
	zParticleF/.style={line width=1pt,postaction={decorate}},
	cscalar/.style={line width=1pt,postaction={decorate},decoration={markings,mark=at position .6 with {\arrow[#1]{latex}}}},
	cscalar2/.style={line width=1pt,postaction={decorate},decoration={markings,mark=at position .8 with {\arrow[#1]{latex}}}},
	photon/.style={line width =.8pt, decorate, decoration={snake, segment length=3pt, amplitude=1.8pt,  pre length=.1cm, post length=.1cm}},
	 mid arrow/.style={postaction={decorate,decoration={
        markings,
        mark=at position .5 with {\arrow[#1]{latex}}}}} ,
        worlddot/.style={dotted, line width=.8pt},
	worlddot2/.style={dotted, line width=1pt}   }

\DeclareFontFamily{OT1}{pzc}{}
\DeclareFontShape{OT1}{pzc}{m}{it}{<-> s * [1.350] pzcmi7t}{}
\DeclareMathAlphabet{\mathpzc}{OT1}{pzc}{m}{it}

\AtBeginDocument{%
  \setlength{\abovedisplayskip}{6pt plus 2pt minus 2pt}%
  \setlength{\belowdisplayskip}{6pt plus 2pt minus 2pt}%
  \setlength{\abovedisplayshortskip}{2pt plus 1pt minus 1pt}%
  \setlength{\belowdisplayshortskip}{4pt plus 1pt minus 1pt}%
}
\def\cO{\mathcal{O}}

\def\eps{\epsilon}

\def\d{\mathrm{d}}
\def\D{\mathrm{D}}
\def\eE{\mathrm{E}}
\def\eK{\mathrm{K}}

\def\dd{\delta\!\!\!{}^-\!}

\def\d{\mathrm{d}}
\def\eps{\epsilon}

\def\nn{\nonumber}

\def\eqn#1{eq.~\eqref{#1}}

\def\Eqn#1{Eq.~\eqref{#1}}

\def\Fig#1{Fig.~{\ref{#1}}}

\def\App#1{Appendix~{\ref{#1}}}
\def\rcite#1{ref.~\cite{#1}}
\def\rcites#1{refs.~\cite{#1}}

\newcommand{\be}{\begin{equation}}
\newcommand{\ee}{\end{equation}}
\newcommand{\ba}{\begin{align}}
\newcommand{\ea}{\end{align}}
\ifx\genfrac\sdflkaj\else\fi
\newcommand{\sfrac}[2]{{\textstyle\frac{#1}{#2}}}

\newcommand*{\vct}[1]{\boldsymbol{#1}}

\usetikzlibrary{calc}
\newcommand{\combSkeleton}{%
  \path (-1.2,0) (4.2,0); 
  \foreach \n/\x in {1/0,2/1.5,3/3}{
    \coordinate (A\n) at (\x,1.7);
    \coordinate (B\n) at (\x,0);
  }
  \draw[dotted] (-0.7,0) -- (3.7,0);
  \draw[dotted] (-0.7,1.7) -- (A1);
  \draw[dotted] (A3) -- (3.7,1.7);
  \foreach \n in {1,2,3}{
    \draw[photon] (B\n) -- (A\n);
    \node[left] at ($(B\n)!.5!(A\n)$) {\small $\ell_{\n}\!\!\uparrow$};
  }
}
\newcommand{\combDots}{%
  \foreach \n in {1,2,3}{
    \draw[fill] (A\n) circle (.08);
    \draw[fill] (B\n) circle (.08);
  }
}
\newcommand{\combExt}[1]{%
  \draw[zParticle={}] (#1) -- ++(0,0.9) node[above]{\small $z_1^\mu(\omega\!\to\!0)$};
}

\begin{document}

\preprint{HU-EP-26/36-RTG}

\title{Soft Factorisation of WQFT and the Spinning 4PM Magnusian}

\author{Riccardo Gonzo\,\orcidlink{0000-0001-7285-6295}}
\affiliation{%
School of Mathematical Sciences and STAG Research Centre, 
University of Southampton, 
Southampton SO17 1BJ, United Kingdom
}
\affiliation{%
School of Mathematical Sciences,
Queen Mary University of London,
London E1 4NS, United Kingdom
}
\author{Gustav Mogull\,\orcidlink{0000-0003-3070-5717}}
\affiliation{%
School of Mathematical Sciences,
Queen Mary University of London,
London E1 4NS, United Kingdom
}
\affiliation{%
Institut f\"ur Physik,
Humboldt-Universit\"at zu Berlin,
10099 Berlin, Germany
}
\affiliation{Max Planck Institut f\"ur Gravitationsphysik,
Am M\"uhlenberg 1, 14476 Potsdam, Germany}

\author{Benjamin Sauer\,\orcidlink{0000-0002-2071-257X}}
\affiliation{%
Institut f\"ur Physik,
Humboldt-Universit\"at zu Berlin,
10099 Berlin, Germany
}
\affiliation{%
Physik Department, James-Franck-Stra{\ss}e 1, Technische Universit\"at M\"unchen,
D--85748 Garching, Germany
}

\begin{abstract}
We derive the soft divergences of $\hat{N}=-i\log\hat{S}$
in the Worldline Quantum Field Theory (WQFT) formalism,
focusing on poles due to single massive-massive (Coulomb) and massive-massless (tail) graviton exchanges.
Our derivation is inspired by the Faddeev-Kulish procedure,
switching the interaction picture to one with an infrared-finite $\hat{N}$-matrix.
Soft divergences emerging from the worldline are correlators of Wilson lines.
As confirmation of this structure,
we demonstrate explicitly how divergences of inclusive observables including
the momentum impulse $\Delta p_i^\mu$ and spin kick $\Delta S_i^\mu$ cancel,
and we reproduce the infrared divergence of the one-loop scattering waveform.
Using the Magnus series representation of $\hat{N}$ and Murua coefficients,
we then calculate the Magnusian $N=\langle0|\hat{N}|0\rangle$ at fourth post-Minkowskian (4PM) order,
including linear-in-spin corrections. For aligned spins, $-\partial N/\partial L$ reproduces the full scattering angle through 3PM order;
at 4PM order, $N$ develops a soft divergence as predicted by our soft factorisation.
Using the Feynman $i0^+$ prescription, we also derive an infrared-finite 4PM Magnusian
which compactly encodes existing results for the conservative momentum impulse 
and spin kick via nested iterations of Poisson brackets.
\end{abstract}
 
\maketitle 

\section{Introduction}

Since the first discovery of gravitational waves produced by a binary
black hole merger event~\cite{LIGOScientific:2016aoc,KAGRA:2021vkt},
the gravitational two-body problem has taken on paramount importance in theoretical physics.
With new gravitational wave detectors due to go online
in the next decade~\cite{LISA:2017pwj, Punturo:2010zz, Ballmer:2022uxx},
we must sharpen our theoretical tools to provide the best predictions possible.
This means harnessing theoretical data from all different perturbative regimes:
post-Newtonian (PN)~\cite{Blanchet:2013haa, Porto:2016pyg, Levi:2018nxp},
post-Minkowskian (PM)~\cite{Kosower:2022yvp, Bjerrum-Bohr:2022blt, Buonanno:2022pgc}
and gravitational self-force (SF)~\cite{Mino:1996nk, Poisson:2011nh, Barack:2018yvs, Gralla:2021qaf}.
Modern techniques involving quantum field theory (QFT), 
and more specifically effective field theory (EFT) for
compact bodies~\cite{Goldberger:2004jt,Kalin:2020mvi,Liu:2021zxr,Aoude:2020onz},
now play a crucial role,
involving both scattering amplitudes and worldlines.

In this context, there has been a broad effort to study two-body scattering problems,
wherein the PM expansion in powers of Newton's constant $G$ is natural.
An advantage of looking at scattering, rather than immediately pursuing bound orbits,
is the emphasis on gauge-invariant asymptotic observables,
defined as differences between a quantity at future and past infinity.
In a QFT-based setup, these observables can be interpreted as in-in expectation values
\cite{Schwinger:1960qe,Keldysh:1964ud,Galley:2009px,
Kosower:2018adc,Cristofoli:2021vyo,Kalin:2022hph,Jakobsen:2022psy}.
Observables computed so far include the momentum impulse $\Delta p_i^\mu$,
from which one may also determine the scattering angle $\theta$
and total radiated momentum $P_{\rm rad}^\mu$ due to gravitational wave emission.
For non-spinning particles, $\Delta p_i^\mu$
has been computed up to 5PM order at first-order SF~\cite{Bern:2019nnu,Bern:2021dqo,Bern:2021yeh,Dlapa:2021npj,Dlapa:2021vgp,Dlapa:2022lmu,Dlapa:2023hsl,Driesse:2024xad,Driesse:2024feo},
plus conservative 2SF contributions~\cite{Bern:2025wyd,Driesse:2026qiz};
with spins, together with the change in spin vectors (spin kick) $\Delta S_i^\mu$,
up to 4PM~\cite{Jakobsen:2022fcj,Jakobsen:2022zsx,Jakobsen:2023ndj,Jakobsen:2023hig}.
There is also the radiated angular momentum
$J_{\rm rad}^\mu$~\cite{Manohar:2022dea,DiVecchia:2022piu,Mogull:2025cfn,Jakobsen:2026ayw},
which was recently extended to 4PM order in the non-spinning case~\cite{Georgoudis:2026zzn,Bohnenblust:2026ytj}.
The gravitational waveform has been calculated up to $\cO(G^3)$ 
\cite{Jakobsen:2021smu,Jakobsen:2021lvp,Mougiakakos:2021ckm,Riva:2022fru,DeAngelis:2023lvf,Brandhuber:2023hhl,Georgoudis:2023lgf,Georgoudis:2023ozp,Brandhuber:2023hhy,Herderschee:2023fxh,Bohnenblust:2023qmy,Bohnenblust:2025gir}.
 
Given our desire to ultimately map to bound orbits~\cite{Kalin:2019rwq,Kalin:2019inp,Cho:2021arx,Saketh:2021sri},
an important question is which physical quantities
most conveniently encode the scattering data.
Seeing as we use QFT,
the $\hat{S}$-matrix is an obvious choice:
in-in quantities, which involve two copies of $\hat{S}$,
are more akin to cross-sections.
Yet the appearance of superclassical contributions
implies that conventional scattering amplitudes ---
matrix elements of $\hat{S}$ between specific in- and out-states ---
are not well-suited to the classical $\hbar\to0$ limit.
A better choice is the eikonal phase~\cite{DiVecchia:2021bdo,
Cristofoli:2021jas,DiVecchia:2022nna,DiVecchia:2022piu,DiVecchia:2023frv},
which resums these superclassical terms.
Seeing as the eikonal phase is an ``in-out'' quantity,
it depends on variables defined at the midpoint of scattering.

More recently, the operator logarithm $\hat{N}=-i\log\hat{S}$
has come to prominence.
The $\hat{N}$-matrix is represented perturbatively by the Magnus series
\cite{Magnus:1954zz,Brandhuber:2025igz},
which plays an equivalent role to the Dyson series for $i\,\hat{T}=\hat{S}-1$.
When bracketed on in-states,
$\hat{N}$ depends only on the incoming kinematics
and generates forward-in-time evolution.
In an amplitudes context, $\hat{N}$ reorganises the $\hat{S}$-matrix
so as to avoid the appearance of superclassical
terms~\cite{Damgaard:2021ipf,Damgaard:2023ttc,Damgaard:2023vnx,Ochirov:2026ixx};
on the worldline, $\hat{N}$ admits a representation as a sum of connected diagrams.
The connection between these two pictures
was recently clarified in the classical limit~\cite{Brandhuber:2026dkd}.
An important recent application of $\hat{N}$ has been the
gravitational Compton amplitude,
which encodes physical data from black hole perturbation theory
and has recently been pushed to 5PM order
\cite{Bautista:2026qse,Bautista:2026fcp,Brunello:2026rdk,Brunello:2026lzf}.

The diagrammatic worldline representation of $\hat{N}$ is
most naturally realised using Worldline Quantum Field Theory
(WQFT)~\cite{Mogull:2020sak,Jakobsen:2021zvh,Jakobsen:2023oow,Sauer:2026eun}.
WQFT involves quantising both bulk (gravitational)
and worldline (particle) degrees of freedom simultaneously,
producing Feynman diagrams with a manifest $\hbar$ power counting
well-suited for classical physics.
So far, WQFT has mostly been used to calculate in-in scattering observables,
using an all-retarded causality flow
derived from the Schwinger-Keldysh doubled path integral~\cite{Jakobsen:2022psy} ---
the aforementioned Compton amplitude is a notable exception
\cite{Bautista:2026qse,Bautista:2026fcp,Brunello:2026rdk,Brunello:2026lzf}.
The relevance of $\hat{N}$ and Magnus to WQFT was first pointed
out in \rcite{Kim:2024svw}, and has since been
elaborated in \rcites{Kim:2025hpn,Kim:2025olv,Guo:2026xaw,Kim:2026wyv}.
In \rcite{Gonzo:2026yha}, by two of the present authors,
the WQFT formalism was reformulated to adopt Hamiltonian canonical quantisation,
formalising the link to Magnus and defining
classical amplitudes as matrix elements of $\hat{N}$ on incoming plane waves.
It was also shown how, when reorganised around the SF expansion,
the $\hat{N}$-matrix can be used to describe bound orbits. Magnusian methods have also been applied to dissipative
bound dynamics~\cite{Blanco:2026prj,Placidi:2026xug}
and to numerical inspiral evolution using neural flows~\cite{DaSilva:2026pdk}.

In this paper, our main focus is the vacuum
element $N=\langle0|\hat{N}|0\rangle$,
which we refer to as the Magnusian.
We calculate the Magnusian at 4PM order --- a three-loop Feynman integral computation ---
up to linear order in spins of the two scattered bodies.
Up to 3PM order, it has been demonstrated by explicit calculation
that $N$ reproduces the angle for aligned spins~\cite{Haddad:2024ebn,Haddad:2025cmw}.
For precessing spins, $N$ serves as an excellent means to encode the full degrees of freedom:
for the conservative dynamics,
observables including both the momentum impulse $\Delta p_{i,\rm cons}^\mu$ and spin kick
$\Delta S_{i,\rm cons}^\mu$ are generated using nested iterations of Poisson brackets.
But 4PM is a crucial threshold:
due to radiation-induced recoil in the centre-of-mass frame,
we find that the angle generated by the aligned-spin Magnusian
no longer agrees with the full scattering angle.
Thus, as the direct correspondence to a physical quantity (the scattering angle) is broken,
the Magnusian develops soft divergences due to long-range gravitational forces.
The discrepancy between these scalars, and the eikonal phase,
was also pointed out recently in \rcite{Kim:2025gis}.

The appearance of infrared (IR) divergences at 4PM order necessitates an
analysis of soft divergences in the $\hat{N}$-matrix,
which carries over to the WQFT formalism more generally.
While in principle gravitational soft divergences are well understood,
being described by the one-loop Weinberg exponentiation~\cite{Weinberg:1965nx},
lack of time ordering in $\hat{N}$ makes their structure more
complicated than those seen in matrix elements of $\hat{S}$.
We are inspired by the approach used by Faddeev and Kulish for QED~\cite{Kulish:1970ut},
and extended to gravity in \rcite{Ware:2013zja},
to define an IR-finite $\hat{S}$-matrix~\cite{Hannesdottir:2019umk}
--- and thus an IR-finite $\hat{N}$-matrix ---
by changing interaction pictures to eliminate soft poles.
This enables us to re-derive Weinberg's soft factorisation at the operator level,
and make predictions for matrix elements of $\hat{N}$.
Given the link to in-in scattering observables,
we can then also make predictions for the impulse and spin kick, whose divergences cancel,
and the gravitational waveform, whose soft divergences are absorbed into a redefinition
of retarded time~\cite{Georgoudis:2023eke,Bini:2024rsy}.

The outline of our paper is as follows.
In Section~\ref{sec:WQFT} we review 
the WQFT formalism for spinning bodies, and explain our notation.
Then, in Section~\ref{sec:N} we explain key aspects of the $\hat{N}$-operator,
building on the discussions in \rcites{Haddad:2025cmw,Gonzo:2026yha}.
We explain in detail the physical interpretation of the Magnusian $N$,
comparing it with the scattering angle $\theta$ and radial action $I_r$.
Section~\ref{sec:soft} contains our main discussion of soft divergences,
deriving the soft factorisation of the $\hat{S}$-matrix from Faddeev-Kulish,
then using it to deduce the soft pole of $N$ at 4PM order.
We also demonstrate that these poles cancel from the momentum impulse and spin kick,
and reproduce the soft divergence of the subleading gravitational waveform.
Finally, in Section~\ref{sec:calculation} we outline our main 4PM spinning Magnusian calculation.
This includes the Magnusian in the probe limit,
for conservative scattering as defined by linear response~\cite{Bini:2012ji,Damour:2020tta},
and the fully dissipative case, confirming our prediction for the soft divergence.
In Section~\ref{sec:conclusions} we conclude.

\section{WQFT for Spinning Bodies}
\label{sec:WQFT}

Our description of spinning black holes (and neutron stars)
up to linear order in spins of either body is captured by the worldline action\footnote{In
    \rcites{Jakobsen:2023ndj,Jakobsen:2023hig} the worldline action used an anti-commuting
    field $\psi_i^a$, with the spin tensor $S_i^{\mu\nu}=-im_i\psi_i^\mu\psi_i^\nu$.
    At the current linear order in spin, this formulation is entirely equivalent
    to the bosonic oscillators $\alpha_i^\mu$ discussed here. 
}
\begin{align}\label{worldlineAction}
	S_i=-m_i\int\!\d\tau_i\left(\frac12g_{\mu\nu}\dot{x}_i^\mu\dot{x}_i^\nu-
	i\bar{\alpha}_{i,a}\frac{\D\alpha_i^a}{\d\tau_i}\right)\,,
\end{align}
where $\frac{\D\alpha_i^a}{\d\tau_i}=\dot\alpha_i^a+\dot x_i^\mu{\omega_\mu}^{ab}\alpha_{i,b}$.
The bosonic oscillators $\alpha_i^\mu$, introduced in \rcites{Haddad:2024ebn,Haddad:2025cmw},
are a convenient means of capturing the spinning degrees of freedom on the worldline,
the spin tensor then being $S_i^{\mu\nu}=-2im_i\bar{\alpha}_i^{[\mu}\alpha^{\nu]}_i$,
and the spin vector $S_i^\mu=m_ia_i^\mu=\frac12{\eps^\mu}_{\nu\rho\sigma}v_i^\nu S_i^{\rho\sigma}$.
These degrees of freedom are coupled to the gravitational field $h_{\mu\nu}(x)$,
which is described by the usual bulk Einstein--Hilbert action:
\begin{align}\label{bulkAction}
	S_{\rm EH}=\int\!\d^Dx\left(-\frac1{16\pi G_D}\sqrt{-g}R+
	(\partial_\nu h^{\mu\nu}-\sfrac12\partial^\mu{h^\nu}_\nu)^2\right),
\end{align}
which also includes the de Donder gauge-fixing term.
We use dimensional regularisation in $D=4-2\eps$ dimensions,
with the metric $g_{\mu\nu}=\eta_{\mu\nu}+\kappa h_{\mu\nu}$
and mostly minus signature $\eta_{\mu\nu}={\rm diag}(+1,-1,-1,-1)$,
$\kappa=\sqrt{32\pi G_D}$.
The $D$-dimensional extension to Newton's constant is
$G_D=(4\pi e^{\gamma_E}R^2)^{-\eps}G$,
$R$ being a length scale that
features in our discussion of soft divergences.

The worldline fields are expanded around their background configurations at past infinity:
\begin{subequations}\label{backgroundExpansions}
\begin{align}
    x_i^\mu(\tau_i)&=b_i^\mu+v_i^\mu\tau_i+z_i^\mu(\tau_i)\,,\label{trajectory}\\
    \alpha_i^\mu(\tau_i)&=\alpha_i^\mu+\alpha_i^{\prime\mu}(\tau_i)\,,\\
    S_i^\mu(\tau_i)&=S_i^\mu+S_i^{\prime\mu}(\tau_i)\,.
\end{align}
\end{subequations}
To avoid a cluttered notation,
we use the time dependence $\tau_i$ for worldline fields to distinguish
the time-dependent values from their values at past infinity, $\tau_i\to-\infty$.
By convention, we set $b\cdot v_1=b\cdot v_2=0$,
where $b^\mu=b_2^\mu-b_1^\mu$ is the relative impact parameter,
and $v_1^2=v_2^2=1$ with our current choice of proper time coordinates $\tau_i$.
We also have $\gamma=v_1\cdot v_2$, which is the Lorentz boost factor
$\gamma=1/\sqrt{1-v^2}$, $v$ being the relative velocity between the
initial frames of particles 1 and 2.
We also introduce the unit-normalised orbital angular momentum vector $L^\mu$:
\begin{align}\label{angMomentum}
    L_\sigma=\frac{\epsilon_{\mu\nu\rho\sigma}b^\mu v_1^\nu v_2^\rho}{|b|\sqrt{\gamma^2-1}}\,.
\end{align}
Finally, we adopt the covariant Tulczyjew--Dixon spin-supplementary condition (SSC)
in which $p_{i,\mu}S_i^{\mu\nu}=0$, limiting the spin to the three degrees
of freedom captured by the spin vectors $\vct{S}_i$ in the inertial frames of $p_i^\mu$.

In the canonical WQFT formalism~\cite{Gonzo:2026yha},
the fields $\hat{h}_{\mu\nu}(x)$, $\hat{z}_i^\mu(\tau)$
and $\hat{\alpha}_i^{\prime\mu}(\tau)$ are all promoted to operators,
denoted by a hat. Throughout this paper, hats are reserved for operators.
From the worldline actions~\eqref{worldlineAction}
and bulk action~\eqref{bulkAction}
one reads off corresponding Feynman rules.
The retarded graviton propagator is
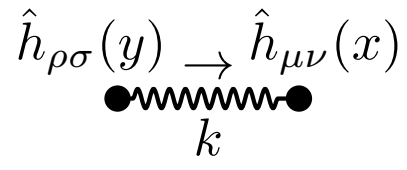
\begin{align}\label{eq:gravProp}
\begin{aligned}
	\begin{tikzpicture}[baseline={(current bounding box.center)}]
	\begin{feynman}
	\coordinate (x) at (-.7,0);
	\coordinate (y) at (0.5,0);
	\draw [photon] (x) -- (y) node [midway, below] {$k$} node [midway, above] {$\rightarrow$} ;
	\draw [fill] (x) circle (.08) node [above, xshift=-5pt, yshift=1pt] {$\hat h_{\rho\sigma}(y)$};
	\draw [fill] (y) circle (.08) node [above, xshift=5pt, yshift=1pt] {$\hat h_{\mu\nu}(x)$};
	\end{feynman}
	\end{tikzpicture}
    &=\Theta(x^0-y^0)[\hat h_{\mu\nu}(x),\hat h_{\rho\sigma}(y)]\\[-7pt]
    &=\int_k e^{-ik\cdot(x-y)}\frac{iP_{\mu\nu;\rho\sigma}}{k^2+k^0\,i0^+}\,,
\end{aligned}
\end{align}
where $\int_k:=\int\d^Dk/(2\pi)^D$ is the off-shell integration measure and
$P_{\mu\nu;\rho\sigma}:=\eta_{\mu(\rho}\eta_{\sigma)\nu}-\sfrac1{D-2}\eta_{\mu\nu}\eta_{\rho\sigma}$.
The worldline propagators are
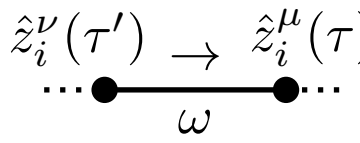
\begin{subequations}
\begin{align}
  \begin{tikzpicture}[baseline={(current bounding box.center)}]
    \coordinate (in) at (-0.6,0);
    \coordinate (out) at (1.4,0);
    \coordinate (x) at (-.2,0);
    \coordinate (y) at (1.0,0);
    \draw [zUndirected] (x) -- (y) node [midway, below] {$\omega$} node [midway, above] {$\rightarrow$};
    \draw [background] (in) -- (x);
    \draw [background] (y) -- (out);
    \draw [fill] (x) circle (.08) node [above, xshift=-5pt, yshift=1pt] {$\hat{z}_i^\nu(\tau')$};
    \draw [fill] (y) circle (.08) node [above, xshift=5pt, yshift=1pt] {$\hat{z}_i^\mu(\tau)$};
  \end{tikzpicture}&=\Theta(\tau-\tau')[\hat{z}^\mu_i(\tau),\hat{z}^\nu_i(\tau')]\\[-7pt]
  &=\int_\omega e^{-i\omega(\tau-\tau')}\frac{-i\eta^{\mu\nu}}{m_i(\omega+i0^+)^2}\,,\nn\\
  
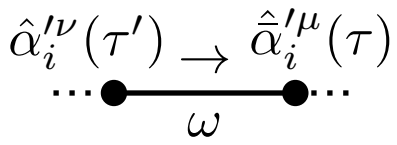
\begin{tikzpicture}[baseline={(current bounding box.center)}]
    \coordinate (in) at (-0.6,0);
    \coordinate (out) at (1.4,0);
    \coordinate (x) at (-.2,0);
    \coordinate (y) at (1.0,0);
    \draw [zUndirected] (x) -- (y) node [midway, below] {$\omega$} node [midway, above] {$\rightarrow$};
    \draw [background] (in) -- (x);
    \draw [background] (y) -- (out);
    \draw [fill] (x) circle (.08) node [above, xshift=-5pt, yshift=1pt] {$\hat{\alpha}^{\prime\nu}_i(\tau')$};
    \draw [fill] (y) circle (.08) node [above, xshift=5pt, yshift=1pt] {$\hat{\bar{\alpha}}^{\prime\mu}_i(\tau)$};
  \end{tikzpicture}&=\Theta(\tau-\tau')[\hat{\bar{\alpha}}^{\prime\mu}_i(\tau),\hat{\alpha}^{\prime\nu}_i(\tau')]\\[-7pt]
  &=\int_\omega e^{-i\omega(\tau-\tau')}\frac{-i\eta^{\mu\nu}}{m_i(\omega+i0^+)}\,.\nn
\end{align}
\end{subequations}
where $\int_\omega:=\int\d\omega/(2\pi)$.
The arrow on the propagators indicates the momentum or energy flow on the retarded propagators.
As for the vertices,
the simplest is the single-graviton emission vertex from a worldline $i$:
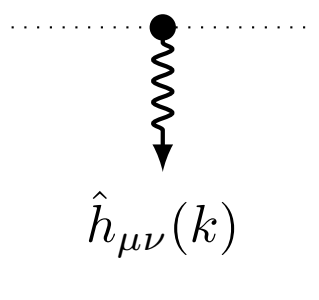
\begin{align}\label{eq:vertexH}
\begin{tikzpicture}[baseline={(current bounding box.center)}]
    \coordinate (in) at (-1,0);
    \coordinate (out) at (1,0);
    \coordinate (x) at (0,0);
    \node (k) at (0,-1.3) {$\hat{h}_{\mu\nu}(k)$};
    \draw [dotted] (in) -- (x);
    \draw [dotted] (x) -- (out);
    \draw [graviton] (x) -- (k);
    \draw [fill] (x) circle (.08);
\end{tikzpicture}&=
-i\frac{m_i\kappa }{2}e^{ik\cdot b_i}\dd(k\cdot v_i) \\[-20pt]
&\quad\times\bigg(v_i^\mu v_i^\nu+ik_\rho \frac{S_i^{\rho(\mu}}{m_i} v_i^{\nu)}+\cO(S^2)\bigg)\,,\nn
\end{align}
where $\dd(\omega):=2\pi\delta(\omega)$.
The higher-spin corrections do not contribute to the soft structure
of the theory.

\section{The $\hat{N}$-Operator}
\label{sec:N}

Our main focus in this paper is the $\hat{N}$-operator,
which is related to the scattering $\hat{S}$-matrix by:
\begin{align}\label{Smatrix}
	\hat{S}=\hat{U}(+\infty,-\infty)=e^{i\hat{N}}\,,
\end{align}
$\hat{U}(t,t_0)=\exp(i\hat{N}(t,t_0))$ being the usual time-evolution operator.
We define the matrix elements
\begin{align}\label{NElements}
	N(k_1^{\sigma_1},k_2^{\sigma_2},\cdots,k_n^{\sigma_n}):=
	{}_{\rm in}\langle k_1^{\sigma_1}k_2^{\sigma_2}\cdots k_n^{\sigma_n}|\hat{N}|0\rangle_{\rm in}\,.
\end{align}
States contracted on the left or right are simply related by $k^\mu_i\to-k^\mu_i$,
and the vacuum element
\begin{align}
    N:={}_{\rm in}\langle0|\hat{N}|0\rangle_{\rm in}
\end{align}
is the Magnusian.\footnote{Our
        use of the term Magnusian differs from that in \rcites{Kim:2025gis,Guo:2026xaw}.
        In their case, radiative degrees of freedom are included by way of an external source to the theory,
        so $N$ captures the information of the full $\hat N$-matrix.
}
External graviton states carrying polarisations $\sigma$
are defined using the raising operators
acting on the incoming vacuum $|0\rangle_{\rm in}$:
\begin{align}
    |k^\sigma\rangle=\hat{a}^{\sigma\dagger}(k)|0\rangle_{\rm in}\,,
\end{align}
the raising and lowering operators satisfying the usual algebra:
\begin{align}\label{raisinglowering}
	[\hat{a}^\sigma(k),\hat{a}^{\sigma'\dagger}(q)]=
    \delta^{\sigma\sigma'}2|\vct{k}|\dd(\vct{k}-\vct{q})\,,
\end{align}
with the notation $\dd(\vct{q}):=(2\pi)^{D-1} \delta^{(D-1)}(\vct{q})$.
In order to produce matrix elements of $\hat N$
that generate forward-in-time evolution,
we bracket with the incoming vacuum $|0\rangle_{\rm in}$,
defined at past infinity $t\to-\infty$, on both sides ---
unlike conventional $\hat{S}$-matrix theory,
wherein one brackets between in- and out-states.
Seeing as we always use the in-vacuum,
we drop the ``in'' subscript from now on.

A perturbative representation of $i\hat{N}$ in terms of the interaction Hamiltonian
$\hat{H}_I(t)$ is provided by the Magnus series~\cite{Magnus:1954zz},
which plays the same role as the Dyson series for $i\hat{T}=\hat{S}-1$.
The first two terms are
\begin{subequations}\label{eq:MagnusTerms}
\begin{align}
	i\,\hat{N}^{(1)}(t,t_0)
	&=
	-i
	\int_{t_0}^{t}\!\mathrm dt_1\,
	\hat H_I(t_1)\,,
	\label{eq:Magnus1}
	\\[0.5em]
	i\,\hat{N}^{(2)}(t,t_0)
	&=
	\frac{(-i)^2}{2}
	\int_{t_0}^{t}\!\mathrm dt_1
	\int_{t_0}^{t_1}\!\mathrm dt_2\,
	[\hat H_I(t_1),\hat H_I(t_2)]\,.
	\label{eq:Magnus2}
\end{align}
\end{subequations}
As was demonstrated in \rcite{Gonzo:2026yha},
the terms of this series can be directly evaluated
to calculate matrix elements of $\hat{N}$.
Yet, as such calculations rapidly grow in complexity,
it is more efficient to simply draw all possible connected, tree-level Feynman diagrams,
carrying all possible routings of internal retarded and advanced propagators.
These diagrams are weighted by the numerical Murua coefficients $\omega$~\cite{murua},
the values of which are reviewed in \App{murua}.
These coefficients are assigned in addition to the usual symmetry factors.

\subsection{Scattering observables from the $\hat{N}$-operator}

The relationship between $\hat{N}$ and physical scattering observables is
provided by the in-in formalism.
The expectation value of any time-dependent observable is given by
a one-point function:
\begin{align}
	\langle\hat\cO(t,\vct{x})\rangle_{\rm in-in}
	:=\langle0|\hat{U}(-\infty,t)\hat\cO(t,\vct{x})\hat{U}(t,-\infty)|0\rangle\,.
\end{align}
When computing scattering observables in the far future $t\to\infty$,
one inserts the $\hat{N}$-operator~\eqref{Smatrix}.
For example, the change in a worldline operator $\hat\lambda^\mu$,
which includes $p_i^\mu$ (the momentum impulse)
or $S_i^\mu$ (the spin kick), is given by
\begin{align}\label{ininObservable}
\begin{aligned}
    \Delta\lambda^\mu&:=
    \langle \hat{\lambda}^\mu(t\to+\infty)\rangle_{\rm in-in}-\langle \hat{\lambda}^\mu(t\to-\infty)\rangle_{\rm in-in}\\
    &\,\,=\langle0|(e^{-i[\hat{N},\bullet]}-1)\hat{\lambda}^\mu|0\rangle\,.
\end{aligned}
\end{align}
The nested brackets in the exponent can be interpreted as causality cuts~\cite{Kim:2024svw}:
differences between retarded and advanced propagators of a field,
whose role is to reverse causality prescriptions on the diagrams involved.
As was illustrated in \rcite{Haddad:2025cmw},
acting with these brackets on $\hat N$ in an iterative manner results in the all-retarded
forward-in-time causality flow associated with %
in-in observables.

One can also derive a version of \eqn{ininObservable} that holds after loop integration,
involving the matrix elements of $\hat{N}$~\eqref{NElements}.
Brackets involving gravitational degrees of freedom
are resolved by inserting a complete set of states:
\begin{align}\label{coherent}
	\hat{N}&=N+
	\sum_\sigma\int_{\vct{k}}\left(N(k^\sigma)\hat{a}^{\sigma\dagger}(k)+
	N(-k^\sigma)\hat{a}^\sigma(k)\right)\nn\\[-6pt]
	&+\frac{1}{2}\sum_{\sigma_1,\sigma_2}\int_{\vct{k}_1,\vct{k}_2}\Big(
		N(k_1^{\sigma_1},k_2^{\sigma_2})\hat{a}^{\sigma_1\dagger}(k_1)\hat{a}^{\sigma_2\dagger}(k_2)\\[-3pt]
        &\qquad\qquad+
        N(-k^{\sigma_1}_1,-k^{\sigma_2}_2)\hat{a}^{\sigma_1}(k_1)\hat{a}^{\sigma_2}(k_2)\nn\\
		&\qquad\qquad+
		2N(k^{\sigma_1}_1,-k^{\sigma_2}_2)\hat{a}^{\sigma_1\dagger}(k_1)\hat{a}^{\sigma_2}(k_2)
	\Big)+\cdots\,,\nn
\end{align}
where $\int_{\vct{k}}:=\int_k\,\Theta(k^0)\dd(k^2)$
is the on-shell integration measure.
Meanwhile, brackets between worldline operators are re-interpreted as Poisson brackets
involving the background constants $b_i^\mu$, $v_i^\mu$ and $S_i^\mu$:
$[\hat{A},\hat{B}]\to i\{A,B\}$.
As was explained in \rcite{Gonzo:2026yha},
this is permitted on grounds of the interaction Hamiltonian depending
on the perturbed worldline fields $z_i^\mu(\tau_i)$ and $\alpha_i^{\prime\mu}(\tau_i)$
via their true dependence on the full fields
$x_i^\mu(\tau_i)$ and $\alpha_i^\mu(\tau_i)$~\eqref{backgroundExpansions}.
The first few terms in \eqn{ininObservable} are then~\cite{Alessio:2025flu,Gonzo:2026yha}
\begin{align}\label{obsFromN}
    \Delta\lambda^\mu&=(e^{\{N,\bullet\}}-1)\lambda^\mu\\
    &\quad-\frac{i}{2!}\sum_\sigma\int_{\vct{k}}
    \Big(N(-k^\sigma)\{N(k^\sigma),\lambda^\mu\}-\rm{c.c.}\Big)+\cdots\,,\nn
\end{align}
higher-order terms in this formula being provided in \rcite{Alessio:2025flu}.
The linear and angular momentum fluxes
$P^\mu_{\rm rad}=-\Delta P^\mu$ and $J^\mu_{\rm rad}=-\Delta J^\mu$ are
\begin{subequations}\label{radFluxes}
\begin{align}
P_{\rm rad}^\mu
=
\frac{i}{2}\sum_\sigma\int_{\vct{k}}
\Big(
N(-k^\sigma)\,\{N(k^\sigma),P^\mu\}
-\text{c.c.}
\Big)
+\cdots\,,
\label{eq:DeltaP_rad}
\\
J_{\rm rad}^\mu
=
\frac{i}{2}\sum_\sigma\int_{\vct{k}}
\Big(
N(-k^\sigma)\,\{N(k^\sigma),J^\mu\}
-\text{c.c.}
\Big)
+\cdots\,,
\label{eq:DeltaJ_rad}
\end{align}
\end{subequations}
where $J^\mu$ is the total angular momentum vector, which also includes spin.
These fluxes vanish when $N(k_1^{\sigma_1},k_2^{\sigma_2},\ldots)=0$.

When applying \eqn{obsFromN} to integrated expressions,
one must be careful to account for constraints involving
the background variables $b_i^\mu$, $v_i^\mu$ and $S_i^\mu$.
We have $b\cdot v_i=0$,
where $b^\mu=b_2^\mu-b_1^\mu$,
$v_i^2=1$ in the proper time gauge, and $v_i\cdot S_i=0$.
The Poisson brackets are~\cite{Alessio:2025flu,Haddad:2025cmw}
\begin{subequations}\label{eq:PBgi}
\begin{align}
\{b^{\mu},\, v_{i}^{\nu}\}\,
&=\, -\,\frac{(-1)^{i}}{m_{i}}\, P^{\mu\nu}\,,\label{bvBracket}
\\
\{S_{i}^{\mu},\, S_{j}^{\nu}\}\,
&=\,\delta_{ij}\,{\varepsilon^{\mu\nu}}_{\rho\sigma}v_i^\rho S_i^\sigma\,,
\\
\{b^{\mu},\, S_{i}^{\nu}\}\,
&=\,\frac{(-1)^{i}}{m_{i}}\, {P^{\mu}}_\rho\, v_i^{\nu}S_i^\rho\,,
\\\label{eq:bbBracket}
\{b^{\mu},\, b^{\nu}\}\,
&=\, 2\, b^{[\mu}\Bigl(\,\frac{\bar{v}_{2}^{\nu]}}{m_{2}}\, -\,\frac{\bar{v}_{1}^{\nu]}}{m_{1}}\,\Bigr)\\
&\qquad+P^{\mu}{}_{\rho}\, P^{\nu}{}_{\sigma}
{\varepsilon^{\rho\sigma}}_{\lambda\tau}
\Bigl(\,\frac{v_1^\lambda S_1^\tau}{m_{1}^{2}}\,
+\,\frac{v_2^\lambda S_2^\tau}{m_{2}^{2}}\,\Bigr)\,,\nn
\end{align}
\end{subequations}
where $P^{\mu\nu}=\eta^{\mu\nu}-v_1^\mu\bar{v}_1^\nu-v_2^\mu\bar{v}_2^\nu$ and
\begin{align}\label{eq:vbarDef}
\bar{v}_{1}^{\mu}\, &=\, -\,\frac{v_{1}^{\mu} - \gamma\, v_{2}^{\mu}}{\gamma^{2}-1}\,,
&
\bar{v}_{2}^{\mu}\, &=\, -\,\frac{v_{2}^{\mu} - \gamma\, v_{1}^{\mu}}{\gamma^{2}-1}\,,
\end{align}
are the dual velocities
satisfying $v_{i}\cdot\bar{v}_{j}\, =\,\delta_{ij}$.
These brackets provide the evolution of the full system
from past to future infinity over the constrained phase space.

Finally, as was pointed out in \rcites{Haddad:2025cmw,Gonzo:2026yha},
$N$ defines a natural conservative scattering prescription,
wherein the impulse $\Delta p_{i,N}^\mu$, spin kick $\Delta S_{i,N}^\mu$
and change in impact parameter $\Delta b_{i,N}^\mu$ are all given by the Poisson brackets in \eqn{obsFromN}:
\begin{align}\label{consN}
    \Delta\lambda_N^\mu
    :=
    \big(e^{\{N,\bullet\}}-1\big)\lambda^\mu\,.
\end{align}
Discarding the radiative matrix elements $N(k_1^{\sigma_1},k_2^{\sigma_2},\ldots)$
implies that the linear and angular momentum losses~\eqref{radFluxes} directly vanish~\cite{Alessio:2025flu,Gonzo:2026yha}. However, $N$ itself contains radiative contributions,
so this prescription does not coincide with other conservative prescriptions ---
notably linear response~\eqref{eq:linResponse} that we will discuss below.

\subsection{Scattering angle from the Magnusian}
\label{sec:magangle}

Let us now focus on the part of the dynamics defined by the Magnusian $N$ alone~\eqref{consN},
which as we have seen above automatically implies zero
momentum and angular momentum losses.
In the centre-of-mass frame, the incoming momenta are $p_1^\mu=(E_1,\vct{p})$ and $p_2^\mu=(E_2,-\vct{p})$.
We write $L=|\vct{b}||\vct{p}|$ and introduce the unit directions
$\vct{e}_b=\vct{b}/|\vct{b}|$,
$\vct{e}_p=\vct{p}/|\vct{p}|$ and
$\vct{e}_L=-\vct{e}_b\times\vct{e}_p$.
The spin projections along this axis are $s_i=\vct{S}_i\cdot\vct{e}_L$.
The momentum generated by \eqn{consN} is
\begin{align}
    \vct{p}_N'=e^{\{N,\bullet\}}\vct{p}\,.
\end{align}
For spins aligned with the orbital angular momentum,
this momentum can also be expressed in terms of a single scattering angle $\theta_N$:
\begin{align}\label{rotMomentum}
\begin{aligned}
    \vct{p}'_N&=
    |\vct{p}|(\cos\theta_N\,\vct{e}_p+\sin\theta_N\,\vct{e}_b)\\
    &=e^{\theta_N(\vct{e}_L\times\bullet)}\vct{p}\,.
\end{aligned}
\end{align}
Comparing these two formulae, we see that the action of nested Poisson brackets
is equivalent to that of a rotation by the angle $\theta_N$:
$\{N,\bullet\}=\theta_N(\vct{e}_L\times\bullet)$.
Evaluating the brackets on $\vct{p}$ then gives
\begin{align}
    \{N,\vct{p}\}
    &=
    -\frac{\partial N}{\partial|\vct{b}|}
    \vct{e}_b
    =
    -\frac{\partial N}{\partial L}\,
    \vct{e}_L\times\vct{p}\,,
\end{align}
so we deduce that the angle in \eqn{rotMomentum} is given by
\begin{align}\label{angleFromN}
    \theta_{N}
    =
    -\frac{\partial N}{\partial L}\,.
\end{align}
All derivatives are taken at fixed energy and spin projections.
The energy, angular momentum magnitude and spin projections are all
conserved under the flow generated by Poisson brackets~\eqref{eq:PBgi}.

We now compare the angle $\theta_N$ defined by the Magnusian~\eqref{angleFromN} with
the full scattering angle $\theta$, including radiation effects,
in the \emph{initial} centre-of-mass frame:
\begin{align}\label{fullAngle}
    \cos\theta
    =
    \frac{\vct{p}\cdot\vct{p}'}
         {|\vct{p}||\vct{p}'|}\,,
\end{align}
where $\vct{p}'$ is the outgoing relative momentum.
To see the precise relationship, we decompose the additional radiative contribution to the impulse as 
\begin{equation}
    \vct{p}'-\vct{p}'_N =\delta p_\parallel\vct{e}_p+\delta p_b\vct{e}_b\,.
\end{equation}
The leading radiative corrections $\delta p_\parallel$ and $\delta p_b$ begin
at 3PM and 4PM order respectively,
i.e.~$\delta p_\parallel^{(3)}$ and $\delta p_b^{(4)}$.
Expanding \eqn{fullAngle} in $G$ gives
\begin{align}\label{eq:3PM-angle-flux}
    \theta-\theta_N
    &=
    \frac{G^4}{|\vct{p}|}
    \left(
        \delta p_b^{(4)}
        -\theta^{(1)}\delta p_\parallel^{(3)}
    \right)
    +\cO(G^5)\,.
\end{align}
So the scattering angle $\theta_N$ agrees with the full angle $\theta$
up to and including 3PM ---
despite the appearance of radiation reaction at this order.
The longitudinal 3PM correction $\delta p_\parallel^{(3)}$ affects the
angle only when combined with the 1PM deflection,
and the radiation-reaction contribution to the 3PM angle is already included in $N^{(3)}$.
For aligned spins, this was verified through fourth order in spins in \rcite{Haddad:2025cmw}.
Yet $N^{(3)}$ does \emph{not} capture the full momentum impulse $\Delta p_1^{(3)\mu}$,
or the full spin kick $\Delta S_1^{(3)\mu}$ for precessing spins:
these also receive the radiative corrections in \eqn{obsFromN} carried by $N(k)$.

At 4PM order, the full scattering angle no longer
agrees with that defined by the Magnusian~\eqref{angleFromN}:
$\theta^{(4)}\neq\theta_N^{(4)}$.
Furthermore, recoil in the centre-of-mass frame
implies that the change in the relative momentum vector develops a non-polynomial
dependence on the masses:
$\Delta \vct{p}^{(4)} = \Delta \vct{p}^{(4)}_1+E_1 \vct{P}_{\rm rad}^{(4)}/E$~\cite{Dlapa:2022lmu,Dlapa:2023hsl},
where $E=\sqrt{m_1^2+m_2^2+2\gamma m_1m_2}$ is the total energy
and $E_1=m_1(m_1+\gamma m_2)/E$ is the initial energy of particle 1.
The complete 4PM scattering angle, including dissipative terms,
has the schematic form~\cite{Jakobsen:2023hig}
\begin{align}\label{eq:angle}
  \frac{\theta^{(4)}}{\Gamma} 
  =
  \frac{M^4}{|b|^4}\left(
  \theta_{\rm{cons},\nu^0}^{(4)}
  \!+\!
  \nu \theta_{\rm{cons},\nu}^{(4)}
  \!+\!
  \frac{\nu}{\Gamma^2}
  \Big(
  \theta_{\rm{rad},\nu}^{(4)}
  \!+\!
  \nu
  \theta_{\rm{rad},\nu^2}^{(4)}
  \Big)\right)\,,
\end{align}
where $\nu=m_1m_2/M^2$ is the symmetric mass ratio, $M=m_1+m_2$
is the total mass and $\Gamma=E/M$.
The conservative/radiative splitting of the angle is here defined by linear response
\cite{Bini:2012ji,Damour:2020tta}:
\begin{align}\label{eq:linResponse}
  \theta_{\rm rad}=-
  \frac12\bigg(\frac{\partial\theta_{\rm cons}}{\partial E}E_{\rm rad}
  +\frac{\partial\theta_{\rm cons}}{\partial J}J_{\rm rad}\bigg)+\cO(\nu^2)\,,
\end{align}
which is valid up to the current first order in self force.
Note that while the conservative part of the 4PM angle enjoys a polynomial dependence on the masses,
the same is not true for the radiative part.
In contrast, mass polynomiality of $\theta_N^{(4)}/\Gamma$ --- including radiative contributions
--- follows directly from the WQFT Feynman rules.

\subsection{The Magnusian and the radial action}
\label{sec:magradaction}

Finally, let us explore the relationship between the Magnusian $N$
and the radial action, again for non-spinning or aligned-spin scattering.
To define the conservative radial action $I_r$
we use the linear response split $\theta=\theta_{\rm cons}+\theta_{\rm rad}$~\eqref{eq:linResponse}.
When the conservative dynamics admits a canonical radial Hamiltonian $H(r,p_r,L,s_1,s_2)$,
the energy constraint $H=E$ determines the radial momentum. The interaction part of the radial action is
\begin{align}\label{eq:radialActionDefinition}
    I_r
    =
    \left[
    2\int_{r_{\rm min}}^\infty\!\d r\,p_{r,\rm CM}
    \right]_{\rm reg}
    +\pi L\,,
\end{align}
with a suitable regularisation of the IR divergences. Using the same IR prescription as for $N$, the Hamilton--Jacobi relation gives the scattering angle
\begin{align}\label{eq:angleFromIr}
    \theta_{\rm cons}=-\frac{\partial I_r}{\partial L}\,,
\end{align}
where the derivative holds $E$ and $s_i$ fixed.

In the integrable scattering scenario~\cite{Gonzo:2024zxo,Kim:2025gis}, since the Magnusian can also be constructed from the same conservative Hamiltonian, we have $\theta_N=\theta_{\rm cons}$ and $N= I_r$;\footnote{
This identification uses the aligned spin kinematics. For generic spin orientations, a single scattering angle typically does not determine the full canonical map, which also includes spin precession and changes of the orbital plane. However, in the integrable scattering case~\cite{Kim:2025gis,Akpinar:2025tct}, we can still identify $N=I_r$, as shown for a spinning particle moving in a Kerr background~\cite{Gonzo:2024zxo}.} however, this fails when radiation-reaction terms are included.
Since the energy loss starts at $G^3$,
only the angular momentum loss contributes to \eqn{eq:linResponse} at 3PM order:
\begin{align}
    \theta^{(3)}
    =
    \theta_N^{(3)}
    =
    \theta_{\rm cons}^{(3)}
    -
    \frac12
    \frac{\partial\theta_{\rm cons}^{(1)}}{\partial J}
    J_{\rm rad}^{(2)}\,,
\end{align}
which implies that
\begin{align}
    -\frac{\partial N^{(3)}}{\partial L} = -\frac{\partial I^{(3)}_r}{\partial L}  -
    \frac12
    \frac{\partial\theta_{\rm cons}^{(1)}}{\partial J} J_{\rm rad}^{(2)}\,.
\end{align}
Thus, despite the agreement between $\theta^{(3)}$ and $\theta_N^{(3)}$,
we see that $N^{(3)} \neq I^{(3)}_r$.

We conclude that, despite their apparent similarity,
the full Magnusian $N$ and radial action $I_r$ are very different objects.
Yet both carry similar perturbative data regarding scattering in the PM regime,
and could for example be used to match an
effective-one-body (EOB) description of the
dynamics~\cite{Buonanno:1998gg,Khalil:2022ylj,Buonanno:2024vkx,Buonanno:2024byg},
and enjoy mappings to bound orbits~\cite{Kalin:2019rwq,Kalin:2019inp,Cho:2021arx,Saketh:2021sri}.
From this perspective, the main advantage of the radial action $I_r$
is its clear link to the conservative scattering angle $\theta_{\rm cons}$,
which is a well-defined physical observable.
The Magnusian $N$, on the other hand, benefits from not invoking
the centre-of-mass frame as part of its definition,
which as we have seen suffers from recoil at higher PM orders.
The Magnusian therefore has a simple polynomial dependence on the masses.
It also provides a convenient encoding of spinning degrees of freedom,
capturing data associated with both the momentum impulse $\Delta p_i^\mu$ \emph{and}
the spin kick $\Delta S_i^\mu$.
Yet, as it is not a physical observable,
the full Magnusian also exhibits soft divergences.
The structure of these divergences is the subject of the next section.

\section{Soft Factorisation of $\hat{N}$}
\label{sec:soft}

In this section we derive the generic 
form of the soft divergences of the $\hat{N}$-matrix~\eqref{symmSoft},
leading to our main prediction for the soft 
divergence of $N^{(4)}$~\eqref{4pmDivergence}.
Then we confirm that these divergences leave the inclusive
in-in matrix elements~\eqref{ininObservable} IR-finite ---
in the same way that IR divergences cancel from QFT cross-sections.
We also re-derive the IR divergence of the subleading gravitational waveform.
While in principle the soft divergences of gravitational
$\hat{S}$-matrices are well understood,
being captured by the one-loop Coulomb phase,
the absence of time ordering in $\hat{N}$ complicates
an understanding of its divergences,
and calls for a more detailed approach.
We achieve this in the WQFT formalism
using the method first applied by Faddeev and Kulish
in QED~\cite{Kulish:1970ut},
in which an IR-finite $\hat{S}$-matrix is obtained by switching
interaction pictures to one that eliminates soft divergences.
Our derivation is closely inspired by \rcites{Hannesdottir:2019umk,Ware:2013zja}.

\subsection{Soft theory}

We begin by isolating the parts of the worldline and bulk theories
responsible for soft divergences in the full WQFT ---
thereby defining the soft theory.
By focusing on the Hamiltonian,
and adopting the canonical quantisation setup developed in \rcite{Gonzo:2026yha},
we gain access to the Magnus expansion for subsequent manipulations.
Asymptotic interactions are defined by their leading power behaviour in the soft region.
For a graviton momentum $q^\mu$ becoming soft, we scale
\begin{align}\label{softScaling}
    &q^\mu \sim \lambda\,, &
    &p_i^\mu,k_a^\mu\sim \lambda^0\,, &
    &\lambda\to0\,,
\end{align}
where $p_i$ and $k_a$ denote hard massive (worldline) and massless (graviton) momenta, respectively. %
For a massive particle, the leading soft interaction follows from the term
linear in the graviton field;
in the bulk, it comes from the three-graviton vertex. 

First we consider the massive worldline Hamiltonian, given in terms of the coordinate time $t$:
\begin{align}
   \hspace{-10pt} H_i(t)=-\frac{1}{2m_i}\int_{-\infty}^\infty\d\tau_i\,\delta(t\!-\!\bar{x}_i^0(\tau_i))
    \pi_{i,\mu}\pi_{i,\nu} g^{\mu\nu}(x_i(\tau_i)),
\end{align}
where $\bar{x}_i^0(\tau_i)=b_i^0+\tau_iv_i^0$.
Spin dependence comes from the canonical momentum
$\pi_i^\mu=p_i^\mu+\sfrac12\omega^{\mu}_{ab}S_i^{ab}$.
At linear order in the soft field as $\lambda\to0$,
the spin connection $\omega^\mu_{ab}$ contains a derivative of the gravitational field,
so the explicit spin coupling is subleading in $q^\mu$ and can therefore be omitted.
Taking the gravitons in momentum space,
and parametrising the worldline trajectory $x^\mu_i(\tau_i)$ as in \eqn{trajectory},
we have
\begin{align}\label{hhh}
    h_{\mu\nu}(x_i(\tau_i))=\int_q e^{-iq\cdot(b_i+\tau_iv_i+z_i(\tau_i))}h_{\mu\nu}(q)\,.
\end{align}
Each insertion of $z_i^\mu$ in the soft coupling carries an additional power of $q^\mu$,
and therefore does not contribute to the leading IR pole because
$q\cdot z_i=O(\lambda)$ implies $e^{-iq\cdot z_i}=1+\cO(\lambda)$.

\begin{figure}[t]
    \centering
    \begin{tikzpicture}[scale=0.9,baseline={(current bounding box.center)}]
        \coordinate (L) at (-1.2,0);
        \coordinate (R) at (1.2,0);
        \coordinate (V) at (0,0);
        \coordinate (S) at (0,1.1);

        \draw[dotted] (L) -- (R);
        \draw[blue, photon] (V) -- (S);
        \fill (V) circle (0.07);

        \node[below] at (-0.8,-0.08) {$p_i$};
        \node[right] at (0.1,0.8) {$q$};
        \node at (0,-0.55) {(a)};
    \end{tikzpicture}
    \hspace{2cm}
    \begin{tikzpicture}[scale=0.9,baseline={(current bounding box.center)}]
        \coordinate (L) at (-1.2,0);
        \coordinate (R) at (1.2,0);
        \coordinate (V) at (0,0);
        \coordinate (S) at (0,1.1);

        \draw[red, photon] (L) -- (V);
        \draw[red, photon] (V) -- (R);
        \draw[blue, photon] (V) -- (S);
        \fill (V) circle (0.07);

        \node[below] at (-0.8,-0.12) {$k$};
        \node[below] at (0.8,-0.12) {$k+q$};
        \node[right] at (0.1,0.8) {$q$};
        \node at (0,-0.55) {(b)};
    \end{tikzpicture}
    \caption{Leading soft couplings to (a) a massive worldline and (b) a hard graviton.}
    \label{fig:asymptotic-soft-vertices}
\end{figure}
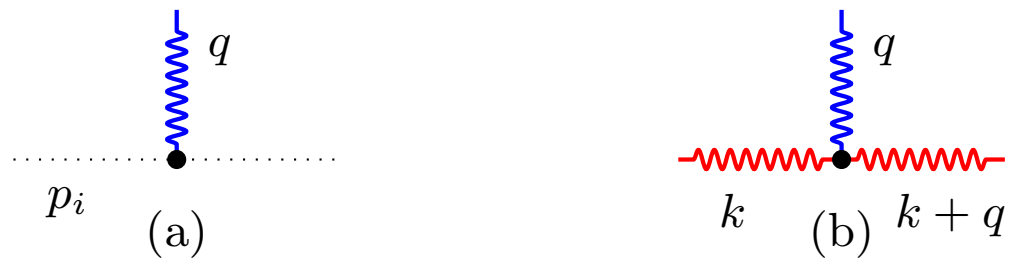

Thus, asymptotic evolution is 
described by a gravitational Wilson line along a straight reference trajectory.
Choosing $b_i^\mu$ such that $b_i^0=0$,
so that $\tau_i=0$ corresponds to $t=0$,
the soft worldline interaction Hamiltonian is
\begin{align}\label{eq:Hwl-soft}
    H_i^s(t)
    &=
    \frac{\kappa}{2m_i}
    \int_{-\infty}^{\infty}\d\tau_i\,
    \delta(t-v_i^0\tau_i)\,
    p_i^\mu p_i^\nu\,
    h_{\mu\nu}(b_i+v_i\tau_i)\,.
\end{align}
Here $h_{\mu\nu}(q)$ is the full Fourier coefficient of the interaction picture field~\eqref{hhh},
as in \rcite{Gonzo:2026yha},
with $h_{\mu\nu}^\dagger(q)=h_{\mu\nu}(-q)$.
This Hamiltonian contains only one coupling:
the single-emission vertex from the worldline, Fig.~\ref{fig:asymptotic-soft-vertices}(a).
The hard momentum is $p_i^\mu=m_i v_i^\mu$, with $p_i^2=m_i^2$ and $p_i^0>0$.

The coupling to a hard graviton follows instead from the Einstein--Hilbert term~\eqref{bulkAction},
by taking one momentum $q^\mu$ soft in the
three-graviton vertex while keeping the hard momentum $k^\mu$ fixed,
Fig.~\ref{fig:asymptotic-soft-vertices}(b).
At leading order in $\lambda$~\eqref{softScaling},
\begin{align}\label{eq:Hrad-soft}
    H_{\rm rad}^{s}(t)
    &= %
    \frac{\kappa}{2}
    \sum_{\sigma}
    \int_{\vct{k}}%
    \int_q \frac{k^\mu k^\nu}{k^0}\,
    h_{\mu\nu}(q)\,\\
    &\quad\times a^{\sigma\dagger}(k+q)
    a^\sigma(k)\,
    e^{i(E_{\vct{k}+\vct{q}}-E_{\vct{k}}-q^0)t}
    +\cO(\lambda)\,, \nn
\end{align}
the energy of a massless particle being $E_{\vct{k}}=|\vct{k}|$.
At higher PM orders than what is considered in this work,
additional multi-soft regions (relevant for tail-of-tail effects)
will also need to be included.
In the following, we write $\hat H_{\rm wl}^{s}=\sum_i\hat H_i^{s}$
and $H_{\rm soft}=H_{\rm wl}^{s}+H_{\rm rad}^{s}$.

\subsection{Change of interaction picture}

Having defined the asymptotic soft Hamiltonian,
we now introduce an IR-finite hard $\hat{S}_H$-matrix for the full WQFT.
To do so, we switch interaction pictures to one in which soft interactions
are subtracted from those of the full theory.
Using $\hat U(t,0)=\hat U_{\rm soft}(t,0)\hat U_H(t,0)$,
with $\hat U_{\rm soft}$ generated by $\hat H_{\rm soft}$ for all $t$, the remaining hard interactions are generated by
\begin{align}\label{eq:change-picture}
    \hat H_H(t)
    =
    \hat U_{\rm soft}^{\dagger}(t,0)
    \bigl[\hat H_I(t)-\hat H_{\rm soft}(t)\bigr]
    \hat U_{\rm soft}(t,0)\,,
\end{align}
where $\hat H_I$ is the original interaction picture Hamiltonian.
Using the common coordinate time variable $t$,
we have chosen $t=0$ as the reference slice separating the past and future evolutions.
The full scattering operator is then factorised as
\begin{align}\label{asympFactorisation}
    e^{i\hat N}
    =
    \hat{\Omega}_+\,
    e^{i\hat N_H}\,
    \hat{\Omega}_-
    =
    e^{i\hat N_+}
    e^{i\hat N_H}
    e^{i\hat N_-}\,,
\end{align}
where $e^{i\hat N_H}\equiv\hat U_H(+\infty,-\infty)$,
with $\hat U_H$ generated by the Hamiltonian in \eqn{eq:change-picture}.
The future factor $\hat\Omega_+$ describes asymptotic evolution from $t=0$ to $+\infty$,
while $\hat\Omega_-$ describes evolution from $-\infty$ to $t=0$:
\begin{subequations}
\begin{align}
    \hat{\Omega}_+
    &\equiv \hat U_{\rm soft}(+\infty,0) =e^{i\hat N_+}\,,
    \\
    \hat{\Omega}_-
    &\equiv \hat U_{\rm soft}(0,-\infty) = e^{i\hat N_-}\,.
\end{align}
\end{subequations}
Thus, $\hat{\Omega}_+$ and $\hat{\Omega}_-$ each account for one
half of the full evolution from $-\infty$ to $+\infty$ in the soft theory.
 
Let us now examine the soft operators $\hat{N}_\pm$ in more detail.
Using \eqn{eq:MagnusTerms}, their Magnus expansions are
\begin{subequations}
\begin{align}
    &i\hat N_+
    =
    -i\int_0^\infty \d t\,\hat H_{\rm soft}(t)\\
    &\!\!\!-\frac12
    \int_0^\infty \d t_1\! \int_0^\infty\d t_2\,
    \Theta(t_1\!-\!t_2)
    \big[
        \hat H_{\rm soft}(t_1),
        \hat H_{\rm soft}(t_2)
    \big]
    +\cdots\,,
    \nn\\
    &i\hat N_-
    =
    -i\int_{-\infty}^0 \d t\,\hat H_{\rm soft}(t)\\
    &\!\!\!-\frac12
    \int_{-\infty}^0  \d t_1\! \int_{-\infty}^0\d t_2\,
    \Theta(t_1\!-\!t_2)
    \big[
        \hat H_{\rm soft}(t_1),
        \hat H_{\rm soft}(t_2)
    \big]
    +\cdots\,.\nn
\end{align}
\end{subequations}
Consider first the terms linear in the soft field.
At first order in Magnus, the worldline contributions are
\begin{subequations}
\begin{align}
    \hat N_{+,1}^{\rm wl}
    &=
    -\frac{\kappa}{2}
    \sum_i \frac{\hat p_i^\mu\hat p_i^\nu}{m_i}
    \int_0^\infty\!\d\tau_i\,\hat{h}_{\mu\nu}(b_i+\tau_i \hat v_i)\,,
    \\
    \hat N_{-,1}^{\rm wl}
    &=
    -\frac{\kappa}{2}
    \sum_i \frac{\hat p_i^\mu\hat p_i^\nu}{m_i}
    \int_{-\infty}^0\!\d\tau_i\,\hat{h}_{\mu\nu}(b_i+\tau_i \hat v_i)\,.
    \label{eq:Npm-linear-wl}
\end{align}
\end{subequations}
The radiative contribution, using 
\begin{equation}
   \Delta E_k(q) \equiv E_{\vct{k}+\vct{q}}-E_{\vct{k}}\,, 
\end{equation}
and the half-line time integrals
\begin{subequations}\label{eq:half-line-integrals}
\begin{align}
    \int_0^\infty \d t\,\,
    e^{i\omega t}
    &=\frac{i}{\omega+i0^+}\,,
    \\
    \int_{-\infty}^{0}\d t\,\,
    e^{i\omega t}
    &=
    -\frac{i}{\omega-i0^+}\,,
\end{align}
\end{subequations}
is
\begin{subequations}
\begin{align}
    \hat N_{+,1}^{\rm rad}
    &=%
    -i\frac{\kappa}{2}
    \sum_\sigma
    \int_{\vct{k}}\int_{q}
    \frac{k^\mu k^\nu}{k^0} 
    \frac{
        \hat h_{\mu\nu}(q)\,
        \hat{a}^{\sigma\dagger}(k+q)\hat{a}^\sigma(k)\,
    }{
        \Delta E_k(q)-q^0+i0^+
    }
    \,,
    \\
    \hat N_{-,1}^{\rm rad}
    &=
    i\frac{\kappa}{2}
    \sum_\sigma
    \int_{\vct{k}}\int_{q}
    \frac{k^\mu k^\nu}{k^0}
    \frac{
        \hat h_{\mu\nu}(q)\,
        \hat{a}^{\sigma\dagger}(k+q)\hat{a}^\sigma(k)\,
    }{
        \Delta E_k(q)-q^0-i0^+
    }
    \,.
\end{align}
\end{subequations}
These terms are linear in $\hat h_{\mu\nu}(q)$ and generate the coherent
soft dressing of the hard degrees of freedom.

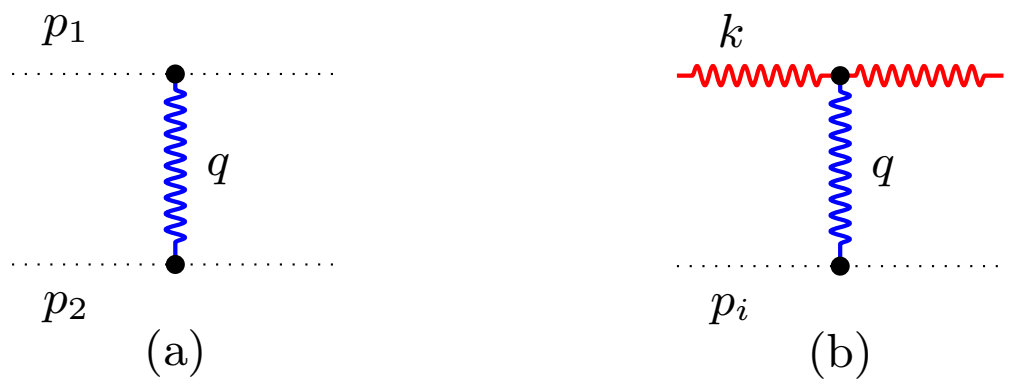
\begin{figure}[t]
    \centering
    \begin{tikzpicture}[scale=0.9,baseline={(current bounding box.center)}]
        \coordinate (A) at (0,0.7);
        \coordinate (B) at (0,-0.7);
        \draw[dotted] (-1.2,0.7) -- (1.2,0.7);
        \draw[dotted] (-1.2,-0.7) -- (1.2,-0.7);
        \draw[blue, photon] (A) -- (B);
        \fill (A) circle (0.07);
        \fill (B) circle (0.07);
        \node[above] at (-0.8,0.78) {$p_1$};
        \node[below] at (-0.8,-0.78) {$p_2$};
        \node[right] at (0.1,0) {$q$};
        \node at (0,-1.35) {(a)};
    \end{tikzpicture}
    \hspace{2cm}
    \begin{tikzpicture}[scale=0.9,baseline={(current bounding box.center)}]
        \coordinate (A) at (0,0.7);
        \coordinate (B) at (0,-0.7);
        \draw[red, photon] (-1.2,0.7) -- (A);
        \draw[red, photon] (A) -- (1.2,0.7);
        \draw[dotted] (-1.2,-0.7) -- (1.2,-0.7);
        \draw[blue, photon] (A) -- (B);
        \fill (A) circle (0.07);
        \fill (B) circle (0.07);
        \node[above] at (-0.8,0.78) {$k$};
        \node[below] at (-0.8,-0.78) {$p_i$};
        \node[right] at (0.1,0) {$q$};
        \node at (0,-1.35) {(b)};
    \end{tikzpicture}
    \caption{The two parts of the soft operator
    $\hat{N}_{\rm soft}=\hat N_{\rm Coul}+\hat N_{\rm tail}$~\eqref{eq:softN}:
    (a) Coulomb pole, where a soft graviton (blue) is exchanged between the worldlines, and
    (b) tail pole, where a soft graviton (blue) is exchanged between a hard graviton (red) and each worldline.}
    \label{fig:soft-Coul-tail}
\end{figure}

The $1/\eps$ poles are generated at second order in the Magnus expansion.
The commutator of the soft Hamiltonian $\hat H_{\rm soft}$ decomposes as 
\begin{align}
    \label{eq:second-Magnus-soft-split}
    &[\hat H_{\rm soft}(t_1),\hat H_{\rm soft}(t_2)]
    ={}
    [\hat H_{\rm wl}^{s}(t_1),\hat H_{\rm wl}^{s}(t_2)]
    \\
    &\qquad+
    [\hat H_{\rm wl}^{s}(t_1),\hat H_{\rm rad}^{s}(t_2)]
    +
    [\hat H_{\rm rad}^{s}(t_1),\hat H_{\rm wl}^{s}(t_2)]
    +\cdots\,.\nonumber
\end{align}
Taking first the massive-massive contribution
$[\hat H_{\rm wl}^{s}(t_1),\hat H_{\rm wl}^{s}(t_2)]$, we obtain ---
up to an overall $1/2$ --- the Coulomb term, illustrated in \Fig{fig:soft-Coul-tail}(a):
\begin{align}\label{eq:Npm-Coul}
    \left.
    \hat N_{+,2}^{\rm wl-wl}
    \right|_{1/\eps}
    =
    \left.
    \hat N_{-,2}^{\rm wl-wl}
    \right|_{1/\eps} \equiv \frac{1}{2} \hat N_{\rm Coul}\,.
\end{align}
Plugging in two copies of $\hat H_i^s(t)$~\eqref{eq:Hwl-soft},
the time-ordered soft field commutator
$\Theta(x^0-y^0)[\hat{h}_{\mu\nu}(x),\hat{h}_{\rho\sigma}(y)]$
generates the retarded graviton propagator~\eqref{eq:gravProp}.
The two remaining half-line time integrals are evaluated using \eqn{eq:half-line-integrals},
leaving\footnote{
    The two worldline orderings give the time-symmetric
    average of the retarded and advanced propagators:
    \begin{align}
    \frac12\left[\frac{1}{q^2+i0^+\,q^0} +\frac{1}{q^2-i0^+\,q^0}\right]
    =\operatorname{PV}\left(\frac{1}{q^2}\right)\,.
    \end{align}
}
\begin{align}\label{coulombPole}
    \hat N_{\rm Coul} &=-\frac{\kappa^2}{2}
    \int_q\,
    \operatorname{PV}\left(\frac{1}{q^2}\right) \nn \\
    &\quad\times\frac{
        \left[
            (\hat p_1\cdot\hat p_2)^2
            -(m_1^2m_2^2)/(D-2)
        \right]
        e^{-i q\cdot(b_1-b_2)}
    }{
        (\hat p_1\cdot q-i0^+)
        (\hat p_2\cdot q+i0^+)
    }
    \Bigg|_{1/\epsilon} \nonumber \\
     &=
     -\frac{Gm_1m_2}{\eps}\,
     F(\hat\gamma)\,,
\end{align}
in agreement with \rcite{Weinberg:1965nx}.
Here we have introduced
\begin{align}\label{Fgamma}
    F(\gamma)&=
    \frac{2\gamma^2-1}{\sqrt{\gamma^2-1}}\,, &
    \hat\gamma
    &\equiv
    \frac{\hat p_1\cdot\hat p_2}{m_1m_2}\,. 
\end{align}
The $1/\eps$ produced in $\hat{N}_{\rm Coul}$ is purely an IR pole.
The ultraviolet (UV) divergence that one might have expected, given the overall $\int\!{\rm d}^Dq/q^4$
power counting, is regulated by the $e^{i q\cdot b}$ phase factor
which drops out in the soft limit $q^\mu\to0$.

The sum of the two mixed worldline-radiative commutators
$[\hat H_{\rm wl}^{s}(t_1),\hat H_{\rm rad}^{s}(t_2)]$ and
$[\hat H_{\rm rad}^{s}(t_1),\hat H_{\rm wl}^{s}(t_2)]$ gives the tail
term, depicted in \Fig{fig:soft-Coul-tail}(b),
with equal contributions on the past and future half-lines:
\begin{align}\label{eq:Npm-tail}
    \left.
    \hat N_{+,2}^{\rm wl-rad}
    \right|_{1/\eps}
    =
    \left.
    \hat N_{-,2}^{\rm wl-rad}
    \right|_{1/\eps} \equiv 
    \frac{1}{2}\hat N_{\rm tail}\,.
\end{align}
Again, the time-ordered commutator gives the retarded graviton propagator~\eqref{eq:gravProp}.
In this case, it is convenient to work in the $i$-th particle's rest frame,
wherein $v_i^0=1$ and we identify $t=\tau_i$~\eqref{eq:Hwl-soft}.
The numerator of the graviton propagator gives $P_{00;\alpha\beta}k^\alpha k^\beta=(k^0)^2$,
and so the tail operator becomes
\begin{align}  \label{taildef}
    &\hat N_{\rm tail}
    =-\frac{\kappa^2}{2}
    \sum_{i,\sigma}m_i\int_{\vct{k}}k^0
    \\
    &\quad\times \int_q\,
    \frac{
        e^{iq\cdot b_i}
        \hat a^{\sigma\dagger}(k+q)
        \hat a^\sigma(k)
    }{
        (q^0+i0^+)
        [q^0-\Delta E_k(q)-i0^+]
    }\,
    \operatorname{PV}\left(\frac1{q^2}\right) \Bigg|_{1/\eps}\,. \nn 
\end{align}
The two energy denominators come from the worldline and
hard graviton time integrals~\eqref{eq:half-line-integrals}.
To obtain the IR pole we keep the leading soft coefficient:
\begin{align}
    e^{iq\cdot b_i} \hat a^{\sigma\dagger}(k+q)
    \hat a^\sigma(k)
    =
    \hat a^{\sigma\dagger}(k)\hat a^\sigma(k)
    +\cO(\lambda)\,,
\end{align}
where $q^\mu\sim\lambda$ at fixed $k^\mu$ and $b_i^\mu$.
Then we can evaluate the $q^0$ integral in \eqn{taildef} directly, obtaining
\begin{align}
    \hat N_{\rm tail}
    &= 
    \frac{\kappa^2\pi}{2}
    \sum_{i,\sigma}\int_{\vct{k}}(k\cdot p_i)\,
    \hat a^{\sigma\dagger}(k)\hat a^\sigma(k) \\
    &\qquad\qquad\times
    \int\!\frac{\d^{D-1}\vct{q}}{(2\pi)^{D-1}}\,
    \frac{\delta(\Delta E_k(q))}
         {|\vct{q}|^2}
    \Bigg|_{1/\eps}
    \nonumber\\
    &=
    -\frac{2G}{\eps}
    \sum_\sigma\int_{\vct{k}}
    (k\cdot\hat P)\,
    \hat a^{\sigma\dagger}(k)\hat a^\sigma(k)\,,
    \label{tailPole}
\end{align}
where $\hat P^\mu=\hat p_1^\mu+\hat p_2^\mu$. It is worth noticing that the momentum integral is the forward integral evaluated in Appendix~\ref{app:tailAmplitude} in the massive particle rest frame.
Moreover, we observe that the tail operator is a function only of the graviton number operator,
and therefore vanishes when acting on the radiative vacuum state.

All $1/\eps$ poles can therefore be collected into
the universal soft operator, depicted in \Fig{fig:soft-Coul-tail}:
\begin{align}\label{eq:softN}
    \hat N_{\rm soft}
    =
    \underbrace{-\frac{Gm_1m_2}{\eps}F(\hat{\gamma})
    \vphantom{\sum_\sigma\int_{\vct{k}}}}_{\hat{N}_{\rm Coul}}
    \underbrace{-\frac{2G}{\eps}
    \sum_\sigma\int_{\vct{k}}
    (k\cdot\hat P)\,
    \hat a^{\sigma\dagger}(k)\hat a^\sigma(k)}_{\hat{N}_{\rm tail}}\,,
\end{align}
where $F(\hat\gamma)$ was introduced in \eqn{Fgamma}.
Noticing that the pole factor enters the two asymptotic factors with the same
sign in the exponent, $\log\hat{\Omega}_\pm|_{1/\eps} = \frac{i}{2}\hat N_{\rm soft}$,
we are prompted to extract the common soft $1/\epsilon$ pole factor by defining
\begin{align}
    \hat{\Omega}_+
    &=
    e^{i\hat N_{\rm soft}/2}\hat{\Omega}'_+\,,
    &
    \hat{\Omega}_-
    &=
    \hat{\Omega}_-^{\prime}
    e^{i\hat N_{\rm soft}/2}\,.
    \label{eq:Omega-soft-factor}
\end{align}
Substituting \eqn{eq:Omega-soft-factor} into
the $\hat{N}$-matrix decomposition~\eqref{asympFactorisation} gives
\begin{align}
    e^{i\hat N}
    &=
    e^{i\hat N_{\rm soft}/2}
    \left(
        \hat{\Omega}^\prime_+\,
        e^{i\hat N_H}\,
        \hat{\Omega}_-^{\prime}
    \right)
    e^{i\hat N_{\rm soft}/2}\,.
\end{align}
We define the operator remaining between the two universal soft factors by
\begin{align}
    e^{i\hat N_{\rm hard}}
    \equiv
    \hat\Omega'_+\,
    e^{i\hat N_H}\,
    \hat\Omega_-^{\prime}\,.
     \label{eq:Nhard-definition}
\end{align}
and therefore obtain
\begin{align}\label{symmSoft}
    e^{i\hat N}
    =
    e^{i\hat N_{\rm soft}/2}
    e^{i\hat N_{\rm hard}}
    e^{i\hat N_{\rm soft}/2}\,.
\end{align}
While the underlying IR factorisation \eqref{asympFactorisation} is systematic,
with $\hat\Omega_\pm$ containing the full Magnus evolution and its higher-order iterations,
in \eqn{symmSoft} we have made explicit only the IR structures required through 4PM order.
Additional soft regions, such as those associated with tail-of-tail effects or
massless-massless divergences, can be incorporated by extending $\hat N_{\rm soft}$~\eqref{eq:softN}.

\subsection{Soft divergences of $\hat{N}$-matrix elements}

Given the structure of the soft divergences in \eqn{symmSoft},
let us now explore the implications for matrix elements of $\hat{N}$.
To compare $\hat{N}$ with $\hat{N}_{\rm hard}$
in \eqn{symmSoft} we use the Baker--Campbell--Hausdorff (BCH) formula twice:
\begin{align}\label{bch}
	i\hat{N}&=i\hat{N}_{\rm soft}+i\hat{N}_{\rm hard}
	+\sfrac1{12}[i\hat{N}_{\rm hard},[i\hat{N}_{\rm hard},i\hat{N}_{\rm soft}]]\nonumber\\
     &\qquad-\sfrac1{24}[i\hat{N}_{\rm soft},[i\hat{N}_{\rm soft},i\hat{N}_{\rm hard}]]+\cdots\,.
\end{align}
The brackets on the right-hand side
are resolved by inserting the coherent-state expansion~\eqref{coherent} for $\hat{N}_{\rm hard}$,
after which we may bracket on specific Fock graviton states.
All remaining brackets must involve the worldline operators,
and so can be re-interpreted as classical Poisson brackets acting
on the background parameters at past infinity~\eqref{eq:PBgi}: $[\hat{A},\hat{B}]\to i\{A,B\}$.
Finally, we insert the PM expansion $N=GN^{(1)}+G^2N^{(2)}+\cdots$.
This procedure is exactly the same as that used to 
derive the observables~\eqref{obsFromN}.

At leading-PM order,
the BCH corrections in \eqn{bch} play no role.
The 1PM Magnusian is given by\footnote{Seeing
    as the $\cO(S)$ spinning corrections to $N^{(1)}$ do not contribute to the soft pole,
    we do not restate them here --- the full expression for $N^{(1)}$
    up to quadratic order in spin is given in \rcite{Jakobsen:2021zvh}.}
\begin{align}\label{1PMdiv}
	&iN^{(1)}=\,\,
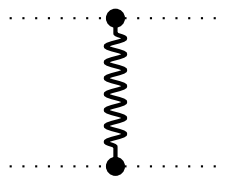
    \begin{tikzpicture}[baseline={(current bounding box.center)},scale=.7]
    \coordinate (inA) at (0,.7);
    \coordinate (outA) at (2,.7);
    \coordinate (inB) at (0,-.7);
    \coordinate (outB) at (2,-.7);
    \coordinate (xA) at (1,.7);
    \coordinate (xB) at (1,-.7);
    \draw [dotted] (inA) -- (outA);
    \draw [dotted] (inB) -- (outB);
    \draw [photon] (xA) -- (xB);
    \draw [fill] (xA) circle (.08);
    \draw [fill] (xB) circle (.08);
  \end{tikzpicture}\\
    &=\bigg(\frac{2R}{|b|}\bigg)^{-2\eps}
    \frac{2ie^{-\eps\gamma_E}\Gamma(-\eps)((D\!-\!2)\gamma^2-1)m_1m_2}{(D\!-\!2)\sqrt{\gamma^2-1}}
    +\cO(S)\nn\\
    &=-i\frac{m_1m_2}{\eps}F(\gamma)+\cO(\eps^0)\,.\nn
\end{align}
This soft divergence is accounted for by the Coulomb pole~\eqref{coulombPole},
confirming that $N^{(1)}_{\rm hard}=N^{(1)}-N^{(1)}_{\rm soft}$ is finite.
Similarly, the leading-PM Compton amplitude
$N(-k_1^{\sigma_1},k_2^{\sigma_2})=\langle k_2^{\sigma_2}|\hat{N}|k_1^{\sigma_1}\rangle$
is given by
\begin{align}
	&iN^{(1)}(-k_1^{\sigma_1},k_2^{\sigma_2})\,\,=\,\,
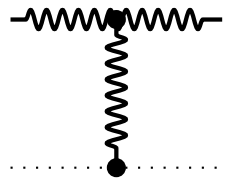
    \begin{tikzpicture}[baseline={(current bounding box.center)},scale=.7]
    \coordinate (inA) at (0,.7);
    \coordinate (outA) at (2,.7);
    \coordinate (inB) at (0,-.7);
    \coordinate (outB) at (2,-.7);
    \coordinate (xA) at (1,.7);
    \coordinate (xB) at (1,-.7);
    \draw [photon] (inA) -- (outA);
    \draw [dotted] (inB) -- (outB);
    \draw [photon] (xA) -- (xB);
    \draw [fill] (xA) circle (.08);
    \draw [fill] (xB) circle (.08);
  \end{tikzpicture}
  \,\,\,\,+\,\,\,\,
  
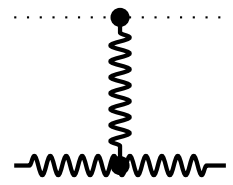
\begin{tikzpicture}[baseline={(current bounding box.center)},scale=.7]
    \coordinate (inA) at (0,.7);
    \coordinate (outA) at (2,.7);
    \coordinate (inB) at (0,-.7);
    \coordinate (outB) at (2,-.7);
    \coordinate (xA) at (1,.7);
    \coordinate (xB) at (1,-.7);
    \draw [dotted] (inA) -- (outA);
    \draw [photon] (inB) -- (outB);
    \draw [photon] (xA) -- (xB);
    \draw [fill] (xA) circle (.08);
    \draw [fill] (xB) circle (.08);
  \end{tikzpicture}
  \,\,\,\,+\,\,\,\,\cdots\nn\\
    &\!\!=-\frac{2i}{\eps}\delta^{\sigma_1\sigma_2}(k_1\cdot P)2 k_1^0 \dd(\vct{k}_1-\vct{k}_2)%
    +\cO(\eps^0)\,,
\end{align}
ignoring diagrams that do not contribute to the soft pole ---
this expression is derived in Appendix~\ref{app:tailAmplitude}.
This agrees with the tail pole~\eqref{tailPole},
the $1/\eps$ this time being produced by soft graviton
exchange between the worldlines and a hard graviton.

At higher-PM orders the brackets in \eqn{bch} start to play a role.
Focusing on the worldline degrees of freedom,
Poisson brackets with the Magnusian vanish:
\begin{align}
	\{N,\gamma\}&=0\,, &
	\{N,P^\mu\}&=0\,.
\end{align}
Using the brackets in \eqn{eq:PBgi},
this statement follows from $N$ depending on $b_i^\mu$ only
via the relative impact parameter $b^\mu=b_2^\mu-b_1^\mu$,
so $\{b^\mu,\gamma\}=0$ and $\{b^\mu,P^\nu\}=0$ immediately imply the above.
For the radiative $\hat{N}$-matrix elements, more generally one can show that
\begin{subequations}\label{bracketsGammaP}
\begin{align}
	\{N(k_1^{\sigma_1},\ldots,k_n^{\sigma_n}),\gamma\}&=
    -\frac{i\sum_{i=1}^nk_i\cdot P}{m_1m_2}N(k_1^{\sigma_1},\ldots,k_n^{\sigma_n})\,,\label{bracketGamma}\\
	\{N(k_1^{\sigma_1},\ldots,k_n^{\sigma_n}),P^\mu\}&=
    \bigg(\!\!-i\sum_{i=1}^nk_i^\mu\bigg)N(k_1^{\sigma_1},\ldots,k_n^{\sigma_n})\,.\label{bracketP}
\end{align}
\end{subequations}
We derive these relationships in \App{bracketsApp}.
So the only way for the soft matrix $\hat{N}_{\rm soft}$ to generate corrections
to $\hat{N}$ is via the radiative matrix elements $N(k_1,k_2,\ldots)$.

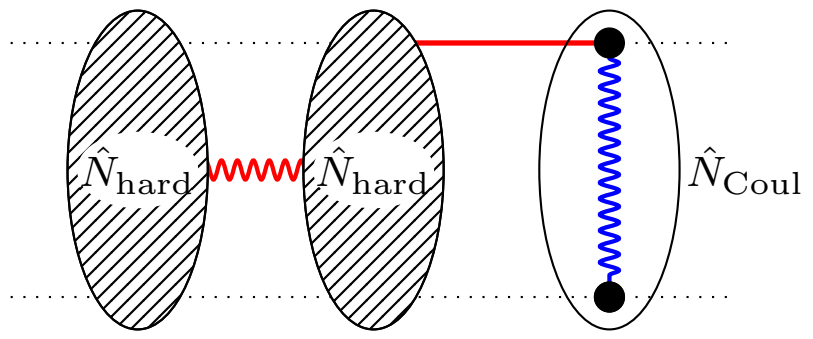
\begin{figure}[t]
 \begin{tikzpicture}[baseline={(current bounding box.center)},scale=1.2]
    \coordinate (inA) at (0,.7);
    \coordinate (outA) at (4,.7);
    \coordinate (inB) at (0,-.7);
    \coordinate (outB) at (4,-.7);
    \coordinate (xA) at (.7,.7);
     \coordinate (yA) at (2,.7);
     \coordinate (zA) at (3.3,.7);
     \coordinate (xB) at (.7,-.7);
    \coordinate (yB) at (2,-.7);
    \coordinate (zB) at (3.3,-.7);
    \coordinate (xM) at (.7,0);
    \coordinate (yM) at (2,0);
    \coordinate (zM) at (3.3,-0);
    \draw [dotted] (inA) -- (outA);
    \draw [dotted] (inB) -- (outB);
    \draw [draw=none] (xA) to[out=40,in=140] (zA);
    \draw [photon] (xA) -- (xM);
    \draw [photon] (xM) -- (xB);
    \draw [blue, photon] (zA) -- (zB);
    \draw [photon] (yB) -- (yM);
    \draw [red, photon] (xM) -- (yM);
    \draw [red, photon] (yM) -- (yA);
    \draw [red, zUndirected] (yA) -- (zA);
    \draw [fill] (xA) circle (.08);
    \draw [fill] (yA) circle (.08);
    \draw [fill] (zA) circle (.08);
    \draw [fill] (xB) circle (.08);
    \draw [fill] (yB) circle (.08);
    \draw [fill] (zB) circle (.08);
    \draw [fill] (xM) circle (.08);
    \draw [fill] (yM) circle (.08);
    \draw [fill] (zB) circle (.08);
    \draw[fill=white] (xM) ellipse (11pt and 25pt);
    \draw[pattern=north east lines] (xM) ellipse (11pt and 25pt);
    \fill[white] (xM) ellipse (9.5pt and 6pt);
    \node[font=\footnotesize] at (xM) {$\hat{N}_{\rm hard}$};
    \draw[fill=white] (yM) ellipse (11pt and 25pt);
    \draw[pattern=north east lines] (yM) ellipse (11pt and 25pt);
    \fill[white] (yM) ellipse (9.5pt and 6pt);
    \node[font=\footnotesize] at (yM) {$\hat{N}_{\rm hard}$};
    \draw[fill=none] (zM) ellipse (11pt and 25pt);
    \node[font=\footnotesize] (middle) at (4.05,0) {$\hat{N}_{\rm Coul}$};
  \end{tikzpicture}
\caption{The Coulomb pole of the 4PM Magnusian.
The two cut lines, highlighted in red,
are produced by causality cuts involving either gravitons or worldline deflection modes.}
\label{fig:Coulomb}
\end{figure}

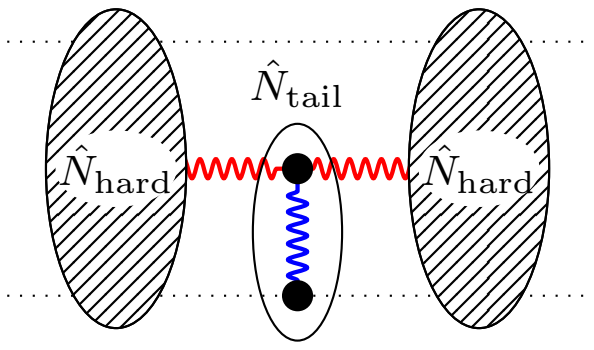
\begin{figure}[t]
 \begin{tikzpicture}[baseline={(current bounding box.center)},scale=1.2]
    \coordinate (inA) at (0.4,.7);
    \coordinate (outA) at (3.6,.7);
    \coordinate (inB) at (0.4,-.7);
    \coordinate (outB) at (3.6,-.7);
    \coordinate (xA) at (1,.7);
     \coordinate (yA) at (2,.7);
     \coordinate (zA) at (3,.7);
     \coordinate (xB) at (1,-.7);
    \coordinate (yB) at (2,-.7);
    \coordinate (zB) at (3,-.7);
    \coordinate (xM) at (1,0);
    \coordinate (yM) at (2,0);
    \coordinate (zM) at (3,-0);
    \draw [dotted] (inA) -- (outA);
    \draw [dotted] (inB) -- (outB);
    \draw [draw=none] (xA) to[out=40,in=140] (zA);
    \draw [photon] (xA) -- (xM);
    \draw [photon] (xM) -- (xB);
    \draw [photon] (zA) -- (zM);
    \draw [photon] (zM) -- (zB);
    \draw [blue, photon] (yB) -- (yM);
    \draw [red, photon] (xM) -- (yM);
    \draw [red, photon] (yM) -- (zM);
    \draw [fill] (xA) circle (.08);
    \draw [fill] (zA) circle (.08);
    \draw [fill] (xB) circle (.08);
    \draw [fill] (yB) circle (.08);
    \draw [fill] (zB) circle (.08);
    \draw [fill] (xM) circle (.08);
    \draw [fill] (yM) circle (.08);
    \draw [fill] (zM) circle (.08);
    \draw[fill=white] (xM) ellipse (11pt and 25pt);
    \draw[pattern=north east lines] (xM) ellipse (11pt and 25pt);
    \fill[white] (xM) ellipse (9.5pt and 6pt);
    \node[font=\footnotesize] at (xM) {$\hat{N}_{\rm hard}$};
    \draw[fill=white] (zM) ellipse (11pt and 25pt);
    \draw[pattern=north east lines] (zM) ellipse (11pt and 25pt);
    \fill[white] (zM) ellipse (9.5pt and 6pt);
    \node[font=\footnotesize] at (zM) {$\hat{N}_{\rm hard}$};
    \draw[fill=none] (2,-.35) ellipse (7pt and 17pt);
    \node[font=\footnotesize] (middle) at (2,.47) {$\hat{N}_{\rm tail}$};
  	\end{tikzpicture}
\caption{The tail pole of the 4PM Magnusian.
In this case, a hard (cut) graviton exchanges a soft graviton with either worldline.}
\label{fig:tail}
\end{figure}

The structure of the soft expansion~\eqref{bch},
particularly the absence of single brackets,
implies that soft corrections to $\hat{N}$ are delayed until quite high PM orders.
For the Magnusian $N$, the first correction produced by
$\frac1{12}[i\hat{N}_{\rm hard},[i\hat{N}_{\rm hard},i\hat{N}_{\rm soft}]]$
appears only at 4PM order:
two copies of $N_{\rm hard}(k)\sim G^{3/2}$
combine with $\hat{N}_{\rm soft}\sim G$ to give an overall $G^4$ ---
see Figs.~\ref{fig:Coulomb} and \ref{fig:tail}.\footnote{While
    technically $N_{\rm hard}(k)$ begins at $G^{1/2}$,
    i.e.~the single-emission vertex from the worldline,
    terms produced by this insertion are \emph{static}:
    they consist of zero-energy momenta on the external line.
}
The double-soft correction
$-\frac1{24}[i\hat{N}_{\rm soft},[i\hat{N}_{\rm soft},i\hat{N}_{\rm hard}]]$
has no effect, seeing as at least one graviton raising or lowering
operator will always remain to be annihilated by the vacuum $|0\rangle$.
So at 2PM and 3PM orders the Magnusian is totally unaffected by soft factorisation:
$N^{(2)}=N^{(2)}_{\rm hard}$ and $N^{(3)}=N^{(3)}_{\rm hard}$.
This is consistent with our observation in Section~\ref{sec:magangle}
that, at these low PM orders, an $L$-derivative of the Magnusian
gives the scattering angle $\theta$~\eqref{fullAngle},
which is a physical quantity and thus IR-safe.
In the case of the one-point function $N(k^\sigma)$,
we take one copy of $N_{\rm hard}(k^\sigma)\sim G^{3/2}$,
one copy of $N_{\rm hard}\sim G$,
and combine with $\hat{N}_{\rm soft}\sim G$ to give an overall $G^{7/2}$.
This agrees with the observation made in \rcite{Georgoudis:2023lgf} that,
at order $G^{5/2}$ involving one-loop Feynman integrals,
$N(k^\sigma)$ is IR-finite.

Focusing now on the 4PM Magnusian,
we begin with the Coulomb pole,
illustrated in Fig.~\ref{fig:Coulomb}.
It is a straightforward exercise to show that
\begin{align}
	&\frac1{12}\langle0|[i\hat{N}_{\rm hard},[i\hat{N}_{\rm hard},i\hat{N}_{\rm Coul}]]|0\rangle\\
    &=\frac{G}{6\eps}F'(\gamma)\sum_\sigma\int_{\vct{k}}(ik\cdot P)
	N_{\rm hard}(-k^\sigma)N_{\rm hard}(k^\sigma)+\cdots\,,\nn
\end{align}
the higher-order terms not being needed at 4PM order.
Seeing as $\hat{N}_{\rm Coul}$~\eqref{coulombPole}
depends on the kinematics only via $\hat{\gamma}$,
the innermost bracket above is evaluated using \eqn{bracketGamma} on $N_{\rm hard}(k)$.
Converting to a Poisson bracket on the incoming data, we have
\begin{align}
    \{N_{\rm hard}(k^{\sigma}),N_{\rm Coul}\}&=
    \frac{G}{\eps}(ik\cdot P)F'(\gamma)N_{\rm hard}(k^{\sigma})\,.
\end{align}
The outermost bracket, which involves the graviton raising and lowering operators~\eqref{raisinglowering},
connects $N_{\rm hard}(k^{\sigma})$ with $N_{\rm hard}(-k^{\sigma})$ via a cut graviton.
The 4PM tail divergence, illustrated in Fig.~\ref{fig:tail}, has a similar derivation:
\begin{align}
	&\frac1{12}\langle0|[i\hat{N}_{\rm hard},[i\hat{N}_{\rm hard},i\hat{N}_{\rm tail}]]|0\rangle\\
    &=-\frac{G}{3\eps}\sum_\sigma\int_{\vct{k}}(ik\cdot P)
	N_{\rm hard}(-k^\sigma)N_{\rm hard}(k^\sigma)+\cdots\,.\nn
\end{align}
In this case, both brackets are evaluated using the graviton
raising and lowering operators~\eqref{raisinglowering}.

To complete our derivation of the 4PM soft pole,
we identify the radiated energy in the initial centre-of-mass frame:
\begin{align}\label{Erad}
E_{\rm rad}=\frac1E\sum_\sigma
\int_{\vct{k}}(k\cdot P)N(-k^\sigma)N(k^\sigma)+\cdots\,,
\end{align}
where $E=|P|=\sqrt{m_1^2+m_2^2+2m_1m_2\gamma}$ is the total incoming energy.
\Eqn{Erad} is derived from the linear momentum flux~\eqref{eq:DeltaP_rad} by taking
the time component in the centre-of-mass frame,
and resolving the Poisson bracket with \eqn{bracketP}.
Identifying $N(k^\sigma)=N_{\rm hard}(k^\sigma)$ at leading-PM order,
we are finally left with
\begin{align}\label{4pmDivergence}
	N^{(4)}=N^{(4)}_{\rm hard}-\frac{E}{3\eps}\bigg(1-\frac12F'(\gamma)\bigg)E^{(3)}_{\rm rad}\,,
\end{align}
where $E^{(3)}_{\rm rad}$ is the 3PM radiated energy,
including linear-in-spin contributions~\cite{Jakobsen:2022zsx}.
This is our main prediction for the soft divergence of the 4PM Magnusian,
which we confirm by explicit calculation in Section~\ref{sec:calculation}.

\subsection{Scattering observables}
\label{sec:irfactorsforobs}

Now we explore what the implications of soft factorisation~\eqref{symmSoft}
are for the computed results of physical scattering observables.
Consider the change in the worldline operator $\hat\lambda^\mu$,
which could be $\hat{p}_i^\mu$ (the momentum impulse) or $\hat{S}_i^\mu$ (the spin kick).
These observables are fully inclusive,
and all soft divergences should cancel out.
To see this explicitly,
we plug the decomposition of $\hat{N}$ into hard and soft parts~\eqref{symmSoft}
into the in-in expectation for $\lambda^\mu+\Delta\lambda^\mu$~\eqref{ininObservable}:
\begin{align}\label{nestedSoft}
\begin{aligned}
	&\lambda^{\mu}+\Delta\lambda^\mu=\langle0|e^{-i\hat{N}}\hat{\lambda}^\mu e^{i\hat{N}}|0\rangle=
    \langle0|e^{-i[\hat{N},\bullet]}\hat{\lambda}^\mu|0\rangle\\
    &=\langle0|e^{-i[\hat{N}_{\rm soft}/2,\bullet]}(e^{-i[\hat{N}_{\rm hard},\bullet]}(e^{-i[\hat{N}_{\rm soft}/2,\bullet]}\hat{\lambda}^\mu))|0\rangle\,.
\end{aligned}
\end{align}
The three nested adjoint operators
come from the three factors in \eqn{symmSoft}.

First we observe that $\hat{N}_{\rm soft}$ does not depend on position.
Thus, $[\hat{N}_{\rm soft},\hat\lambda^\mu]=0$
and so the innermost term in \eqn{nestedSoft} simplifies to
$e^{-i[\hat{N}_{\rm soft}/2,\bullet]}\hat{\lambda}^\mu=\hat{\lambda}^\mu$.
In the outermost soft factor,
the contribution from $\hat{N}_{\rm tail}$~\eqref{tailPole} is
eliminated by $\hat{a}^\sigma(k)$ or $\hat{a}^{\sigma\dagger}(k)$
acting on the vacuum $|0\rangle$.
Each insertion of $\hat{N}_{\rm tail}$ introduces
a pair of raising and lowering operators,
of which a nested bracket only eliminates one --- not both.
We are left with
\begin{align}
	\lambda^\mu+\Delta\lambda^\mu
    &=e^{\{N_{\rm Coul}/2,\bullet\}}\langle0|e^{-i[\hat{N}_{\rm hard},\bullet]}\hat{\lambda}^\mu|0\rangle\,.\nn
\end{align}
Given that all remaining brackets involving $N_{\rm Coul}$ must now involve
the worldline degrees of freedom,
we have taken the liberty of extracting this divergence
and replacing $[\hat{A},\hat{B}]\to i\{A,B\}$,
the Poisson brackets on background variables~\eqref{eq:PBgi}.
Yet, seeing as $N_{\rm Coul}$ depends on the kinematic data only via $\gamma$,
and $\{b^\mu,\gamma\}=\{v_i^\mu,\gamma\}=\{S_i^\mu,\gamma\}=0$,
this outer factor also cancels.

We therefore conclude that
\begin{align}
	\lambda^\mu+\Delta\lambda^\mu&=
    \langle0|e^{-i[\hat{N}_{\rm hard},\bullet]}\hat{\lambda}^\mu|0\rangle\,.
\end{align}
So only the IR-finite $\hat{N}_{\rm hard}$ is relevant
for the momentum impulse $\Delta p_i^\mu$ and spin kick $\Delta S_i^\mu$,
and these observables are manifestly free of soft divergences.
The cancellation of divergences does \emph{not} apply to the change
in the impact parameters $\Delta b_i^\mu$,
which diverges at leading-PM order~\cite{Bini:2024hme,Mogull:2025cfn,Jakobsen:2026ayw,Bohnenblust:2026ytj}.

\subsection{Gravitational waveform}

\begin{figure}[t]
 \begin{tikzpicture}[baseline={(current bounding box.center)},scale=1.2]
    \coordinate (hA) at (.7,.7);
    \coordinate (hB) at (.7,-.7);
    \coordinate (hM) at (.7,0);
    \coordinate (tM) at (1.7,0);
    \coordinate (tB) at (1.7,-.7);
    \coordinate (emission) at (2.7,0);
    \draw[dotted] (0,.7) -- (2.7,.7);
    \draw[dotted] (0,-.7) -- (2.7,-.7);
    \draw[photon] (hA) -- (hM);
    \draw[photon] (hM) -- (hB);
    \draw[red,photon] (hM) -- (tM);
    \draw[red,photon] (tM) -- (emission);
    \draw[blue,photon] (tM) -- (tB);
    \fill (hA) circle (.08);
    \fill (hB) circle (.08);
    \fill (tM) circle (.08);
    \fill (tB) circle (.08);
    \draw[fill=white] (hM) ellipse (11pt and 25pt);
    \draw[pattern=north east lines] (hM) ellipse (11pt and 25pt);
    \fill[white] (hM) ellipse (9.5pt and 6pt);
    \node[font=\footnotesize] at (hM) {$\hat{N}_{\rm hard}$};
    \draw[fill=none] (1.7,-.35) ellipse (7pt and 17pt);
    \node[font=\footnotesize] at (1.7,.47) {$\hat{N}_{\rm tail}$};
 \end{tikzpicture}
\caption{The tail pole of the one-loop gravitational waveform.
The emitted hard graviton, shown in red, exchanges a soft graviton,
shown in blue, with either worldline.}
\label{fig:waveformTail}
\end{figure}
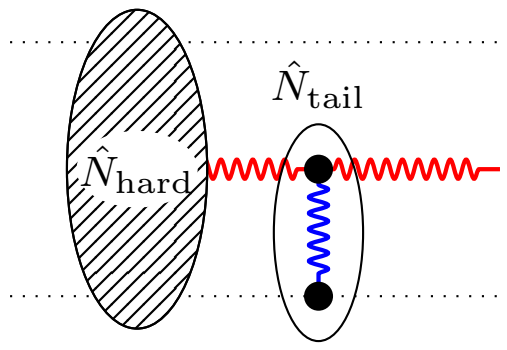

Finally, let us also re-derive the soft divergences of the gravitational waveform at $\cO(G^3)$,
involving one-loop integrals.
Our starting point is
\begin{align}\label{nestedWaveform}
	&\langle\hat{a}^\sigma(k)\rangle_{\rm in-in}=
    \langle0|e^{-i[\hat{N},\bullet]}\hat{a}^\sigma(k)|0\rangle\\
    &=\langle0|e^{-i[\hat{N}_{\rm soft}/2,\bullet]}(e^{-i[\hat{N}_{\rm hard},\bullet]}(e^{-i[\hat{N}_{\rm soft}/2,\bullet]}\hat{a}^\sigma(k)))|0\rangle\,.\nn
\end{align}
Unlike in our discussion of $\Delta p_i^\mu$ and $\Delta S_i^\mu$,
the soft divergences do not cancel.
On the innermost bracket,
only the tail divergence~\eqref{tailPole} plays a role,
as it depends on the graviton raising and lowering operators.
Using
\begin{align}
    [\hat{N}_{\rm tail},\hat{a}^\sigma(k)]=\frac{2G}{\epsilon}(k\cdot\hat{P})\hat{a}^\sigma(k)\,,
\end{align}
we see that the inner factor in \eqn{nestedWaveform} exponentiates:
\begin{align}\label{tailExponent}
    e^{-i[\hat{N}_{\rm tail}/2,\bullet]}\hat{a}^\sigma(k)=
    e^{-(G/\eps)(ik\cdot\hat{P})}\hat{a}^\sigma(k)\,.
\end{align}
For the one-loop pole, we need only the term linear in $G$ in this exponential,
acting on the leading hard emission. \Eqn{bracketP} gives
\begin{align}
    \{k\cdot P,N_{\rm hard}(k^\sigma)\}
    =
    ik^2N_{\rm hard}(k^\sigma)
    =
    0.
\end{align}
The tail phase can therefore be taken outside the hard expectation value through $\cO(G^{5/2})$
in the one-point function, giving the contribution illustrated in Fig.~\ref{fig:waveformTail}.
A similar pattern of soft factors manifesting as supertranslations
in this way was also seen in \rcites{Elkhidir:2024izo,Elkhidir:2026ahd}.

\begin{figure}[t]
 \begin{tikzpicture}[baseline={(current bounding box.center)},scale=1.2]
    \coordinate (cA) at (.7,.7);
    \coordinate (cB) at (.7,-.7);
    \coordinate (cM) at (.7,0);
    \coordinate (hA) at (2,.7);
    \coordinate (hB) at (2,-.7);
    \coordinate (hM) at (2,0);
    \coordinate (emission) at (3,0);
    \draw[dotted] (0,.7) -- (3,.7);
    \draw[dotted] (0,-.7) -- (3,-.7);
    \draw[blue,photon] (cA) -- (cB);
    \draw[red,zUndirected] (cA) -- (hA);
    \draw[red,photon] (hA) -- (hM);
    \draw[photon] (hM) -- (hB);
    \draw[red,photon] (hM) -- (emission);
    \fill (cA) circle (.08);
    \fill (cB) circle (.08);
    \fill (hA) circle (.08);
    \fill (hB) circle (.08);
    \draw[fill=none] (cM) ellipse (11pt and 25pt);
    \node[font=\footnotesize] at (-.05,0) {$\hat{N}_{\rm Coul}$};
    \draw[fill=white] (hM) ellipse (11pt and 25pt);
    \draw[pattern=north east lines] (hM) ellipse (11pt and 25pt);
    \fill[white] (hM) ellipse (9.5pt and 6pt);
    \node[font=\footnotesize] at (hM) {$\hat{N}_{\rm hard}$};
 \end{tikzpicture}
\caption{The Coulomb pole of the one-loop gravitational waveform,
with the emitted (hard) graviton drawn in red.
Here $\hat{N}_{\rm hard}$ is connected to the Coulomb blob by a worldline cut,
also shown in red.}
\label{fig:waveformCoulomb}
\end{figure}
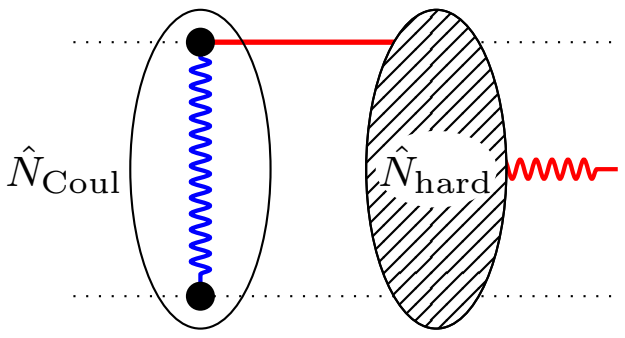

Meanwhile, on the outermost factor only the Coulomb pole is relevant,
as the tail pole annihilates the vacuum,
exactly as in the discussion of Section~\ref{sec:irfactorsforobs}.
So we have
\begin{align}
\begin{aligned}
	&\langle\hat{a}^\sigma(k)\rangle_{\rm in-in}\\
    &=e^{-(G/\eps)(ik\cdot P)}e^{\{N_{\rm Coul}/2,\bullet\}}
    \langle0|e^{-i[\hat{N}_{\rm hard},\bullet]}\hat{a}^\sigma(k)|0\rangle\,.
\end{aligned}
\end{align}
The exponentiated action of $\{N_{\rm Coul}/2,\bullet\}$ produces iterations
of $-\frac{G}{2\eps}(ik\cdot P)F'(\gamma)$, which exponentiate in the same manner
as those of the tail pole~\eqref{tailExponent}. Indeed, using \eqn{bracketGamma},
\begin{align}
  \hspace{-8pt}  \{N_{\rm Coul},N_{\rm hard}(k^\sigma)\}
    =
    -\frac{G}{\eps}
    (ik\cdot P)F'(\gamma)N_{\rm hard}(k^\sigma).
\end{align}
This gives the Coulomb contribution to the one-loop IR pole, illustrated in Fig.~\ref{fig:waveformCoulomb}.

Putting the pieces together, we are finally left with
\begin{align}\label{waveformPole}
	\langle\hat{a}^\sigma(k)\rangle_{\rm in-in} &=\exp\left(-\sfrac{G}{\eps}(ik\cdot P)\left(1+\sfrac12F'(\gamma)\right)\right)\\
    &\quad \times \langle0|e^{-i[\hat{N}_{\rm hard},\bullet]}\hat{a}^\sigma(k)|0\rangle +\cO(G^{7/2})\,.\nn
\end{align}
The explicit exponential gives the Coulomb and single tail contributions to the one-loop  waveform, in agreement with the literature~\cite{Georgoudis:2023lgf,Georgoudis:2023ozp,Brandhuber:2023hhy,Herderschee:2023fxh,Bohnenblust:2023qmy,Bohnenblust:2025gir}.

For the waveform we use the canonical Bondi-Metzner-Sachs (BMS) frame,
with vanishing Bondi shear at $u\to-\infty$,
and exclude exactly zero frequency on-shell states from the intermediate-state
sums~\cite{Bini:2024rsy,Bohnenblust:2025gir}.
The same soft operators enter the 4PM Magnusian pole~\eqref{4pmDivergence},
and indeed the divergence in \eqn{waveformPole} is almost identical ---
the only difference being the sign on $F'(\gamma)$.
At $\cO(G^4)$ in the waveform, the double-soft BCH term in \eqn{bch} also contributes. It must be combined with higher iterations of \eqn{nestedWaveform} and additional soft regions, including those associated with tail-of-tail effects.

\section{4PM Magnusian}
\label{sec:calculation}

In this section we explain our procedure for calculating the 4PM Magnusian $N^{(4)}$,
both in the probe limit $N^{(4,0)}$ and at 1SF order $N^{(4,1)}$,
and present our results ---
confirming the soft factorisation proposed in the previous section.
The PM-expanded Magnusian is
\begin{align}\label{eq:PM-SF-expansion}
    N=\sum_iG^i N^{(i)}=
    \sum_{i,j} \frac{G^i M^{i+1}}{|b|^{i-1}} \nu^{j+1} N^{(i,j)}\,,
\end{align}
where $i$ counts PM orders in $G$.
The polynomial dependence of $N$ on the masses $m_i$ ---
a consequence of the WQFT Feynman rules ---
implies a clean SF counting with respect to $j$ in powers of $\nu$.
The coefficients $N^{(i,j)}$ are dimensionless.

In principle, 
we should draw all vacuum 4PM Feynman diagrams,
summed over all possible internal causality flows and weighted
by Murua coefficients --- see \App{murua}.
These should then be integrated by standard procedure:
integration-by-parts (IBP) reduction to master integrals
\cite{Chetyrkin:1981qh,Maierhofer:2017gsa,Lange:2025fba},
and then insertion of expressions for those masters.
Yet we are able to make use of several important simplifications,
and recycle work done to calculate scattering observables ---
the momentum impulse $\Delta p_1^\mu$ and spin kick $\Delta S_1^\mu$ ---
by two of the present authors up to linear order in spin~\cite{Jakobsen:2023ndj,Jakobsen:2023hig}.

\subsection{Magnusian from the impulse}

Our strategy for calculating the Magnusian $N$
is to first obtain an integrand without
the $i0^+$ prescriptions assigned, and determine them later on.
To do so, we use the momentum impulse $\Delta p_1^\mu$.
From \eqn{obsFromN} we note that the impulse is related to $N$ by
\begin{align}\label{leadingDp1}
	\Delta p_1^\mu=\{N,p_1^\mu\}+\cdots\,.
\end{align}
The higher-order terms $\cdots$ involve iterations
of lower-PM $\hat{N}$-matrix elements joined together by causality cuts.
At the integrand level, these can be interpreted as
differences between retarded and advanced propagators,
both for the worldlines and for the gravitons~\cite{Haddad:2025cmw}.
Suppressing internal $i0^+$ prescriptions at the integrand level
identifies retarded with advanced propagators,
causing these causality cuts to vanish
and making \eqn{leadingDp1} a precise equality.

Meanwhile, the Magnusian $N$ is given by a Fourier transform:
\begin{align}
\begin{aligned}
	N&=\int_q e^{iq\cdot b}\dd(q\cdot v_1)\dd(q\cdot v_2)N(q)\\
	\implies\{N,p_1^\mu\}&=\int_q e^{iq\cdot b}\dd(q\cdot v_1)\dd(q\cdot v_2)N(q)(iq^\mu)\,.
\end{aligned}
\end{align}
$N(q)$ is the integrand of $N$ prior to Fourier transform
with $q\cdot v_1=q\cdot v_2=0$, $q^\mu$ being the total exchanged momentum.
In the second step, we have acted with the Poisson bracket \eqref{bvBracket},
whose action is a derivative with respect to the impact parameter $b^\mu$.
Comparing this with \eqn{leadingDp1},
we see that $N(q)$ is isolated by taking the $q^\mu$-component
of the integrand for $\Delta p_1^\mu$.

This property can also be understood in terms of diagrams.
The Feynman diagrams of the impulse and Magnusian are almost the same:
besides the $i0^+$ prescriptions of internal propagators,
which we are currently ignoring,
the only other difference is the outgoing deflection mode
on the first worldline in $\Delta p_1^\mu$,
which raises the tensor rank of the integrand.
We recall that worldline vertices satisfy a
recursive property~\cite{Mogull:2020sak,Jakobsen:2021zvh}:
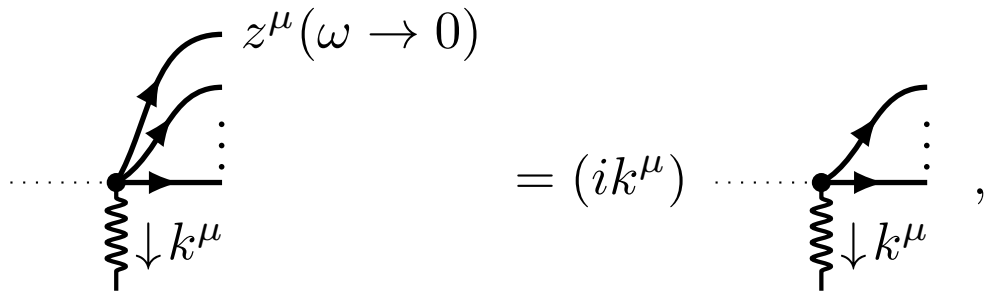
\begin{align}
  \begin{tikzpicture}[baseline={([yshift=-.5ex]x)},scale=.7]
  \coordinate (in) at (-1,0);
  \coordinate (out1) at (1,0);
  \coordinate (out2) at (1,0.5);
  \coordinate (out3) at (1,0.9);
  \coordinate (out4) at (1,1.4);
  \coordinate (x) at (0,0);
  \coordinate (y) at (0,-.6);
  \draw (y) node [right] {$\downarrow\!k^\mu$};
    \node (k1) at (-.8,-1.3) {};
    \node (k2) at (0,-1.2) {};
    \node (k3) at (.8,-1.3) {};
  \draw (out2) node [right] {$\!\!\!\vdots$};
  \draw (out4) node [right] {$z^\mu(\omega\to0)$};
  \draw [dotted] (in) -- (x);
  \draw [zParticle] (x) -- (out1);
  \draw [zParticle] (x) to[out=30,in=180] (out3);
  \draw [zParticle] (x) to[out=60,in=180] (out4);
  \draw [photon] (x) -- (k2);
  \draw [fill] (x) circle (.08);
  \end{tikzpicture}&=
  (ik^\mu)\,\,\,
  \begin{tikzpicture}[baseline={([yshift=-.5ex]x)},scale=.7]
    \coordinate (in) at (-1,0);
    \coordinate (out1) at (1,0);
    \coordinate (out2) at (1,0.5);
    \coordinate (out3) at (1,0.9);
    \coordinate (x) at (0,0);
    \coordinate (y) at (0,-.6);
  \draw (y) node [right] {$\downarrow\!k^\mu$};
     \node (k1) at (-.8,-1.3) {};
    \node (k2) at (0,-1.2) {};
    \node (k3) at (.8,-1.3) {};
    \draw (out1) node [right] {};
    \draw (out2) node [right] {$\!\!\!\vdots$};
    \draw (out3) node [right] {};
    \draw [dotted] (in) -- (x);
    \draw [zParticle] (x) -- (out1);
    \draw [zParticle] (x) to[out=30,in=180] (out3);
     \draw [photon] (x) -- (k2);
    \draw [fill] (x) circle (.08);
    \end{tikzpicture}\,,
\end{align}
where one of the deflection modes carries zero energy ($\omega=0$).
So, taking the comb graphs at 3PM order in the probe limit as an example,
this implies that the relevant graph integrands are related
to those in the Magnusian by:
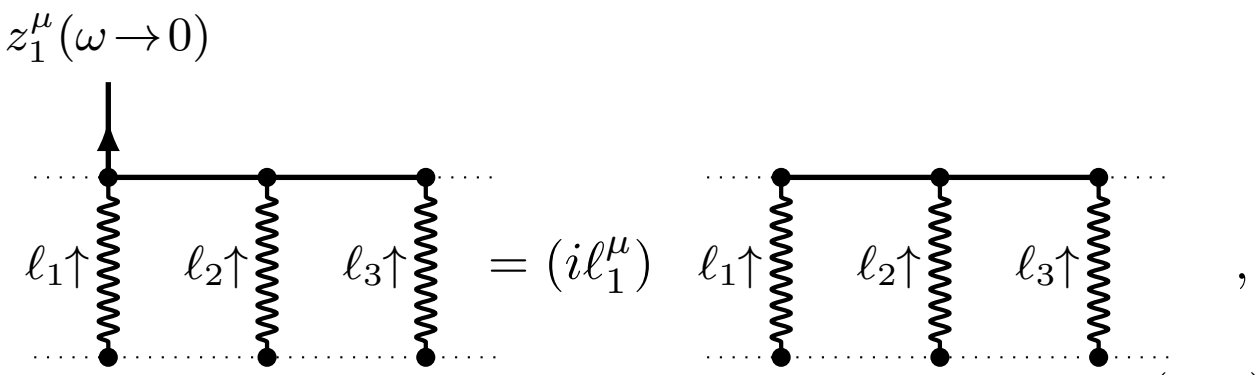
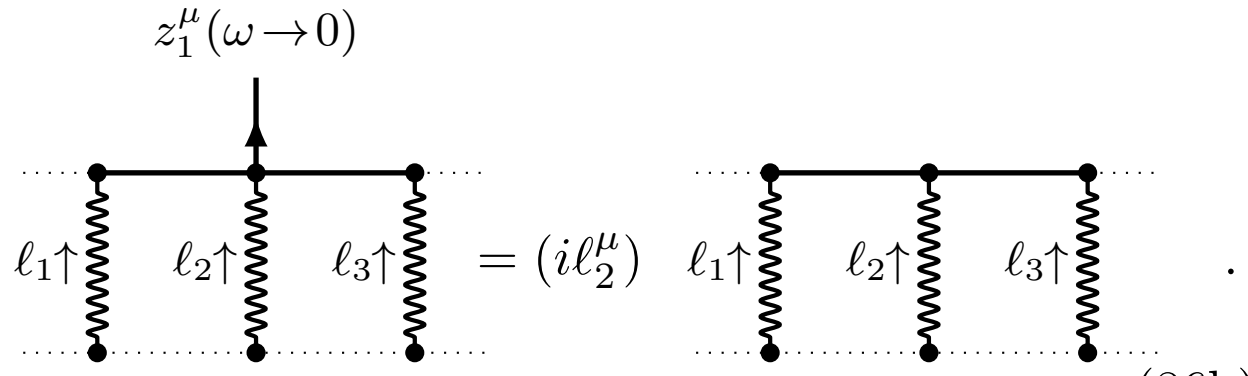
\begin{subequations}
\begin{align}
\begin{tikzpicture}[scale=.7,baseline={([yshift=-.5ex]combMid)}]
  \combSkeleton
  \coordinate (combMid) at ($(A1)!.5!(B1)$);
  \draw[zUndirected={}] (A2) -- (A1);
  \draw[zUndirected={}] (A3) -- (A2);
  \combExt{A1}
  \combDots
\end{tikzpicture}\!\!\!\!\!\!\!\!\!&=(i\ell_1^\mu)
\begin{tikzpicture}[scale=.7,baseline={([yshift=-.5ex]combMid)}]
  \combSkeleton
  \coordinate (combMid) at ($(A1)!.5!(B1)$);
  \draw[zUndirected] (A1) -- (A3);
  \combDots
\end{tikzpicture}\,,\\
\begin{tikzpicture}[scale=.7,baseline={([yshift=-.5ex]combMid)}]
  \combSkeleton
  \coordinate (combMid) at ($(A1)!.5!(B1)$);
  \draw[zUndirected={}] (A1) -- (A2);
  \draw[zUndirected={}] (A3) -- (A2);
  \combExt{A2}
  \combDots
\end{tikzpicture}\!\!\!\!\!\!\!\!\!&=(i\ell_2^\mu)
\begin{tikzpicture}[scale=.7,baseline={([yshift=-.5ex]combMid)}]
  \combSkeleton
  \coordinate (combMid) at ($(A1)!.5!(B1)$);
  \draw[zUndirected] (A1) -- (A3);
  \combDots
\end{tikzpicture}\,.
\end{align}
\end{subequations}
Adding up all such contributions we get $iq^\mu=i\ell_1^\mu+i\ell_2^\mu+i\ell_3^\mu$,
i.e.~the total momentum exchanged between the two worldlines.
This argument works for all possible diagrams in the problem.

We have now derived a complete integrand for $N(q)$,
yet we have also lost track of internal $i0^+$ prescriptions.
Our next observation is that the Murua coefficients (see Appendix~\ref{murua})
may be assigned to our integrand \emph{after}
reduction to a basis of master integrals,
using IBP identities and tensor reduction.
This is due to the edge contraction rule
satisfied by these coefficients~\eqref{eq:edgecontraction}:
for every diagram with a given internal $i0^+$ prescription on a propagator,
there is a counterpart diagram with the opposite prescription ---
and the corresponding Murua coefficient.
When that propagator is cancelled,
whether due to a tensor reduction identity or IBPs,
the Murua coefficients of the pair combine as per \eqn{eq:edgecontraction}.
This leaves a single diagram,
whose Murua coefficient now lacks the cancelled propagator.
This process may be repeated as many times as required,
meaning that ultimately Murua coefficients can be assigned
to the master integrals directly.
The same observation was also made in \rcite{Bautista:2026fcp},
in the context of the WQFT-Compton amplitude.

\begin{figure*}[t]
	\centering
	\begin{subfigure}{.15\textwidth}
	\centering
    \begin{tikzpicture}[baseline={(current bounding box.center)},scale=.7]
    \coordinate (inA) at (0.4,.7);
    \coordinate (outA) at (3.6,.7);
    \coordinate (inB) at (0.4,-.7);
    \coordinate (outB) at (3.6,-.7);
    \coordinate (xA) at (1,.7);
     \coordinate (yA) at (2,.7);
     \coordinate (zA) at (3,.7);
     \coordinate (xB) at (1,-.7);
    \coordinate (yB) at (2,-.7);
    \coordinate (zB) at (3,-.7);
    \coordinate (xM) at (1,0);
    \coordinate (yM) at (2,0);
    \coordinate (zM) at (3,-0);
    \draw [dotted] (inA) -- (outA);
    \draw [dotted] (inB) -- (outB);
    \draw [draw=none] (xA) to[out=40,in=140] (zA);
    \draw [photon] (xA) -- (xM);
    \draw [photon] (xM) -- (xB);
    \draw [photon] (zA) -- (zM);
    \draw [photon] (zM) -- (zB);
    \draw [photon] (yB) -- (yM);
    \draw [purple, photon] (xM) -- (yM);
    \draw [purple, photon] (yM) -- (zM);
    \draw [fill] (xA) circle (.08);
    \draw [fill] (zA) circle (.08);
    \draw [fill] (xB) circle (.08);
    \draw [fill] (yB) circle (.08);
    \draw [fill] (zB) circle (.08);
    \draw [fill] (xM) circle (.08);
    \draw [fill] (yM) circle (.08);
    \draw [fill] (zM) circle (.08);
  	\end{tikzpicture}\\[8pt]
	$\omega\left(
	\begin{tikzpicture}[baseline={([yshift=-.5ex](0,0))}]
        \path [draw=black, zUndirected] (0,0) -- (.5,0);
		\path [draw=black, zUndirected] (1.0,0) -- (.5,0);
        \filldraw[fill=black] (0,0) circle (.06);
		\filldraw[fill=black] (0.5,0) circle (.06);
		\filldraw[fill=black] (1.0,0) circle (.06);
		\draw[white] (0,0.3536) circle (.06);
		\draw[white] (0,-0.3536) circle (.06);
    \end{tikzpicture}
	\right)$
	\caption{}
	\end{subfigure}
	\begin{subfigure}{.19\textwidth}
	\centering
    \begin{tikzpicture}[baseline={(current bounding box.center)},scale=.7]
    \coordinate (inA) at (0.4,.7);
    \coordinate (outA) at (3.6,.7);
    \coordinate (inB) at (0.4,-.7);
    \coordinate (outB) at (3.6,-.7);
    \coordinate (xA) at (1,.7);
     \coordinate (yA) at (2,.7);
     \coordinate (zA) at (3,.7);
     \coordinate (xB) at (1,-.7);
    \coordinate (yB) at (2,-.7);
    \coordinate (zB) at (3,-.7);
    \coordinate (xM) at (1,0);
    \coordinate (yM) at (2,0);
    \coordinate (zM) at (3,-0);
    \draw [dotted] (inA) -- (outA);
    \draw [dotted] (inB) -- (outB);
    \draw [draw=none] (xA) to[out=40,in=140] (zA);
    \draw [photon] (xA) -- (xM);
    \draw [photon] (xM) -- (xB);
    \draw [photon] (zA) -- (zB);
    \draw [photon] (yB) -- (yM);
    \draw [purple, photon] (xM) -- (yM);
    \draw [purple, photon] (yM) -- (yA);
    \draw [purple, zUndirected] (yA) -- (zA);
    \draw [fill] (xA) circle (.08);
    \draw [fill] (yA) circle (.08);
    \draw [fill] (zA) circle (.08);
    \draw [fill] (xB) circle (.08);
    \draw [fill] (yB) circle (.08);
    \draw [fill] (zB) circle (.08);
    \draw [fill] (xM) circle (.08);
    \draw [fill] (yM) circle (.08);
    \draw [fill] (zB) circle (.08);
  \end{tikzpicture}\\[8pt]
	$\omega\left(
	\begin{tikzpicture}[baseline={([yshift=-.5ex](0,0))}]
        \path [draw=black, zUndirected] (0,0) -- (0.5,0);
		\path [draw=black, zUndirected] (1.0,0) -- (0.5,0);
		\path [draw=black, zUndirected] (1.0,0) -- (1.5,0);
        \filldraw[fill=black] (0,0) circle (.06);
		\filldraw[fill=black] (0.5,0) circle (.06);
		\filldraw[fill=black] (1.0,0) circle (.06);
		\filldraw[fill=black] (1.5,0) circle (.06);
		\draw[white] (0,0.3536) circle (.06);
		\draw[white] (0,-0.3536) circle (.06);
    \end{tikzpicture}
	\right)$
  \caption{}
	\end{subfigure}
	\begin{subfigure}{.21\textwidth}
	\centering
  \begin{tikzpicture}[baseline={(current bounding box.center)},scale=.7]
    \coordinate (inA) at (0.4,.7);
    \coordinate (outA) at (4.6,.7);
    \coordinate (inB) at (0.4,-.7);
    \coordinate (outB) at (4.6,-.7);
    \coordinate (xA) at (1,.7);
     \coordinate (yA) at (2,.7);
     \coordinate (zA) at (3,.7);
     \coordinate (wA) at (4,.7);
    \coordinate (xB) at (1,-.7);
    \coordinate (yB) at (2,-.7);
    \coordinate (zB) at (3,-.7);
    \coordinate (wB) at (4,-.7);
    \coordinate (xM) at (1,0);
    \coordinate (yM) at (2,0);
    \coordinate (zM) at (3,-0);
    \draw [dotted] (inA) -- (outA);
    \draw [dotted] (inB) -- (outB);
    \draw [draw=none] (xA) to[out=40,in=140] (zA);
    \draw [photon] (xA) -- (xB);
    \draw [purple, photon] (yA) -- (yB);
    \draw [photon] (zA) -- (zB);
    \draw [photon] (wA) -- (wB);
    \draw [purple, zUndirected] (xB) -- (yB);
    \draw [purple, zUndirected] (zA) -- (wA);
    \draw [purple, zUndirected] (yA) -- (zA);
    \draw [fill] (xA) circle (.08);
    \draw [fill] (yA) circle (.08);
    \draw [fill] (zA) circle (.08);
    \draw [fill] (xB) circle (.08);
    \draw [fill] (yB) circle (.08);
    \draw [fill] (zB) circle (.08);
    \draw [fill] (wB) circle (.08);
    \draw [fill] (wA) circle (.08);
  \end{tikzpicture}\\[8pt]
	$\omega\left(
	\begin{tikzpicture}[baseline={([yshift=-.5ex](0,0))}]
        \path [draw=black, zUndirected] (0,0) -- (0.5,0);
		\path [draw=black, zUndirected] (1.0,0) -- (0.5,0);
		\path [draw=black, zUndirected] (1.0,0) -- (1.5,0);
		\path [draw=black, zUndirected] (1.5,0) -- (2.0,0);
        \filldraw[fill=black] (0,0) circle (.06);
		\filldraw[fill=black] (0.5,0) circle (.06);
		\filldraw[fill=black] (1.0,0) circle (.06);
		\filldraw[fill=black] (1.5,0) circle (.06);
		\filldraw[fill=black] (2.0,0) circle (.06);
		\draw[white] (0,0.3536) circle (.06);
		\draw[white] (0,-0.3536) circle (.06);
    \end{tikzpicture}
	\right)$
  \caption{}
\end{subfigure}
	\begin{subfigure}{.22\textwidth}
	\centering
 \begin{tikzpicture}[baseline={(current bounding box.center)},scale=.7]
    \coordinate (inA) at (0.4,.7);
    \coordinate (outA) at (4.6,.7);
    \coordinate (inB) at (0.4,-.7);
    \coordinate (outB) at (4.6,-.7);
    \coordinate (xA) at (1,.7);
     \coordinate (yA) at (2,.7);
     \coordinate (zA) at (3,.7);
     \coordinate (wA) at (4,.7);
    \coordinate (xB) at (1,-.7);
    \coordinate (yB) at (2,-.7);
    \coordinate (zB) at (3,-.7);
    \coordinate (wB) at (4,-.7);
    \coordinate (xM) at (1,0);
    \coordinate (yM) at (2,0);
    \coordinate (zM) at (3,-0);
    \draw [dotted] (inA) -- (outA);
    \draw [dotted] (inB) -- (outB);
    \draw [draw=none] (xA) to[out=40,in=140] (zA);
    \draw [photon] (xA) -- (xB);
    \draw [purple, photon] (yA) -- (yB);
    \draw [purple, photon] (zA) -- (zB);
    \draw [photon] (wA) -- (wB);
    \draw [purple, zUndirected] (xA) -- (yA);
    \draw [purple, zUndirected] (zA) -- (wA);
    \draw [purple, zUndirected] (yB) -- (zB);
    \draw [fill] (xA) circle (.08);
    \draw [fill] (yA) circle (.08);
    \draw [fill] (zA) circle (.08);
    \draw [fill] (xB) circle (.08);
    \draw [fill] (yB) circle (.08);
    \draw [fill] (zB) circle (.08);
    \draw [fill] (wB) circle (.08);
    \draw [fill] (wA) circle (.08);
  \end{tikzpicture}\\[8pt]
	$\omega\left(
	\begin{tikzpicture}[baseline={([yshift=-.5ex](0,0))}]
        \path [draw=black, zUndirected] (0,0) -- (0.5,0);
		\path [draw=black, zUndirected] (1.0,0) -- (0.5,0);
		\path [draw=black, zUndirected] (1.0,0) -- (1.5,0);
		\path [draw=black, zUndirected] (1.5,0) -- (2.0,0);
		\path [draw=black, zUndirected] (2.5,0) -- (2.0,0);
        \filldraw[fill=black] (0,0) circle (.06);
		\filldraw[fill=black] (0.5,0) circle (.06);
		\filldraw[fill=black] (1.0,0) circle (.06);
		\filldraw[fill=black] (1.5,0) circle (.06);
		\filldraw[fill=black] (2.0,0) circle (.06);
		\filldraw[fill=black] (2.5,0) circle (.06);
		\draw[white] (0,0.3536) circle (.06);
		\draw[white] (0,-0.3536) circle (.06);
    \end{tikzpicture}
	\right)$
  \caption{}
\end{subfigure}
\begin{subfigure}{.19\textwidth}
	\centering
  \begin{tikzpicture}[baseline={(current bounding box.center)},scale=.7]
    \coordinate (inA) at (0.4,.7);
    \coordinate (outA) at (4.6,.7);
    \coordinate (inB) at (0.4,-.7);
    \coordinate (outB) at (4.6,-.7);
    \coordinate (xA) at (1,.7);
    \coordinate (xyAB) at (1.5,0) ;
    \coordinate (yB) at (2,-.7);
     \coordinate (yA) at (2,.7);
     \coordinate (zA) at (3,.7);
     \coordinate (wA) at (4,.7);
    \coordinate (xB) at (1,-.7);
    \coordinate (zB) at (3,-.7);
    \coordinate (wB) at (4,-.7);
    \coordinate (xM) at (1,0);
    \coordinate (yM) at (2,0);
    \coordinate (zM) at (3,-0);
    \draw [dotted] (inA) -- (outA);
    \draw [dotted] (inB) -- (outB);
    \draw [draw=none] (xA) to[out=40,in=140] (zA);
    \draw [photon] (xA) -- (yB);
    \filldraw[white] (xyAB) circle (6pt) ;
    \draw [photon] (yA) -- (xB);
    \draw [purple, photon] (zA) -- (zB);
    \draw [photon] (wA) -- (wB);
    \draw [purple, zUndirected] (zB) -- (yB);
    \draw [purple, zUndirected] (zA) -- (wA);
    \draw [purple, zUndirected] (yA) -- (zA);
    \draw [fill] (xA) circle (.08);
    \draw [fill] (yA) circle (.08);
    \draw [fill] (zA) circle (.08);
    \draw [fill] (xB) circle (.08);
    \draw [fill] (yB) circle (.08);
    \draw [fill] (zB) circle (.08);
    \draw [fill] (wB) circle (.08);
    \draw [fill] (wA) circle (.08);
  \end{tikzpicture}\\[8pt]
	$\omega\left(
	\begin{tikzpicture}[baseline={([yshift=-.5ex](0,0))}]
		\path [draw=black, zUndirected] (-0.5,0) -- (0,0);
        \path [draw=black, zUndirected] (0.0,0) -- (.5,0);
		\path [draw=black, zUndirected] (.5,0) -- (.8536,0.3536);
		\path [draw=black, zUndirected] (.5,0) -- (.8536,-0.3536);
        \filldraw[fill=black] (-0.5,0) circle (.06);
		\filldraw[fill=black] (0,0) circle (.06);
		\filldraw[fill=black] (0.5,0) circle (.06);
		\filldraw[fill=black] (.8536,0.3536) circle (.06);
		\filldraw[fill=black] (.8536,-0.3536) circle (.06);
    \end{tikzpicture}\,
	\right)$
  \caption{}
\end{subfigure}
  \caption{Top-sector diagrams of the 4PM-1SF Magnusian,
  with the associated category of Murua coefficient $\omega$ drawn underneath.
  Active propagators for which the $i0^+$ prescription plays a role,
  including those on the worldline, are highlighted.
  The four planar top sectors (a)--(d) are associated with linear Murua
  coefficients~\eqref{linearMurua};
  the nonplanar top-sector (e) carries a branching Murua coefficient~\eqref{muruaNP4}.}\label{fig:topsectors}
\end{figure*}
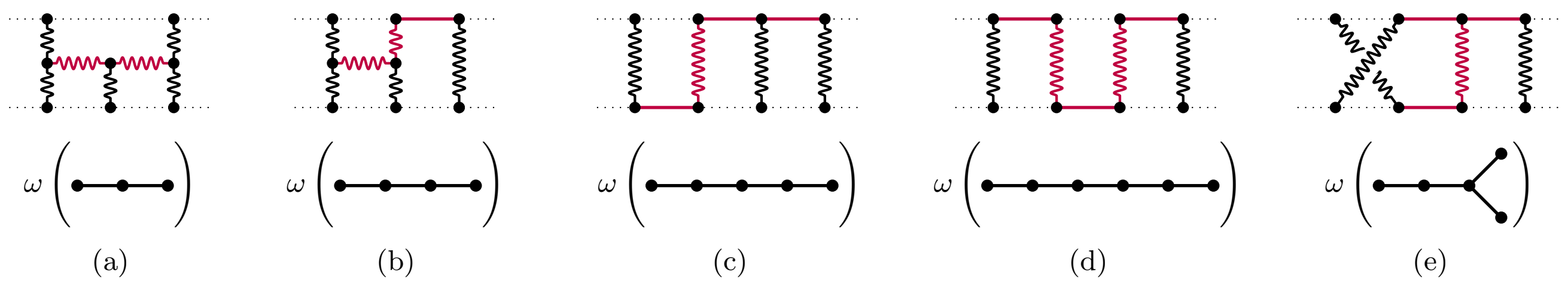

Crucially,
when applying this methodology to assign weightings to
diagrams after IBP reduction,
we may not ``planarise'' the integrals contributing to $N(q)$.
The planarisation procedure, introduced and explained in \rcite{Driesse:2024xad},
involves using partial fraction identities on worldline propagators
to systematically eliminate nonplanar Feynman integrals in favour of a single planar family.
This is possible for all 1SF integrals at all PM orders;
at 2SF nonplanar integral families are required~\cite{Driesse:2026qiz}.
In the present context,
wherein we assign Murua coefficients after IBP and tensor reduction,
using partial fraction identities invalidates our ability to uniquely assign these
coefficients as we are no longer merely cancelling propagators.
In our current 4PM calculation,
nonplanar Feynman integrals are associated with the appearance
of branching Murua coefficients (Fig.~\ref{fig:topsectors})
which could not be applied if we only had planar Feynman integrals.
In principle, we could still apply Murua coefficients to our integrand \emph{before}
applying partial fractions identities,
however we find our approach to be more efficient overall.

Our procedure for calculating the Magnusian may therefore be summarised as follows.
Take the integrand coefficient of $\Delta p_1^\mu$ with $q^\mu$,
after reduction to a basis of master integrals and without using partial fraction identities,
and re-assign $i0^+$ prescriptions by weighting with the Murua coefficients.
In the context of the momentum impulse $\Delta p_1^\mu$,
the relevant subset of integrals multiplying $q^\mu$ have been referred to as $b$-type,
as after Fourier transform they contribute to terms multiplying the impact parameter $b^\mu$.
These integrals have an even number of worldline propagators,
unlike the so-called $v$-type integrals that carry an odd number.
Given the novelty of this approach,
and in order to provide additional verification,
before proceeding to the 1SF dynamics
we begin by reproducing the 0SF probe-limit scattering.

\subsection{Probe limit}

In the probe limit $\nu\to0$, the scattering dynamics is entirely conservative
and described by the motion of a test particle in the potential of a larger body. 
All gravitons are therefore restricted to the potential region of
integration, and the $\gamma$-dependence of these integrals is trivial ---
see \rcite{Jakobsen:2023ndj} for the integral family.
Thus, $N^{(4,0)}$ captures the entire scattering dynamics,
and in the aligned spin case reproduces the radial action,
as per our discussion in Section~\ref{sec:magradaction}.
So $N^{(4,0)}$ does not contain any new physics.
However, by calculating $N^{(4,0)}$ using the Murua coefficients (see Appendix~\ref{murua})
we gain a non-trivial overall consistency check on the validity of our method.
In particular, our results are highly sensitive to the values of the Murua coefficients chosen.

After plugging in results for the integrals,
it turns out that the entire 4PM probe limit result is captured by a single master integral:
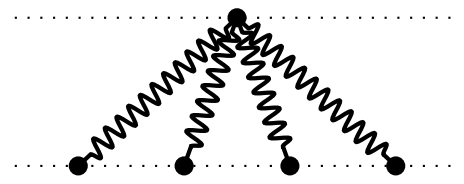
\begin{align}
&\begin{tikzpicture}[baseline={(current bounding box.center)},scale=.7]
    \coordinate (inA) at (0.4,.7);
    \coordinate (outA) at (4.6,.7);
    \coordinate (inB) at (0.4,-.7);
    \coordinate (outB) at (4.6,-.7);
    \coordinate (yzA) at (2.5,.7);
    \coordinate (xB) at (1,-.7);
    \coordinate (yB) at (2,-.7);
    \coordinate (zB) at (3,-.7);
    \coordinate (wB) at (4,-.7);
    \draw [dotted] (inA) -- (outA);
    \draw [dotted] (inB) -- (outB);
    \draw [photon] (yzA) -- (xB);
    \draw [photon] (yzA) -- (yB);
    \draw [photon] (yzA) -- (zB);
    \draw [photon] (yzA) -- (wB);
    \draw [fill] (yzA) circle (.08);
    \draw [fill] (xB) circle (.08);
    \draw [fill] (yB) circle (.08);
    \draw [fill] (zB) circle (.08);
    \draw [fill] (wB) circle (.08);
  \end{tikzpicture}=
  \int_{\ell_1\ell_2\ell_3}
  \frac{\dd(\ell_1\!\cdot\!v_2)\dd(\ell_2\!\cdot\!v_2)\dd(\ell_3\!\cdot\!v_2)}{\ell_1^2\ell_2^2\ell_3^2
  (\ell_1\!+\!\ell_2\!+\!\ell_3\!+\!q)^2}\nn\\
  &\qquad=\frac{(4\pi)^{-\frac92+3\eps}\Gamma^4(\frac12-\eps)\Gamma(-\frac12+3\eps)}{\Gamma(2-4\eps)}
  |q|^{1-6\eps}\,.
\end{align}
Contributions from the other four master integrals cancel amongst themselves.
This is quite different from the probe-limit impulse calculation,
wherein more master integrals appear in the final result.
The result for the probe-limit Magnusian is
\begin{align}\label{probeMagnusian}
  N^{(4,0)}&=\frac{35\pi\,(33\gamma^4-18\gamma^2+1)}{64\,(\gamma^2-1)^{3/2}}\\
  &+\frac{21\pi\gamma\,(33\gamma^4-30\gamma^2+5)}{128\,|b|\,(\gamma^2-1)^2}
  \big(13\,L\cdot a_++3\,\delta\,L\cdot a_-\big)\,,\nn
\end{align}
where
$M = m_1+m_2$,
$\nu = m_1m_2/M^2$,
$\delta = (m_1-m_2)/M$,
$a_\pm = a_1 \pm a_2$ and $L^\mu$ is the unit-normalised angular momentum vector~\eqref{angMomentum}.
This result is checked as part of our confirmation of the full conservative result,
which also includes 1SF contributions --- see Section~\ref{sec:check} below.

\subsection{Comparable mass calculation}

The comparable-mass calculation is physically and computationally more complex.
The integrals can be put into two families, which were
also used in the computation of the impulse~\cite{Jakobsen:2023ndj,Jakobsen:2023hig}.
The first family, ${\rm P}$, is planar:
\begin{subequations}\label{eq:jFamily}
  \begin{align}
    {\rm P}^{(\sigma_1,\sigma_2,\sigma_3,\sigma_4,\sigma_5)}_{n_1,n_2,\ldots,n_{12}}
    :=
    \int_{\ell_1,\ell_2,\ell_3}\!\!\!\!
    \frac{
      \dd(\ell_1\!\cdot\! v_{1})
      \dd(\ell_2\!\cdot\! v_{1})
      \dd(\ell_3\!\cdot\! v_{2})
    }{
      D_1^{n_1}
      D_2^{n_2}\cdots
      D_{12}^{n_{12}}
    }\,,
  \end{align}
  with ($j=1,2,3$, $k=1,2$)
  \begin{align}
    &
    D_k
    =
    \ell_{k}\cdot v_{2} +\sigma_{k}i0^+
    \,,\,\,
    D_3
    =
    \ell_{3}\cdot v_{1} +\sigma_{3}i0^+\,,
    \nn\\
    &
    D_{3+k}
    =
    (\ell_k-\ell_3)^2+\sigma_{3+k}\,(\ell_k^0-\ell_3^0)\,i0^+
    \,,\,\,\\
    &D_{6}
    =
    (\ell_1-\ell_2)^2
    \,,
   D_{6+j}
   =\ell_j^2
   \,,\,\,
   D_{9+j}
   =(\ell_j+q)^2
   \,,\nn
  \end{align}
\end{subequations}
and $q\cdot v_i=0$. This family covers the top sectors
(a)--(d) in Fig.~\ref{fig:topsectors}, which are associated with linear Murua coefficients.
The second family, $\rm NP$, is in principle reducible through partial-fraction
identities; as discussed above, we refrain from using them here. It is
defined as
\begin{subequations}
  \label{eq:iFamily}
  \begin{align}
    {\rm NP}^{(\sigma_1,\ldots,\sigma_6)}_{n_1,n_2,...,n_{12}}
    :=
    \int_{\ell_1,\ell_2,\ell_3}\!\!\!\!\!\!
    \frac{
      \dd(\ell_1\cdot v_{2})
      \dd(\ell_2\cdot v_{1})
      \dd(\ell_3\cdot v_{1})
    }{
      D_1^{n_1}
      D_2^{n_2}
      ...
      D_{12}^{n_{12}}
    }\,,
  \end{align}
  with the propagators ($j=1,2,3$ and $k=1,2$)
  \begin{align}
    &D_{1}
    =
    \ell_1\cdot v_{1} +\sigma_1i0^+
    \,,\,\,
    D_{1+k}
    =
    \ell_{1+k}\cdot v_2 +\sigma_{1+k}i0^+\,,\nn
    \\
    &D_{4}
    =
    (\ell_1+\ell_2+\ell_3+q)^2+\sigma_{4}\,(\ell_1^0+\ell_2^0+\ell_3^0)\,i0^+
    \,,\,\, \nn\\
    &D_{5}
    =
    (\ell_1+\ell_2+q)^2+\sigma_{5}\,(\ell_1^0+\ell_2^0)\,i0^+
    \,,
    \\
    &D_{6}\!
    =
    (\ell_1+\ell_3)^2+\sigma_{6}\,(\ell_1^0+\ell_3^0)\,i0^+
    \,,\,\,\nn\\
    &D_{7}\!
    =
    (\ell_2+\ell_3)^2
    \,,\,\,
    D_{7+j}
    =
    \ell_{j}^2
    \,,\,\,
    D_{10+k}\!
    =
    (\ell_{k}+q)^2\,,\nn
  \end{align}
\end{subequations}
where again $q\cdot v_i=0$.
This family covers the top sector (e) in Fig.~\ref{fig:topsectors},
which is a nonplanar sector associated to a branching Murua coefficient.

The integrals are evaluated with the method of differential
equations~\cite{Kotikov:1990kg,Remiddi:1997ny,Gehrmann:1999as,Henn:2013pwa}. Given a set of master integrals $\vec{I}$, IBP reduction
yields a system of differential equations of the form
\begin{align}
    \frac{\d \vec I}{\d x}=M(x, \epsilon)\, \vec I\,,
\end{align}
where $x=\gamma-\sqrt{\gamma^2-1}$. We then construct a transformation
$T(x,\epsilon)$ to a canonical basis $\vec{J}=T\vec{I}$ satisfying
\begin{align}
    \frac{\d \vec J}{\d x}=\epsilon\, \hat M(x)\, \vec J\,.
\end{align}
The integrals evaluate to multiple polylogarithms and to iterated
integrals over squares of complete elliptic integrals of the first kind ---
see \rcites{Jakobsen:2023ndj,Jakobsen:2023hig,Dlapa:2021vgp,Dlapa:2022lmu,Dlapa:2023hsl}
for details of the canonicalisation of the integrals. We emphasise that, in our setup, the
differential equations are entirely independent of the $i0^+$ prescription of
the integrals, so that we can reuse the exact differential equations employed
in the impulse computation. We solve them in terms of boundary integrals with
arbitrary $i0^+$ prescriptions, in which the integrand is then expressed. The
full Magnusian is finally obtained by replacing each boundary integral with
the sum over all of its $i0^+$ prescriptions, weighted by the corresponding
Murua coefficients.

\subsection{Regions of integration}

As is common in multi-loop PM calculations, we fix boundary integrals in the
small-velocity limit $v\rightarrow 0$ using the method of
regions~\cite{Beneke:1997zp}.
Besides fixing the boundary constants, this
naturally splits the result into regions,
depending on whether individual gravitons are potential (P) $\ell\sim (v,1)$
or radiative (R) $\ell\sim (v,v)$.
In both the planar~\eqref{eq:jFamily}
and nonplanar~\eqref{eq:iFamily} integral families there are two ``active'' graviton propagators
that are kinematically able to go on-shell over the phase space of integration ---
denoted by coloured lines in \Fig{fig:topsectors}.
The remaining ``passive'' gravitons are necessarily potential modes (P),
and their $i0^+$ prescriptions are inconsequential.
Our result therefore splits into three regions:
$\rm PP$, $\rm PR$ and $\rm RR$.

At 4PM order, terms even in the velocity receive contributions from both the
potential $({\rm PP})$ region, scaling as $v^{0\epsilon}$, and the
$(\rm RR)$ region, scaling as $v^{-4\epsilon}$; the latter is associated with
the tail effect. In the impulse, each of these two regions individually
exhibits a $1/\epsilon$ pole. These poles cancel in the sum
(as shown more generically in Section~\ref{sec:irfactorsforobs}) but leave behind
$\log v$ terms at $\mathcal{O}(\epsilon^0)$. In the Magnusian computation, by
contrast, the poles do not cancel and instead leave a soft divergence:
\begin{subequations}
\begin{align}
    N^{(4)}_{\rm{PP}}&=- \dfrac{E}{2 \epsilon}E_{\rm rad}^{(3)}+\mathcal{O}(\epsilon^0)\,,\\
 \,\,
    N^{(4)}_{\rm{RR}}&= v^{-4 \epsilon}\dfrac{E}{6 \epsilon}E_{\rm rad}^{(3)}+\mathcal{O}(\epsilon^0)\,.
\end{align}
\end{subequations}
Compared with the impulse,
the graviton prescription inherited from the Murua coefficients changes the
prefactor of the $\rm RR$ region, so that the sum reproduces the first term
of the soft divergence in eq.~\eqref{4pmDivergence}. This soft tail pole
therefore originates from the $\rm RR$ region. The reason becomes clear from
\Fig{fig:tail}, in which
a worldline exchanges a soft graviton with a hard graviton. The hard graviton
is cut and hence on-shell. Since in the small-$v$ limit only radiative gravitons can
be on-shell, contributions such as those in \Fig{fig:tail} must lie in
the $\rm RR$ region.

The ${\rm PR}\sim v^{-2 \epsilon}$ region produces no divergence in the
impulse at this PM order. In the Magnusian computation, however, it gives
the second term of the soft divergence, which is the Coulomb pole:
\begin{align}
    N^{(4)}_{\rm{PR}}= v^{-2 \epsilon}\dfrac{F'(\gamma)}{2}\dfrac{E}{3 \epsilon}E_{\rm rad}^{(3)}+\mathcal{O}(\epsilon^0)\,.
\end{align}
Again, the connection to \Fig{fig:Coulomb} is immediate. The cut on the
graviton line puts this graviton on-shell and thus forces it into the
radiative region. The cut on the worldline propagator, on the other hand,
forces all remaining gravitons to have vanishing energy component in
the rest frame of one of the two black holes, which makes them potential.
Contributions of the type shown in Fig.~\ref{fig:Coulomb} must therefore lie
in the $\rm PR$ region, and are odd-in-$v$.

\subsection{Complete result}

The full Magnusian at 4PM order up
to linear order in spin takes the form:
\begin{align}\label{eq:N4structure}
 \frac{|b|^3}{M^5\nu} N^{(4)}
  &=
  N^{(4,0)}_{0}
  +\frac{L\cdot a_+}{|b|}\,N^{(4,0)}_{+}
  +\delta\frac{L\cdot a_-}{|b|}\,N^{(4,0)}_{-}
  \nn\\
  &\!\!\!\!\!\!
  +\nu\Big(
  N^{(4,1)}_{0}
  +\frac{L\cdot a_+}{|b|}\,N^{(4,1)}_{+}
  +\delta\frac{L\cdot a_-}{|b|}\,N^{(4,1)}_{-}
  \Big)\,.
\end{align}
The probe-limit 0SF terms were presented in \eqn{probeMagnusian};
the 1SF terms are ($\gamma_\pm = \gamma \pm 1$)
\begin{align}\label{1SFresult}
\begin{aligned}
  N^{(4,1)}_{X}
  &=
  \pi\sum_i p_{X,i}(\gamma)\,F_i(\gamma)\\
  &\quad
  -\frac{\, E_{{\rm rad},X}}{3}
  \bigg[
 \frac{1}{\epsilon}\bigg(\dfrac{R}{2|b|}\bigg)^{-8\epsilon} \bigg(1-\frac{F'(\gamma)}{2}\bigg)
  \\
  &\qquad\qquad\qquad\,\,\,
  +\log\!\Big[\frac{\gamma_-}{2}\Big]\bigg(1+\frac{F'(\gamma)}{2}\bigg)
  \bigg]\,,
\end{aligned}
\end{align}
where $X=0$ denotes the non-spinning part of the answer and $X \in \{+,- \}$
are the linear-in-spin contributions.
Our result includes components of the 3PM radiated energy:
\begin{align}\label{radE3PM}
\begin{aligned}
  E^{(3)}_{\rm rad}
  &=\frac{M^5 \nu^2}{E|b|^3}
  \bigg[
   E_{{\rm rad},0}\\
  &\qquad+\frac{L\cdot a_+}{|b|}\, E_{{\rm rad},+}
  +\delta\frac{L\cdot a_-}{|b| }\, E_{{\rm rad},-}
  \bigg]\,,
\end{aligned}
\end{align}
where expressions for ${E_{\rm rad,X}}$ are given in Appendix \ref{radE}.
The algebraic functions $p_{X,i}(\gamma)$, depending only on $\gamma$,
are given both in Appendix~\ref{pol} and in the ancillary file
attached to the \texttt{arXiv} submission of this paper.
The 14 linearly independent functions $F_i(\gamma)$ are
\begin{align}\label{eq:functionBases}
  &F_{1,\ldots, 5}
  \!=\!
  \bigg\{1,
  \dfrac{
    \log[x]}{\sqrt{\gamma^2-1}},
  \log\! \left[\frac{\gamma_{+}}{2}\right],
  \log^2[x],
  \dfrac{\log[x]\log\!\left[\frac{\gamma_+}{2}\right]}{\sqrt{\gamma^2-1}}
  \bigg\}\,,
  \nn
  \\
  &F_{6,\ldots,9}=
  \bigg\{
  \log[\gamma],
  \log^2\! \left[\frac{\gamma_{+}}{2}\right],
  \text{Li}_{2}\left [\frac{\gamma_{-}}{\gamma_{+}}\right],
  \text{Li}_{2}\left [-\frac{\gamma_{-}}{\gamma_{+}}\right]
  \bigg\}\,,
  \nn\\
  &F_{10,11}
  =
  \bigg\{
  \frac{1}{\sqrt{\gamma^2-1}}\chi_2\left[\sqrt{\frac{\gamma_-}{\gamma_+}}\,\right],\\
&\qquad\qquad\qquad  \dfrac{\text{Li}_2[-x]-\text{Li}_2\left[-\frac{1}{x}\right]+\log[4] \log[x]}
    {\sqrt{\gamma^2-1}}
    \bigg\}\,,
  \nn\\
  &F_{12,13,14}
  =
  \bigg\{
     \eE^2\left [ \frac{\gamma_{-}}{\gamma_{+}} \right ],
   \eK^2\left [ \frac{\gamma_{-}}{\gamma_{+}} \right ],
   \eE\left [ \frac{\gamma_{-}}{\gamma_{+}} \right ]
   \eK\left [ \frac{\gamma_{-}}{\gamma_{+}} \right ]
    \bigg\}\,,\nn
\end{align}
where $\chi_2[z]=\frac{1}{2}\big(\text{Li}_2[z]-\text{Li}_2[-z]\big)$.
These are a subset of the function basis used for the impulse calculation
\cite{Jakobsen:2023pvx}.
We have also introduced $\eK$ and $\eE$,
which are the complete elliptic integrals of the first
and second kind respectively.

In the last line of our result~\eqref{1SFresult} we have exposed $\log[\gamma_-/2]$,
which arises from the $v^{-n \epsilon}$ scalings of the single regions.
These terms are responsible for the nonlocality of the 4PM Magnusian,
and develop an imaginary part for bound-orbit kinematics $\gamma<1$ ---
unlike all of the functions $F_i$~\eqref{eq:functionBases}.
Like the soft $1/\eps$ poles,
which agree with the prediction of our soft analysis~\eqref{4pmDivergence},
these nonlocalities are proportional to the 3PM radiated energy~\eqref{radE3PM}.
Due to partial cancellation of the poles of the $\rm PP$ and $\rm RR$ regions,
we see a relative sign flip in the $F'(\gamma)$ term multiplying $\log[\gamma_-/2]$
compared with that of the $1/\epsilon$ pole.

\subsection{Conservative result}
\label{sec:check}

In addition to calculating the complete 4PM Magnusian~\eqref{1SFresult},
we also calculate its conservative part
using a Feynman $i0^+$ prescription for graviton propagators.
Besides producing a new result,
this also serves as a check on our calculation of $N^{(4)}$:
the conservative Magnusian $N^{(4)}_{\rm cons}$
fully encodes the conservative 4PM momentum impulse and spin kick
computed in \rcites{Bern:2021yeh,Dlapa:2021vgp,Jakobsen:2023ndj}
using the same $i0^+$ prescription.
Using the same integrand as in the full calculation of $N^{(4)}$
we compute even-in-$v$ contributions only,
which means evaluating boundary integrals in the PP and RR but not PR regions.
Integrals are evaluated with Feynman graviton propagators
replacing retarded, and taking the real part.
This means that we are still sensitive to the Murua coefficients
used to determine causality flow for worldline propagators.
In practice, as graviton propagators do not go on-shell in the PP region,
use of the Feynman $i0^+$ prescription only has an
impact on the boundary integrals of the RR region.

Our 1SF conservative result,
also included in the ancillary file on \texttt{arXiv},
has the schematic form
\begin{align}\label{eq:N4coefficients}
  N^{(4,1)}_{{\rm cons},X}&=
  \pi\sum_{i\neq 11} p^{\rm cons}_{X,i}(\gamma)\,F_i(\gamma) -\, E_{{\rm rad},X}\,
  \log\!\Big[\frac{\gamma_-}{2}\Big],
\end{align}
the probe limit 0SF result $N^{(4,0)}$ being the same as that in \eqn{probeMagnusian}.
For aligned spins,
the scattering angle implied by this prescription matches the
one defined by linear response~\eqref{eq:linResponse},
and the Magnusian reproduces the radial action: $N_{\rm cons}^{\rm aligned}=I_r$.\footnote{The Feynman $i0^+$ prescription no longer reproduces linear response  at 5PM order, as has been discussed in \rcites{Porto:2026fsd,Bini:2025vuk,Sauer:2026eun,Georgoudis:2026zzn}. At the present 4PM order this is of no concern.}
We also fully agree with the non-spinning conservative terms
of the 4PM $\hat{N}$-matrix computed in \rcite{Damgaard:2023ttc}.

Given that this is a conservative scattering definition,
\eqn{consN} may be used to determine the
scattering observables $\Delta\lambda_{\rm cons}^\mu$.
Expanding in $G$ up to 4PM order, we have
\begin{align}\label{check4PMCons}
    \Delta\lambda^{(4)\mu}_{\rm cons}&=
    \{N_{\rm cons}^{(4)},\lambda^\mu\}\nn\\
    &\quad+
    \sfrac12\{N^{(2)},\{N^{(2)},\lambda^\mu\}\}\nn\\
    &\quad+
    \sfrac12\{N_{\rm cons}^{(3)},\{N^{(1)},\lambda^\mu\}\}\nn\\
    &\quad+
    \sfrac12\{N^{(1)},\{N_{\rm cons}^{(3)},\lambda^\mu\}\}\\
    &\quad+
    \sfrac1{3!}\{N^{(2)},\{N^{(1)},\{N^{(1)},\lambda^\mu\}\}\}\nn\\
    &\quad+
    \sfrac1{3!}\{N^{(1)},\{N^{(2)},\{N^{(1)},\lambda^\mu\}\}\}\nn\\
    &\quad+
    \sfrac1{3!}\{N^{(1)},\{N^{(1)},\{N^{(2)},\lambda^\mu\}\}\}\nn\\
    &\quad+
    \sfrac1{4!}\{N^{(1)},\{N^{(1)},\{N^{(1)},\{N^{(1)},\lambda^\mu\}\}\}\}\,.\nn
\end{align}
At 1PM and 2PM orders gravitons cannot go on-shell,
so the Feynman prescription $i0^+$ is inconsequential:
$N^{(1)}=N^{(1)}_{\rm cons}$
and $N^{(2)}=N^{(2)}_{\rm cons}$,
see \rcite{Jakobsen:2021zvh} for explicit results.
At 3PM order, the Magnusian up to quartic order in spins was
calculated in \rcites{Damgaard:2021ipf,Kim:2024svw,Haddad:2025cmw}.
$N^{(3)}_{\rm cons}$ is identified according to parity $v\to-v$:
we take the terms evaluated in the potential region (P),
and exclude contributions from the radiative region (R).
Inserting these results into \eqn{check4PMCons},
and applying the brackets~\eqref{eq:PBgi},
we can confirm that our new result for $N^{(4)}_{\rm cons}$ is
consistent with the existing 4PM literature~\cite{Bern:2021yeh,Dlapa:2021vgp,Jakobsen:2023ndj}.
This serves as a full check on $N^{(4)}$ in both the probe limit and potential region.

\section{Conclusions}
\label{sec:conclusions}

The Magnus series provides a powerful framework to revisit the classical gravitational dynamics
of compact binaries, both for scattering and bound orbits,
and offers a natural generating functional for observables: the $\hat{N}$-matrix.
In this work we have shown how,
using the canonical WQFT framework~\cite{Gonzo:2026yha},
we may deduce the leading soft interactions of gravitational scattering.
These interactions naturally generate gravitational
Wilson lines along the past and future asymptotic trajectories,
thereby deriving the Wilson line description of the leading IR behaviour directly within WQFT.
This connects the soft Hamiltonian description to gravitational Wilson line exponentiation~\cite{Naculich:2011ry,White:2011yy,Melville:2013qca,Ware:2013zja} and to the generalised Wilson line formulation of gravitational scattering~\cite{Bonocore:2021qxh,Bonocore:2024uxk,Pirsch:2026zqa}. Keeping these soft factors at the operator level, the Magnus expansion makes explicit how the Coulomb and single tail phases contribute to $\hat N$ through commutators with the hard evolution, leading to the prediction for its IR pole at 4PM order~\eqref{4pmDivergence}.

We have then performed a complete three-loop calculation of the full 4PM Magnusian
$N=\langle0|\hat{N}|0\rangle$ through linear order in the spins.
We calculated both a complete expression for the Magnusian,
whose soft pole agrees with our earlier prediction~\eqref{4pmDivergence},
and an IR-finite conservative expression,
the dissipative part being defined by linear response~\cite{Bini:2012ji,Damour:2020tta}.
The region analysis also identifies the origin of the divergence: the incomplete cancellation between potential--potential and radiation--radiation regions produces the single-tail pole, while the mixed potential--radiation region gives the Coulomb contribution.
Our conservative result also provides a compact generator for the corresponding dynamics:
with the prescription specified in Section~\ref{sec:check}, its nested Poisson brackets reproduce the known 4PM momentum impulse and spin kick for generic spin orientations~\cite{Jakobsen:2023ndj}.
Our setup provides a natural explanation of the cancellation of IR
divergences in scattering observables such as the momentum impulse and spin kick,
while at the same time predicting the ones for the scattering waveform.

At 4PM, radiative effects also modify the relation between the Magnusian and the full scattering angle,
so that the full vacuum Magnusian must be distinguished from a radial action reconstructed from that angle,
even for aligned spins.
The Magnusian, in particular, retains its polynomial mass dependence,
making it an ideal object for studying analytic properties of
scattering and bound observables in the PM and self-force expansions and for future implementations in resummation models.
This is unlike the 4PM scattering angle, which develops a non-polynomial mass dependence
due to recoil in the centre-of-mass frame.
The trade-off for this simplicity, as we have seen, is the appearance of soft divergences.

Our results open the door to many further extensions besides the
immediate extension to higher-PM and spin orders,
which would require us to extend our power counting to additional soft regions,
including those associated with tail-of-tail effects.
Furthermore, retaining the zero-frequency modes and the full time dependence of the asymptotic dressings would allow us to study BMS supertranslation charges and gravitational memory within WQFT~\cite{Strominger:2014pwa,Moult:2025njc,Oertel:2026wsm} --- see also \rcites{DiVecchia:2022owy,DiVecchia:2022piu,Riva:2023xxm} for an amplitude analysis. A concrete goal would be to identify the Bondi frame selected by our operator construction and derive the action of supertranslations on the waveform~\cite{Veneziano:2022zwh,Bini:2024rsy,Elkhidir:2024izo}, keeping track of the asymptotic shear and its change during scattering alongside the IR shift of retarded time.  Finally, the soft sector of perturbative QCD offers a further application of the present Magnus analysis, building on the established soft Hamiltonian and coherent state
constructions~\cite{Catani:1984dp,Ciafaloni:1984zr,Catani:1985xt,DelDuca:1989jt,Gonzo:2019fai}.
Comparing the resulting scattering generator with the web organisation of Wilson line correlators~\cite{Gardi:2010rn,Gardi:2013ita} could clarify how the corresponding colour correlations enter the operator factorisation and its radiative matrix elements.
We leave these exciting prospects for future work.

\sec{Acknowledgements}
We thank Gustav Jakobsen, Donal O'Connell,
Jan Plefka and Rodolfo Russo for valuable discussions.
R.G.~and G.M.~acknowledge OpenAI for providing access to advanced AI models
through the ``ChatGPT for Academic Researchers'' programme.
AI tools were used during the final stages of manuscript preparation only to assist with revisions and literature searches, but not to generate scientific ideas.
This work is supported by The Royal Society under grant URF\textbackslash R1\textbackslash 231578,
``Gravitational Waves from Worldline Quantum Field Theory'' (G.M.)
and grant URF\textbackslash R1\textbackslash 261173, 
``High-precision effective field theory for extreme-mass-ratio inspirals'' (R.G.).
G.M.~is also supported by the STFC Consolidated
Grant ST/X00063X/1, ``Amplitudes, Strings \& Duality''.
B.S. is supported by grant RF\textbackslash ERE\textbackslash 231084, by the 
Excellence Cluster ORIGINS funded by the Deutsche
Forschungsgemeinschaft (DFG, German Research Foundation) under
Germany’s Excellence Strategy – EXC-2094-390783311
and by the Deutsche Forschungsgemeinschaft (DFG, German Research Foundation)
Projektnummer 417533893/GRK2575 ``Rethinking Quantum Field Theory''.

\appendix

\section{Murua coefficients}\label{murua}

Tree-level diagrams contributing to matrix elements of $\hat{N}$
with different internal causality flows
are weighted by the Murua coefficients~\cite{murua,Kim:2024svw,Brandhuber:2025igz}:

\begin{subequations}\label{muruaPlanar}
\begin{align}
	\omega\left(
	\begin{tikzpicture}[baseline={([yshift=-.5ex](0,0))}]
        \path [draw=black, zParticle] (0,0) -- (.8,0);
        \filldraw[fill=black] (0,0) circle (.06);
		\filldraw[fill=black] (.8,0) circle (.06);
    \end{tikzpicture}
	\right)&=\frac12\,,\label{murua2pt}\\
	\omega\left(
	
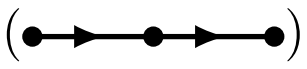
\begin{tikzpicture}[baseline={([yshift=-.5ex](0,0))}]
        \path [draw=black, zParticle] (0,0) -- (.8,0);
		\path [draw=black, zParticle] (.8,0) -- (1.6,0);
        \filldraw[fill=black] (0,0) circle (.06);
		\filldraw[fill=black] (.8,0) circle (.06);
		\filldraw[fill=black] (1.6,0) circle (.06);
    \end{tikzpicture}
	\right)&=\frac13\,,\,\,
	\omega\left(
	
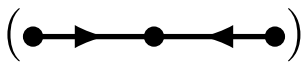
\begin{tikzpicture}[baseline={([yshift=-.5ex](0,0))}]
        \path [draw=black, zParticle] (0,0) -- (.8,0);
		\path [draw=black, zParticle] (1.6,0) -- (.8,0);
        \filldraw[fill=black] (0,0) circle (.06);
		\filldraw[fill=black] (0.8,0) circle (.06);
		\filldraw[fill=black] (1.6,0) circle (.06);
    \end{tikzpicture}
	\right)=\frac16\,,\label{murua3pt}\\
	\omega\left(
	
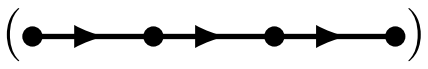
\begin{tikzpicture}[baseline={([yshift=-.5ex](0,0))}]
        \path [draw=black, zParticle] (0,0) -- (0.8,0);
		\path [draw=black, zParticle] (0.8,0) -- (1.6,0);
		\path [draw=black, zParticle] (1.6,0) -- (2.4,0);
        \filldraw[fill=black] (0,0) circle (.06);
		\filldraw[fill=black] (0.8,0) circle (.06);
		\filldraw[fill=black] (1.6,0) circle (.06);
		\filldraw[fill=black] (2.4,0) circle (.06);
    \end{tikzpicture}
	\right)&=\frac14\,,\label{murua4pt}\\
	\omega\left(
	
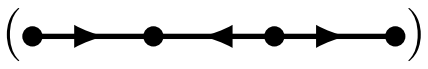
\begin{tikzpicture}[baseline={([yshift=-.5ex](0,0))}]
        \path [draw=black, zParticle] (0,0) -- (0.8,0);
		\path [draw=black, zParticle] (1.6,0) -- (0.8,0);
		\path [draw=black, zParticle] (1.6,0) -- (2.4,0);
        \filldraw[fill=black] (0,0) circle (.06);
		\filldraw[fill=black] (0.8,0) circle (.06);
		\filldraw[fill=black] (1.6,0) circle (.06);
		\filldraw[fill=black] (2.4,0) circle (.06);
    \end{tikzpicture}
	\right)&=
	\omega\left(
	
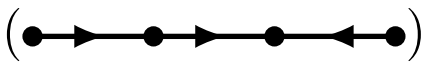
\begin{tikzpicture}[baseline={([yshift=-.5ex](0,0))}]
        \path [draw=black, zParticle] (0,0) -- (0.8,0);
		\path [draw=black, zParticle] (0.8,0) -- (1.6,0);
		\path [draw=black, zParticle] (2.4,0) -- (1.6,0);
        \filldraw[fill=black] (0,0) circle (.06);
		\filldraw[fill=black] (0.8,0) circle (.06);
		\filldraw[fill=black] (1.6,0) circle (.06);
		\filldraw[fill=black] (2.4,0) circle (.06);
    \end{tikzpicture}
	\right)=\frac1{12}\,.\nn
\end{align}
\end{subequations}
These linear Murua coefficients,
whose distinguishing feature is the absence of branches,
are given by binomial coefficients~\cite{Bautista:2026fcp}:
\begin{align}\label{linearMurua}
	\omega(n,k)=\frac1{(n+1)\binom{n}{k}}
	=\frac{k!(n-k)!}{(n+1)!}\,,
\end{align}
where $n$ denotes the total number of internal propagators,
of which $k$ are advanced instead of retarded.
These linear Murua coefficients are associated with planar Feynman diagrams.
The branching Murua coefficients, associated with nonplanar diagrams,
up to three internal propagators are given by
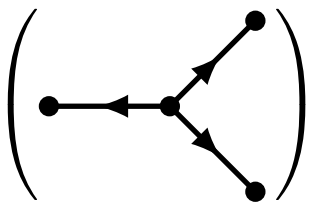
\begin{subequations}\label{muruaNP3}
\begin{align}
	\omega\left(
	\begin{tikzpicture}[baseline={([yshift=-.5ex](0,0))}]
        \path [draw=black, zParticle] (1,0) -- (0.2,0);
		\path [draw=black, zParticle] (1,0) -- (1.5656,0.5656);
		\path [draw=black, zParticle] (1,0) -- (1.5656,-.5656);
        \filldraw[fill=black] (0.2,0) circle (.06);
		\filldraw[fill=black] (1,0) circle (.06);
		\filldraw[fill=black] (1.5656,0.5656) circle (.06);
		\filldraw[fill=black] (1.5656,-.5656) circle (.06);
    \end{tikzpicture}\,
	\right)&=0\,,\\
	\omega\left(
	
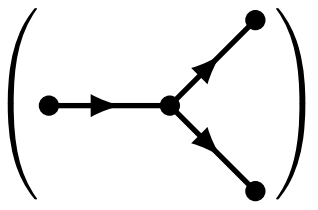
\begin{tikzpicture}[baseline={([yshift=-.5ex](0,0))}]
        \path [draw=black, zParticle] (0.2,0) -- (1,0);
		\path [draw=black, zParticle] (1,0) -- (1.5656,0.5656);
		\path [draw=black, zParticle] (1,0) -- (1.5656,-.5656);
        \filldraw[fill=black] (0.2,0) circle (.06);
		\filldraw[fill=black] (1,0) circle (.06);
		\filldraw[fill=black] (1.5656,0.5656) circle (.06);
		\filldraw[fill=black] (1.5656,-.5656) circle (.06);
    \end{tikzpicture}\,
	\right)&=\frac16\,.
\end{align}
\end{subequations}
With four propagators, we have
\begin{subequations}\label{muruaNP4}
\begin{align}
	\omega\left(
	\begin{tikzpicture}[baseline={([yshift=-.5ex](0,0))}]
		\path [draw=black, zParticle] (-0.6,0) -- (0.2,0);
        \path [draw=black, zParticle] (0.2,0) -- (1,0);
		\path [draw=black, zParticle] (1,0) -- (1.5656,0.5656);
		\path [draw=black, zParticle] (1,0) -- (1.5656,-.5656);
		\filldraw[fill=black] (-0.6,0) circle (.06);
        \filldraw[fill=black] (0.2,0) circle (.06);
		\filldraw[fill=black] (1,0) circle (.06);
		\filldraw[fill=black] (1.5656,0.5656) circle (.06);
		\filldraw[fill=black] (1.5656,-.5656) circle (.06);
    \end{tikzpicture}\,
	\right)&=\frac3{20}\,,\\
	\omega\left(
	\begin{tikzpicture}[baseline={([yshift=-.5ex](0,0))}]
		\path [draw=black, zParticle] (-0.6,0) -- (0.2,0);
        \path [draw=black, zParticle] (0.2,0) -- (1,0);
		\path [draw=black, zParticle] (1.5656,0.5656) -- (1,0);
		\path [draw=black, zParticle] (1,0) -- (1.5656,-.5656);
		\filldraw[fill=black] (-0.6,0) circle (.06);
        \filldraw[fill=black] (0.2,0) circle (.06);
		\filldraw[fill=black] (1,0) circle (.06);
		\filldraw[fill=black] (1.5656,0.5656) circle (.06);
		\filldraw[fill=black] (1.5656,-.5656) circle (.06);
    \end{tikzpicture}\,
	\right)&=\frac1{10}\,,\\
	\omega\left(
	\begin{tikzpicture}[baseline={([yshift=-.5ex](0,0))}]
		\path [draw=black, zParticle] (-0.6,0) -- (0.2,0);
        \path [draw=black, zParticle] (0.2,0) -- (1,0);
		\path [draw=black, zParticle] (1.5656,0.5656) -- (1,0);
		\path [draw=black, zParticle] (1.5656,-.5656) -- (1,0);
		\filldraw[fill=black] (-0.6,0) circle (.06);
        \filldraw[fill=black] (0.2,0) circle (.06);
		\filldraw[fill=black] (1,0) circle (.06);
		\filldraw[fill=black] (1.5656,0.5656) circle (.06);
		\filldraw[fill=black] (1.5656,-.5656) circle (.06);
    \end{tikzpicture}\,
	\right)&=-\frac1{60}\,,\\
	\omega\left(
	\begin{tikzpicture}[baseline={([yshift=-.5ex](0,0))}]
		\path [draw=black, zParticle] (0.2,0) -- (-0.6,0);
        \path [draw=black, zParticle] (0.2,0) -- (1,0);
		\path [draw=black, zParticle] (1,0) -- (1.5656,0.5656);
		\path [draw=black, zParticle] (1,0) -- (1.5656,-.5656);
		\filldraw[fill=black] (-0.6,0) circle (.06);
        \filldraw[fill=black] (0.2,0) circle (.06);
		\filldraw[fill=black] (1,0) circle (.06);
		\filldraw[fill=black] (1.5656,0.5656) circle (.06);
		\filldraw[fill=black] (1.5656,-.5656) circle (.06);
    \end{tikzpicture}\,
	\right)&=\frac1{60}\,,\\
	\omega\left(
	\begin{tikzpicture}[baseline={([yshift=-.5ex](0,0))}]
		\path [draw=black, zParticle] (0.2,0) -- (-0.6,0);
        \path [draw=black, zParticle] (0.2,0) -- (1,0);
		\path [draw=black, zParticle] (1.5656,0.5656) -- (1,0);
		\path [draw=black, zParticle] (1,0) -- (1.5656,-.5656);
		\filldraw[fill=black] (-0.6,0) circle (.06);
        \filldraw[fill=black] (0.2,0) circle (.06);
		\filldraw[fill=black] (1,0) circle (.06);
		\filldraw[fill=black] (1.5656,0.5656) circle (.06);
		\filldraw[fill=black] (1.5656,-.5656) circle (.06);
    \end{tikzpicture}\,
	\right)&=\frac1{15}\,,\\
	\omega\left(
	\begin{tikzpicture}[baseline={([yshift=-.5ex](0,0))}]
		\path [draw=black, zParticle] (0.2,0) -- (-0.6,0);
        \path [draw=black, zParticle] (0.2,0) -- (1,0);
		\path [draw=black, zParticle] (1.5656,0.5656) -- (1,0);
		\path [draw=black, zParticle] (1.5656,-.5656) -- (1,0);
		\filldraw[fill=black] (-0.6,0) circle (.06);
        \filldraw[fill=black] (0.2,0) circle (.06);
		\filldraw[fill=black] (1,0) circle (.06);
		\filldraw[fill=black] (1.5656,0.5656) circle (.06);
		\filldraw[fill=black] (1.5656,-.5656) circle (.06);
    \end{tikzpicture}\,
	\right)&=\frac1{60}\,.
\end{align}
\end{subequations}
These numbers were generated using the \texttt{Mathematica} code
attached to \rcite{Kim:2024svw}. More generally, for a tree $T$ with $n$ propagators, choose vertex
labels $1,\ldots,n+1$ increasing along its arrows. Then~\cite{Brandhuber:2026dkd}
\begin{align}\label{branchingMurua}
    \omega(T)
    =
    \frac{1}{n+1}
    \sum_{\sigma\in\operatorname{LE}(T)}
    \frac{(-1)^{d(\sigma)}}{\binom{n}{d(\sigma)}}\,.
\end{align}
$\operatorname{LE}(T)$ consists of permutations $\sigma=(\sigma_1,\ldots,\sigma_{n+1})$ of the vertex labels in which $i$ appears before $j$ whenever there is an arrow $i\to j$. The descent number $d(\sigma)=\#\{j:\sigma_j>\sigma_{j+1}\}$ counts decreases between adjacent entries. For example, the tree with arrows $1\to2$ and $1\to3$ allows the orderings $(1,2,3)$ and $(1,3,2)$, with zero and one descents respectively, giving
$\omega(T)=\frac13(1-\frac12)=\frac16$.

All Murua coefficients satisfy the \emph{edge contraction rule}:
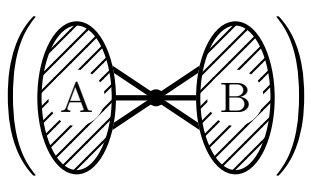
\begin{align}\label{eq:edgecontraction}
\begin{aligned}
\omega\left(\begin{tikzpicture}[scale=.7,baseline={([yshift=-.5ex]current bounding box.center)},thick]
        \draw[zUndirected] (0,0) -- (0.75,0.5);%
        \draw[zUndirected] (0,0) -- (0.75,0.0);%
        \draw[zUndirected] (0,0) -- (0.75,-0.5);%
        \draw[zUndirected] (0,0) -- (-.75,0.5);%
        \draw[zUndirected] (0,0) -- (-.75,0.0);%
        \draw[zUndirected] (0,0) -- (-.75,-0.5);%
        \draw[fill=black] (0,0) circle (.06);
        \draw[fill=white,color=white] (-.75,0) ellipse (10pt and 20pt);
        \draw[fill=white,color=white] (0.75,0) ellipse (10pt and 20pt);
        \draw[pattern=north east lines] (-.75,0) ellipse (10pt and 20pt);
        \draw[pattern=north east lines] (0.75,0) ellipse (10pt and 20pt);
        \filldraw[fill=white,color=white] (-.75,0) circle (7pt);
        \filldraw[fill=white,color=white] (0.75,0) circle (7pt);
        \node (middle) at (-.75,0) {\footnotesize A}; 
        \node (middle) at (0.75,0) {\footnotesize B}; 
\end{tikzpicture}\right)\,=\,\,\,
&\omega\left(
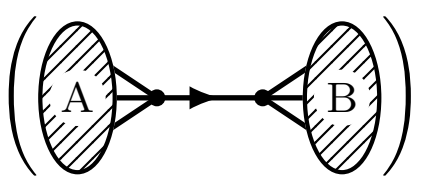
\begin{tikzpicture}[scale=.7,baseline={([yshift=-.5ex]current bounding box.center)},thick]
        \draw[zParticle] (0,0) -- (1,0);
        \draw[zUndirected] (1,0) -- (1.75,0.5);%
        \draw[zUndirected] (1,0) -- (1.75,0.0);%
        \draw[zUndirected] (1,0) -- (1.75,-0.5);%
        \draw[zUndirected] (0,0) -- (-.75,0.5);%
        \draw[zUndirected] (0,0) -- (-.75,0.0);%
        \draw[zUndirected] (0,0) -- (-.75,-0.5);%
        \draw[fill=black] (0,0) circle (.06);
        \draw[fill=black] (1,0) circle (.06);
        \draw[fill=white,color=white] (-.75,0) ellipse (10pt and 20pt);
        \draw[fill=white,color=white] (1.75,0) ellipse (10pt and 20pt);
        \draw[pattern=north east lines] (-.75,0) ellipse (10pt and 20pt);
        \draw[pattern=north east lines] (1.75,0) ellipse (10pt and 20pt);
        \filldraw[fill=white,color=white] (-.75,0) circle (7pt);
        \filldraw[fill=white,color=white] (1.75,0) circle (7pt);
        \node (middle) at (-.75,0) {\footnotesize A}; 
        \node (middle) at (1.75,0) {\footnotesize B}; 
\end{tikzpicture}\right)\\
+\,
&\omega\left(
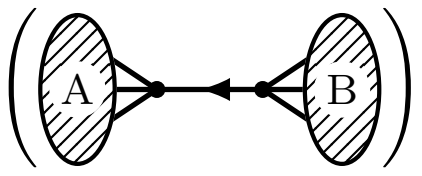
\begin{tikzpicture}[scale=.7,baseline={([yshift=-.5ex]current bounding box.center)},thick]
        \draw[zParticle] (1,0) -- (0,0);
        \draw[zUndirected] (1,0) -- (1.75,0.5);%
        \draw[zUndirected] (1,0) -- (1.75,0.0);%
        \draw[zUndirected] (1,0) -- (1.75,-0.5);%
        \draw[zUndirected] (0,0) -- (-.75,0.5);%
        \draw[zUndirected] (0,0) -- (-.75,0.0);%
        \draw[zUndirected] (0,0) -- (-.75,-0.5);%
        \draw[fill=black] (0,0) circle (.06);
        \draw[fill=black] (1,0) circle (.06);
        \draw[fill=white,color=white] (-.75,0) ellipse (10pt and 20pt);
        \draw[fill=white,color=white] (1.75,0) ellipse (10pt and 20pt);
        \draw[pattern=north east lines] (-.75,0) ellipse (10pt and 20pt);
        \draw[pattern=north east lines] (1.75,0) ellipse (10pt and 20pt);
        \filldraw[fill=white,color=white] (-.75,0) circle (7pt);
        \filldraw[fill=white,color=white] (1.75,0) circle (7pt);
        \node (middle) at (-.75,0) {\footnotesize A}; 
        \node (middle) at (1.75,0) {\footnotesize B}; 
\end{tikzpicture}\right)\,.
\end{aligned}
\end{align}
The sum of any two coefficients,
whose only differing feature is the direction of a single internal arrow,
is the same coefficient with that edge contracted.
Indeed, one can easily check
that $\omega(n,k)=\omega(n+1,k)+\omega(n+1,k+1)$
is satisfied by \eqn{linearMurua};
corresponding identities are also obeyed by the branching (nonplanar)
coefficients~\eqref{branchingMurua}.
As a consequence of unitarity $\hat{N}=\hat{N}^\dagger$,
reversing all internal lines in a Murua coefficient also leaves its value unchanged.
Due to the edge contraction rule,
we need only include the \emph{active propagators} when weighting
diagrams with Murua coefficients,
i.e.~propagators where the $i0^+$ prescription plays a meaningful role
due to the possibility of the propagators going on-shell.

\section{Tail amplitude}\label{app:tailAmplitude}

In this Appendix we derive the soft divergence of the tail term. Writing $q=k_2-k_1$, the leading forward singularity of the connected Compton term is
\begin{align}\label{eq:N1k1k2}
    N^{(1)}(-k_1^{\sigma_1},k_2^{\sigma_2})
    &=
    -16\pi\delta^{\sigma_1\sigma_2}
    \sum_i m_i(k_1\cdot v_i)^2
    \nonumber\\
    &\quad\times
    e^{iq\cdot b_i}
    \frac{\dd(q\cdot v_i)}{q^2}
    +\cdots\,,
\end{align}
where the sum runs over the two worldlines and the omitted terms do not contribute to the IR pole.

In the rest frame of particle $i$, the delta function in \eqn{eq:N1k1k2} fixes $k_2^0=k_1^0=\omega$,
giving $q^2=-2\omega^2(1-\cos\vartheta)$,
where $\vct{k}_1\cdot\vct{k}_2=\omega^2\cos\vartheta$.
We then insert \eqn{eq:N1k1k2} into the quadratic term in \eqn{coherent}.
The pole comes from $k^\mu_2\to k^\mu_1$,
so any factor smooth in this limit can be evaluated at $k^\mu_1$ under the integral, giving
\begin{align}
    &\hat N_{\rm tail}
    =
    G\sum_{\sigma_1,\sigma_2}
    \int_{\vct{k}_1,\vct{k}_2}
    N^{(1)}(-k_1^{\sigma_1},k_2^{\sigma_2}) 
    \hat a^{\sigma_2\dagger}(k_2)
    \hat a^{\sigma_1}(k_1)
    \Bigg|_{1/\eps}
    \nonumber\\
    &=
    \frac{8\pi^2G}{(2\pi)^{3-2\eps}}
    \sum_{i,\sigma}m_i \!
    \int_{\vct{k}} \!
    \omega^{1-2\eps}
    \hat a^{\sigma\dagger}(k)\hat a^\sigma(k) 
    \!\int\!\frac{\d\Omega_{2-2\eps}}{1-\cos\vartheta}
    \Bigg|_{1/\eps}
    \nonumber\\
    &=
    -\frac{2G}{\eps}
    \sum_\sigma\int_{\vct{k}}
    (k\cdot\hat P)
    \hat a^{\sigma\dagger}(k)\hat a^\sigma(k)\,,
\end{align}
where we used
\begin{align}
\begin{aligned}
    \int\frac{\d\Omega_{2-2\eps}}{1-\cos\vartheta}
    &=
    \Omega_{1-2\eps}\,2^{-2\eps}
    B(-\eps,1-\eps)\\
    &=-\frac{2\pi}{\eps}+\cO(\eps^0)\,.
\end{aligned}
\end{align}
$B(x,y)$ is the Euler beta function and $\Omega_d=2\pi^{(d+1)/2}/\Gamma((d+1)/2)$. In the final expression we used $m_i\omega=p_i\cdot k$ and restored the worldline momentum operators, with $\hat P^\mu=\hat p_1^\mu+\hat p_2^\mu$. Equivalently, at fixed worldline momenta the Compton pole is
\begin{align}
    \left.
    N^{(1)}(-k_1^{\sigma_1},k_2^{\sigma_2})
    \right|_{1/\eps}
    &\hspace{-5pt}=
    -\frac{2}{\eps}\delta^{\sigma_1\sigma_2}
    (k_1\cdot P)\,
    2k_1^0\dd(\vct{k}_1-\vct{k}_2)\,,
\end{align}
where $\dd(\vct{q})=(2\pi)^{D-1}\delta^{(D-1)}(\vct{q})$.

\section{Brackets of $\hat{N}$-matrix elements}\label{bracketsApp}

In this Appendix we derive the two identities in \eqn{bracketsGammaP}
involving Poisson brackets of radiative $\hat{N}$-matrix elements
with $\gamma$ and the total momentum $P^\mu$.
Any $\hat N$-matrix element can be written as a Fourier transform
with respect to the momenta emitted by each worldline $q_i^\mu$:
\begin{align}
\begin{aligned}
    &N(k_1^{\sigma_1},\ldots,k_n^{\sigma_n})\\
    &=
    \int_{q_1,q_2}
    \dd(q_1\cdot v_1)\dd(q_2\cdot v_2)
    \dd\Big(q_1^\mu+q_2^\mu-\sum_ik_i^\mu\Big)\\[-5pt]
    &\qquad\quad\times
    e^{i(q_1\cdot b_1+q_2\cdot b_2)}
    I^{\sigma_1\cdots\sigma_n}(q_1,q_2,\{k_i\})\,.
\end{aligned}
\end{align}
This exposes the dependence on the impact parameters $b_i^\mu$.
For the Magnusian $N$, $\sum_ik_i^\mu=0$, so $q_1^\mu=-q_2^\mu$
and $N$ depends only on the relative impact parameter $b^\mu=b_2^\mu-b_1^\mu$.
Taking the bracket with $v_1^\mu$,
\begin{align}
\begin{aligned}
    &\{N(k_1^{\sigma_1},\ldots,k_n^{\sigma_n}),v_1^\mu\}\\
    &=
    \int_{q_1,q_2}
    \dd(q_1\cdot v_1)\dd(q_2\cdot v_2)
    \dd\Big(q_1^\mu+q_2^\mu-\sum_ik_i^\mu\Big)\\[-5pt]
    &\qquad\times
    e^{i(q_1\cdot b_1+q_2\cdot b_2)}\frac{(-iq_1^\mu)}{m_1}
    I^{\sigma_1\cdots\sigma_n}(q_1,q_2,\{k_i\})\,.
\end{aligned}
\end{align}
We contract this result with $v_2^\mu$ and add the corresponding
term with $v_1^\mu\leftrightarrow v_2^\mu$:
\begin{align}
\begin{aligned}
    \{N(k_1^{\sigma_1},\ldots,k_n^{\sigma_n}),\gamma\}&=
    v_{2,\mu}\{N(k_1^{\sigma_1},\ldots,k_n^{\sigma_n}),v_1^\mu\}\\
    &\,+
    v_{1,\mu}\{N(k_1^{\sigma_1},\ldots,k_n^{\sigma_n}),v_2^\mu\}
\end{aligned}
\end{align}
Using the constraints $q_1\cdot v_1=q_2\cdot v_2=0$
plus momentum conservation $q_1^\mu+q_2^\mu=\sum_{i=1}^nk_i^\mu$
under the Fourier integrals, we rearrange this into the
equation stated in the main text~\eqref{bracketGamma}.
Similarly, using
\begin{align}
\begin{aligned}
    \{N(k_1^{\sigma_1},\ldots,k_n^{\sigma_n}),P^\mu\}\!&=
    \!m_1\{N(k_1^{\sigma_1},\ldots,k_n^{\sigma_n}),v_1^\mu\}\\
    &+
    m_2\{N(k_1^{\sigma_1},\ldots,k_n^{\sigma_n}),v_2^\mu\}
\end{aligned}
\end{align}
we obtain the other relationship involving the total momentum $P^\mu$~\eqref{bracketP}.

\onecolumngrid

\section{Components of the 3PM radiated energy}\label{radE}
Here we list the spin components ($0$, $+$, $-$) of the 3PM radiated energy~\eqref{radE3PM}:
\begingroup
\small
\begin{align*}
E_{{\rm rad},0} &= \frac{\pi}{48\gamma_-^2\,\gamma_+^2}\bigg[\sqrt{\gamma^2-1}\,\Big(1151-3336\gamma+3148\gamma^2-912\gamma^3+339\gamma^4-552\gamma^5+210\gamma^6\Big)\\
&\quad +6\sqrt{\gamma^2-1}\,\gamma_-\gamma_+\big(5-76\gamma+150\gamma^2-60\gamma^3-35\gamma^4\big)\log\!\Big[\frac{\gamma_+}{2}\Big]
+3\gamma\big(33-112\gamma^2+165\gamma^4-70\gamma^6\big)\log[x]\bigg]\,,\displaybreak[0]\\[2pt]
E_{{\rm rad},+} &= \frac{\pi}{32\gamma_-\,\gamma_+^2}\bigg[2597-8984\gamma+10125\gamma^2-2977\gamma^3-330\gamma^4-635\gamma^5+420\gamma^6\\
&\quad +6\gamma_-\gamma_+\big(28-89\gamma+247\gamma^2-203\gamma^3-63\gamma^4\big)\log\!\Big[\frac{\gamma_+}{2}\Big]
-27\gamma^2\gamma_+\big(9-27\gamma^2+14\gamma^4\big)\frac{\log[x]}{\sqrt{\gamma^2-1}}\bigg]\,,\displaybreak[0]\\[2pt]
E_{{\rm rad},-} &= \frac{\pi}{32\gamma_-\,\gamma_+^2}\bigg[107-498\gamma+661\gamma^2-277\gamma^3+108\gamma^4-77\gamma^5\\
&\quad +6\gamma_-\gamma_+\big(4-9\gamma+23\gamma^2-27\gamma^3-7\gamma^4\big)\log\!\Big[\frac{\gamma_+}{2}\Big]
-3\gamma^2\gamma_+\big(9-27\gamma^2+14\gamma^4\big)\frac{\log[x]}{\sqrt{\gamma^2-1}}\bigg]\,.
\end{align*}
\endgroup
\section{Components of the 4PM Magnusian}\label{pol}
Here we list all algebraic functions $p_{X,i}(\gamma)$ appearing in the
4PM linear-in-spin Magnusian~\eqref{1SFresult}, with $\gamma_\pm=\gamma\pm1$:
\providecommand{\sq}{\sqrt{\gamma^2-1}}
\begingroup
\footnotesize
\medmuskip=2mu plus 1mu minus 1mu
\thickmuskip=3mu plus 2mu minus 1mu
\setlength{\jot}{2pt}
\begin{align*}
p_{0,1} &= -\frac{1}{1728\gamma^7(\gamma^2-1)^3}\bigg[1224\gamma^7-195885\gamma^8+556524\gamma^9-491038\gamma^{10}-66528\gamma^{11}+449743\gamma^{12}-286044\gamma^{13}\\
&\quad -13992\gamma^{14}+25224\gamma^{15}+17316\gamma^{16}+\log2\big(-82872\gamma^8+240192\gamma^9-171408\gamma^{10}-94464\gamma^{11}+126696\gamma^{12} -4032\gamma^{13}+1152\gamma^{14}\\
&\quad -26496\gamma^{15}+10080\gamma^{16}\big)+\sq\,\bigl(-90+414\gamma^2-2942\gamma^4+26698\gamma^6+117667\gamma^7 -391271\gamma^8+291007\gamma^9+289417\gamma^{10}-602555\gamma^{11}\\
&\quad +223177\gamma^{12}+95421\gamma^{13}-44603\gamma^{14}-17316\gamma^{15} +14400\gamma^{16}+\pi^2\big(-864\gamma^7+5472\gamma^8-7560\gamma^9-6624\gamma^{10}+19872\gamma^{11}-3168\gamma^{12}\\
&\quad -15408\gamma^{13}+4320\gamma^{14}+5760\gamma^{15} -1800\gamma^{17}\big)\bigr)\bigg],\displaybreak[0]\\
p_{0,2} &= -\frac{1}{576(\gamma^2-1)^3}\bigg[4320\gamma-6975\gamma^2-37440\gamma^3+36186\gamma^4+96480\gamma^5-73395\gamma^6-92160\gamma^7+58692\gamma^8+28800\gamma^9\\
&\quad -15852\gamma^{10}+\log2\big(-2376\gamma^2+9648\gamma^4-17256\gamma^6+12960\gamma^8-3360\gamma^{10}\big)+\sq\,\bigl(-2474-24186\gamma +92414\gamma^2-78758\gamma^3-57324\gamma^4\\
&\quad +91862\gamma^5+23176\gamma^6-45234\gamma^7-17392\gamma^8+17532\gamma^9\bigr)\bigg],\displaybreak[0]\\
p_{0,3} &= -\frac{1}{288(\gamma^2-1)^3}\bigg[1488+10008\gamma-52980\gamma^2+59029\gamma^3+38280\gamma^4-86860\gamma^5-828\gamma^6+40713\gamma^7-360\gamma^8\\
&\quad -8346\gamma^9 +\log2\big(360\gamma-5472\gamma^2+10200\gamma^3+4800\gamma^4-20280\gamma^5+3552\gamma^6+11400\gamma^7-2880\gamma^8-1680\gamma^9\big) +\sq\,\bigl(-14912\\
&\quad +27888\gamma-37274\gamma^2-26768\gamma^3+71950\gamma^4-9392\gamma^5-28110\gamma^6+8272\gamma^7+8346\gamma^8\bigr)\bigg],\displaybreak[0]\\
p_{0,4} &= \frac{\gamma\sq}{48(\gamma^2-1)^4}\big[152+99\gamma-336\gamma^2-402\gamma^3+96\gamma^4+719\gamma^5+208\gamma^6-540\gamma^7-120\gamma^8+140\gamma^9\big],\displaybreak[0]\\
p_{0,5} &= -\frac{\gamma}{96(\gamma^2-1)^3}\bigg[99\gamma-402\gamma^3+719\gamma^5-540\gamma^7+140\gamma^9+\sq\,\bigl(-6+1990\gamma^2-3934\gamma^4+2370\gamma^6-420\gamma^8\bigr)\bigg],\displaybreak[0]\\
p_{0,6} &= \frac{\gamma\sq}{9(\gamma^2-1)^2}\big[852+283\gamma^2+140\gamma^4-75\gamma^6\big],\displaybreak[0]\\
p_{0,7} &= \frac{1}{24(\gamma^2-1)^2}\bigg[33\gamma-112\gamma^3+165\gamma^5-70\gamma^7+\sq\,\bigl(-22+76\gamma+82\gamma^2-16\gamma^3-130\gamma^4-60\gamma^5+70\gamma^6\bigr)\bigg],\displaybreak[0]\\
p_{0,8} &= \frac{1}{12(\gamma^2-1)^2}\bigg[81\gamma-228\gamma^2+216\gamma^3-28\gamma^4-75\gamma^5+120\gamma^6-70\gamma^7+\sq\,\bigl(-94+304\gamma-308\gamma^2-64\gamma^3+272\gamma^4 -240\gamma^5\\
&\quad +180\gamma^6-50\gamma^8\bigr)\bigg],\displaybreak[0]\\
p_{0,9} &= -\frac{\sq}{6(\gamma^2-1)}\big[20+111\gamma^2+30\gamma^4-25\gamma^6\big],\displaybreak[0]\\
p_{0,10} &= \frac{\gamma}{24(\gamma^2-1)^3}\bigg[99\gamma-402\gamma^3+719\gamma^5-540\gamma^7+140\gamma^9+\sq\,\bigl(-354+1368\gamma-1390\gamma^2-1200\gamma^3+2854\gamma^4-888\gamma^5\\
&\quad -970\gamma^6+720\gamma^7-140\gamma^8\bigr)\bigg],\displaybreak[0]\\
p_{0,11} &= \frac{\gamma}{48(\gamma^2-1)^3}\bigg[99\gamma-402\gamma^3+719\gamma^5-540\gamma^7+140\gamma^9+\sq\,\bigl(-66+290\gamma^2-554\gamma^4+470\gamma^6-140\gamma^8\bigr)\bigg],\displaybreak[0]\\
p_{0,12} &= -\frac{7\sq}{32\gamma_-^2\,\gamma_+}\big[169+380\gamma^2\big],\displaybreak[0]\\
p_{0,13} &= -\frac{\sq}{16(\gamma^2-1)^2}\big[834+2095\gamma+1200\gamma^2\big],\displaybreak[0]\\
p_{0,14} &= \frac{\sq}{16(\gamma^2-1)^2}\big[1183+2929\gamma+2660\gamma^2+1200\gamma^3\big],\displaybreak[0]\\
p_{+,1} &= -\frac{1}{2304\gamma^8\,\gamma_-^3\,\gamma_+^4}\bigg[630+630\gamma-4400\gamma^2-4400\gamma^3+20772\gamma^4+20772\gamma^5-84344\gamma^6-84344\gamma^7-64528\gamma^8\\
&\quad +2732862\gamma^9-2163610\gamma^{10}-4308922\gamma^{11}+5648246\gamma^{12}-185118\gamma^{13}-3500678\gamma^{14}+2288986\gamma^{15} -57544\gamma^{16}-762066\gamma^{17}\\
&\quad +293656\gamma^{18}+389800\gamma^{19}-88200\gamma^{20}-88200\gamma^{21}+\pi^2\big(4032\gamma^8-15975\gamma^9 +20313\gamma^{10}+20610\gamma^{11}-92286\gamma^{12}+50895\gamma^{13}\\
&\quad +131535\gamma^{14}-127620\gamma^{15}-115524\gamma^{16}+111015\gamma^{17} +90855\gamma^{18}-49950\gamma^{19}-49950\gamma^{20}+11025\gamma^{21}+11025\gamma^{22}\big)+\sq\,\bigl(-49360\gamma^8\\
&\quad -1128058\gamma^9 +3824572\gamma^{10}-3329162\gamma^{11}-813482\gamma^{12}+2805436\gamma^{13}-1677542\gamma^{14}+12296\gamma^{15}+170356\gamma^{16}+96144\gamma^{17}\\
&\quad +\log2\big(-373968\gamma^9+1293696\gamma^{10}-1208688\gamma^{11}-433776\gamma^{12}+1019520\gamma^{13}-194352\gamma^{14}-92160\gamma^{15} -60960\gamma^{16}\\
&\quad +40320\gamma^{17}\big)\bigr)\bigg],\displaybreak[0]\\
p_{+,2} &= \frac{1}{384(\gamma^2-1)^3}\bigg[160-49606\gamma+267034\gamma^2-477184\gamma^3+200566\gamma^4+344250\gamma^5-343062\gamma^6-57564\gamma^7\\
&\quad +115298\gamma^8+36840\gamma^9-36732\gamma^{10}+\sq\,\bigl(480\gamma+25920\gamma^2-28643\gamma^3-125760\gamma^4+106839\gamma^5 +168320\gamma^6-108696\gamma^7-64000\gamma^8\\
&\quad +33372\gamma^9+\log2\big(-5832\gamma^3+21384\gamma^5-20736\gamma^7+6048\gamma^9\big)\bigr)\bigg],\displaybreak[0]\\
p_{+,3} &= -\frac{1}{192\gamma_-^3\,\gamma_+^4}\bigg[18982-133326\gamma-33588\gamma^2+73998\gamma^3-209542\gamma^4+102402\gamma^5+270758\gamma^6-84182\gamma^7\\
&\quad -64136\gamma^8+41108\gamma^9+17526\gamma^{10}+\sq\,\bigl(1200+22329\gamma-101765\gamma^2+155474\gamma^3+9455\gamma^4-217676\gamma^5 +76630\gamma^6+95835\gamma^7\\
&\quad -23308\gamma^8-17526\gamma^9+\log2\big(2016\gamma-6408\gamma^2+14424\gamma^3-3936\gamma^4-32832\gamma^5 +20088\gamma^6+19416\gamma^7-9744\gamma^8-3024\gamma^9\big)\bigr)\bigg],\displaybreak[0]\\
p_{+,4} &= -\frac{1}{32(\gamma^2-1)^3}\big[56+560\gamma^2+243\gamma^3-1008\gamma^4-891\gamma^5+112\gamma^6+864\gamma^7+280\gamma^8-252\gamma^9\big],\displaybreak[0]\\
p_{+,5} &= -\frac{9\gamma^2}{64(\gamma^2-1)^3}\bigg[138+430\gamma^2-1134\gamma^4+650\gamma^6-84\gamma^8+\sq\,\bigl(-27\gamma+99\gamma^3-96\gamma^5+28\gamma^7\bigr)\bigg],\displaybreak[0]\\
p_{+,6} &= \frac{1}{96(\gamma^2-1)^2}\big[11200+74445\gamma^2+10342\gamma^4+13720\gamma^6-11750\gamma^8+3675\gamma^{10}\big],\displaybreak[0]\\
p_{+,7} &= \frac{1}{16(\gamma^2-1)^2}\bigg[28-54\gamma+308\gamma^2+234\gamma^3-196\gamma^4-306\gamma^5-140\gamma^6+126\gamma^7+\sq\,\bigl(-81\gamma^2+243\gamma^4-126\gamma^6\bigr)\bigg],\displaybreak[0]\\
p_{+,8} &= \frac{1}{64\gamma_-^2\,\gamma_+^3}\bigg[896-4225\gamma+4735\gamma^2+5773\gamma^3-10355\gamma^4-746\gamma^5+1046\gamma^6+2298\gamma^7+6778\gamma^8-4325\gamma^9-4325\gamma^{10}\\
&\quad +1225\gamma^{11}+1225\gamma^{12}+\sq\,\bigl(672\gamma-3432\gamma^2+4184\gamma^3+440\gamma^4-1576\gamma^5+1232\gamma^6-1008\gamma^7\bigr)\bigg],\displaybreak[0]\\
p_{+,9} &= -\frac{\gamma}{64(\gamma^2-1)}\big[2385+4884\gamma^2+2670\gamma^4-3100\gamma^6+1225\gamma^8\big],\displaybreak[0]\\
p_{+,10} &= \frac{3\gamma}{16(\gamma^2-1)^3}\bigg[168-918\gamma+1736\gamma^2-450\gamma^3-2408\gamma^4+2562\gamma^5-56\gamma^6-1110\gamma^7+560\gamma^8-84\gamma^9 +\sq\,\bigl(-81\gamma^2\\
&\quad +297\gamma^4-288\gamma^6+84\gamma^8\bigr)\bigg],\displaybreak[0]\\
p_{+,11} &= -\frac{9\gamma^2}{32(\gamma^2-1)^3}\bigg[18-90\gamma^2+154\gamma^4-110\gamma^6+28\gamma^8+\sq\,\bigl(27\gamma-99\gamma^3+96\gamma^5-28\gamma^7\bigr)\bigg],\displaybreak[0]\\
p_{+,12} &= -\frac{5\gamma}{32\gamma_-^2\,\gamma_+}\big[3539+3509\gamma^2+24\gamma^4\big],\displaybreak[0]\\
p_{+,13} &= -\frac{1}{32(\gamma^2-1)^2}\big[5167+22164\gamma+31705\gamma^2+13740\gamma^3+120\gamma^4\big],\displaybreak[0]\\
p_{+,14} &= \frac{1}{32(\gamma^2-1)^2}\big[5167+35390\gamma+53869\gamma^2+35090\gamma^3+13860\gamma^4+240\gamma^5\big],\displaybreak[0]\\
p_{-,1} &= \frac{1}{2304\gamma^8\,\gamma_-^3\,\gamma_+^4}\bigg[210+210\gamma-400\gamma^2-400\gamma^3-436\gamma^4-436\gamma^5+3608\gamma^6+3608\gamma^7+17252\gamma^8-159378\gamma^9\\
&\quad +71946\gamma^{10}+172282\gamma^{11}-84070\gamma^{12}+340190\gamma^{13}-385738\gamma^{14}-745802\gamma^{15}+679228\gamma^{16}+691326\gamma^{17} -389800\gamma^{18}-389800\gamma^{19}\\
&\quad +88200\gamma^{20}+88200\gamma^{21}+\pi^2\big(-576\gamma^8+999\gamma^9-1881\gamma^{10}+7038\gamma^{11}+16254\gamma^{12} -43983\gamma^{13}-48591\gamma^{14}+90756\gamma^{15}+86724\gamma^{16}\\
&\quad -93735\gamma^{17}-90855\gamma^{18}+49950\gamma^{19}+49950\gamma^{20}-11025\gamma^{21} -11025\gamma^{22}\big)+\sq\,\bigl(5440\gamma^8+45718\gamma^9-180192\gamma^{10}+181842\gamma^{11}\\
&\quad -38646\gamma^{12}-84336\gamma^{13}+148722\gamma^{14} -53128\gamma^{15}-16492\gamma^{16}+\log2\big(15408\gamma^9-71712\gamma^{10}+84912\gamma^{11}+7920\gamma^{12}-47904\gamma^{13}\\
&\quad +15504\gamma^{14} -10368\gamma^{15}+7392\gamma^{16}\big)\bigr)\bigg],\displaybreak[0]\\
p_{-,2} &= \frac{1}{384(\gamma^2-1)^3}\bigg[112-682\gamma+13322\gamma^2-34464\gamma^3+19430\gamma^4+22870\gamma^5-23430\gamma^6-5284\gamma^7+7202\gamma^8+3736\gamma^9\\
&\quad -2812\gamma^{10}+\sq\,\bigl(336\gamma+2880\gamma^2-3227\gamma^3-12480\gamma^4+9343\gamma^5+16640\gamma^6-9112\gamma^7-6400\gamma^8+2812\gamma^9 +\log2\big(-648\gamma^3+2376\gamma^5\\
&\quad -2304\gamma^7+672\gamma^9\big)\bigr)\bigg],\displaybreak[0]\\
p_{-,3} &= -\frac{1}{192\gamma_-^3\,\gamma_+^4}\bigg[1338-7818\gamma-764\gamma^2+1726\gamma^3-17774\gamma^4+11034\gamma^5+20742\gamma^6-9126\gamma^7-4948\gamma^8+4184\gamma^9\\
&\quad +1406\gamma^{10}+\sq\,\bigl(72+847\gamma-6983\gamma^2+12742\gamma^3-993\gamma^4-17318\gamma^5+7892\gamma^6+7763\gamma^7-2544\gamma^8-1406\gamma^9 +\log2\big(288\gamma-648\gamma^2\\
&\quad +1176\gamma^3-864\gamma^4-3072\gamma^5+2808\gamma^6+1944\gamma^7-1296\gamma^8-336\gamma^9\big)\bigr)\bigg],\displaybreak[0]\\
p_{-,4} &= -\frac{1}{32(\gamma^2-1)^3}\big[8+48\gamma^2+27\gamma^3-80\gamma^4-99\gamma^5-16\gamma^6+96\gamma^7+40\gamma^8-28\gamma^9\big],\displaybreak[0]\\
p_{-,5} &= -\frac{\gamma^2}{64(\gamma^2-1)^3}\bigg[138+430\gamma^2-1134\gamma^4+650\gamma^6-84\gamma^8+\sq\,\bigl(-27\gamma+99\gamma^3-96\gamma^5+28\gamma^7\bigr)\bigg],\displaybreak[0]\\
p_{-,6} &= \frac{1}{96(\gamma^2-1)^2}\big[448+5837\gamma^2-5018\gamma^4+13720\gamma^6-11750\gamma^8+3675\gamma^{10}\big],\displaybreak[0]\\
p_{-,7} &= \frac{1}{16(\gamma^2-1)^2}\bigg[4-6\gamma+28\gamma^2+26\gamma^3-12\gamma^4-34\gamma^5-20\gamma^6+14\gamma^7+\sq\,\bigl(-9\gamma^2+27\gamma^4-14\gamma^6\bigr)\bigg],\displaybreak[0]\\
p_{-,8} &= -\frac{1}{64\gamma_-\,\gamma_+^3}\bigg[128-385\gamma-130\gamma^2+1163\gamma^3+1176\gamma^4-1746\gamma^5-4924\gamma^6+318\gamma^7+6200\gamma^8+1875\gamma^9-2450\gamma^{10}\\
&\quad -1225\gamma^{11}+\sq\,\bigl(96\gamma-264\gamma^2+80\gamma^3+8\gamma^4-96\gamma^5+112\gamma^6\bigr)\bigg],\displaybreak[0]\\
p_{-,9} &= -\frac{\gamma}{64(\gamma^2-1)}\big[337-236\gamma^2+2670\gamma^4-3100\gamma^6+1225\gamma^8\big],\displaybreak[0]\\
p_{-,10} &= \frac{\gamma}{16(\gamma^2-1)^3}\bigg[72-306\gamma+456\gamma^2-150\gamma^3-552\gamma^4+854\gamma^5-216\gamma^6-370\gamma^7+240\gamma^8-28\gamma^9+\sq\,\bigl(-27\gamma^2+99\gamma^4\\
&\quad -96\gamma^6+28\gamma^8\bigr)\bigg],\displaybreak[0]\\
p_{-,11} &= -\frac{\gamma^2}{32(\gamma^2-1)^3}\bigg[18-90\gamma^2+154\gamma^4-110\gamma^6+28\gamma^8+\sq\,\bigl(27\gamma-99\gamma^3+96\gamma^5-28\gamma^7\bigr)\bigg],\displaybreak[0]\\
p_{-,12} &= -\frac{\gamma}{32\gamma_-^2\,\gamma_+}\big[2113+1135\gamma^2-120\gamma^4\big],\displaybreak[0]\\
p_{-,13} &= -\frac{1}{32\gamma_-^2\,\gamma_+}\big[733+1735\gamma+780\gamma^2-120\gamma^3\big],\displaybreak[0]\\
p_{-,14} &= \frac{1}{32\gamma_-^2\,\gamma_+}\big[733+3493\gamma+1490\gamma^2+780\gamma^3-240\gamma^4\big].
\end{align*}
\endgroup

\twocolumngrid
\bibliographystyle{JHEP}
\bibliography{refs}

\end{document}